\documentclass[reprint,amsmath,amssymb,prb,showpacs]{revtex4-1}
\usepackage{bm,bbm}
\usepackage{graphicx}
\usepackage{hyperref}
\usepackage[caption=false]{subfig}

\newcommand{\me}{\mathrm{e}}
\newcommand{\mi}{\mathrm{i} \,}
\newcommand{\dif}{\mathrm{d}}

\begin{document}
\title{Coulomb interactions, Dirac sea polarization and $SU(4)$ symmetry 
breaking of the integer quantum Hall states of graphene}

\author{Vinu Lukose}
\email{vinu@imsc.res.in}
\author{R. Shankar}
\email{shankar@imsc.res.in}
\affiliation{The Institute of Mathematical Sciences, \\
C.I.T. Campus, Taramani, Chennai 600 113, India}

\date{\today}

\begin{abstract}

We investigate effects of the filled Dirac sea on the $SU(4)$ symmetry
breaking in the integer quantum Hall states of graphene with long-ranged
Coulomb interactions. Our model also includes Hubbard and nearest neighbour
repulsive interactions with strengths $U$ and $V$ respectively. We find that
the symmetry breaking of the $n=0$ Landau levels induces an $SU(4)$
polarization of the Dirac sea. This results in several phases which are absent
when the effects of the Dirac sea are neglected. We compute the phase diagram
in the $U$-$V$ space. We also calculate the excitation gaps in tilted magnetic
fields for all the phases. We compare our model results with experiments and
find a range of $U$ and $V$ that are consistent with them.

\end{abstract}


\maketitle

\section{Introduction} 

Graphene\cite{geim1} has given access to a truly two dimensional electronic
gas. The low energy physics for graphene is governed by the four species of
Dirac quasi-particles leading to an emergent $SU(4)$ symmetry. Dirac particles
of the low energy and the long wavelength theory were predicted\cite{gusynin}
to have an unusual sequence of quantum Hall plateaus for graphene, $\sigma_{H}
= 4(n + 1/2) e^{2}/h$ and it was observed\cite{geim, kim}. The charge neutral
point of graphene in a magnetic field is the filled Dirac sea of quartets of
Landau levels with index $n<0$ and half-filled quartet of the $n=0$ Landau
level. The orbital degeneracy of each member of the quartet is always
saturated in this article. The Dirac sea manifests in the non-interacting
model in the unusual form of the Hall conductivity mentioned above.

High mobility graphene samples on boron nitride unveiled plateaus at all
integer values\cite{kim.nphys} indicating that the $SU(4)$ symmetry is
spontaneously broken. The theory for the $SU(4)$ symmetry breaking near the
charge neutral point has been addressed in past\cite{nomura, kun.yang, alicea,
gusynin1, fuchs, goerbig, herbut, maxim, yacoby, jung, herbut2}. The long-ranged
Coulomb interaction was shown to break the $SU(4)$ symmetry
spontaneously\cite{nomura} and it was predicted that the Hall plateaus should
reveal for all integer values. The interactions can be written as a sum of 
$SU(4)$ symmetric long-ranged Coulomb part and short-ranged symmetry breaking
parts. The long-ranged Coulomb interaction results in the spontaneous breaking
of the $SU(4)$ symmetry. The pattern of symmetry breaking is $U(4) \rightarrow
U(m) \times U(4-m)$\cite{kun.yang}, where $m<4$ is the number of members of
the quartet of occupied Landau level with the highest energy. The ground state
manifold is the coset space $U(4)/(U(m) \times U(4-m)$. 

The symmetry breaking terms pick out a state (or states) in the ground state
manifold.  The short-ranged interaction and the Zeeman terms play an important
role in deciding the nature of the ground state at Hall conductivities
$\sigma_{H} = 0, \pm 1$. The ground state at $\sigma_{H} = \pm 1$ was shown to
be a valley-spin polarized state\cite{alicea}. The ground state at $\sigma_{H}
= 0$, there are various possibilities: charge ordered state\cite{alicea,
herbut, maxim, yacoby}, anti-ferromagnetic state\cite{herbut, maxim}, canted
anti-ferromagnetic state\cite{maxim, herbut2} and Kekule ordering\cite{maxim}. The
filled Dirac sea of Landau levels with the short-ranged interaction terms were
first included in the computation of ground state energies by
Herbut\cite{herbut}.

In all these works, the effects of the long-ranged Coulomb interaction of the
Dirac sea and the consequences for the $SU(4)$ symmetry breaking have not been
investigated. We expect the long-ranged Coulomb interactions to significantly
affect the structure of the Dirac sea. In this paper we take up this
technically challenging problem of including the Dirac sea with the Coulomb
interactions. 

We start with a realistic model on the lattice that includes the Coulomb
interaction between the point charges as well as the on-site and the
nearest-neighbour repulsion to account for the finite extent of the
wavefunctions. We perform a systematic long wavelength expansion and derive
the effective low energy continuum model. This model consists of an $SU(4)$
symmetric part and $SU(4)$ breaking terms. The symmetric part includes the
kinetic term and the long-ranged Coulomb interactions. The symmetry breaking
part consists of short-ranged interaction terms and the Zeeman term and these
are suppressed by a factor of $a/\ell_{c}$ with respect to the symmetric
terms, where $a$ is the graphene lattice constant and $\ell_{c} \equiv
\sqrt{\hbar / e B}$, is the magnetic length. 

Since we are interested in the regime where $a / \ell_{c} \ll 1$, we adopt the
strategy of approximately solving the symmetric part using the variational
method and then treating the symmetry breaking terms in first order
perturbation theory. The variational state we choose is the ground state of a
quartet of Dirac particles in the background of a mass matrix. The sixteen
parameters that specify the hermitain mass matrix are our variational
parameters. We develop a method, based on the `heat kernel' representation of
the Dirac propagator, which makes the computation of the Coulomb energy
tractable.

The picture that emerges from our calculations is the following: consider the
non-interacting system with partially filled quartet for the $n=0$ Landau
level and fully filled quartet for the $n<0$ Landau levels. We will refer to
the $n<0$ occupied Landau levels as the Dirac sea. The Dirac sea is a $SU(4)$
singlet and has all four charges uniformly distributed on the sites. The
partially filled quartet of the $n=0$ Landau level breaks the $SU(4)$ symmetry
and the four charges are distributed unequally between the two sublattices of
the honeycomb lattice. When the Coulomb interactions are turned on, this
induces a redistribution of the charges in the sea, leading to what we call a
staggered polarization of the Dirac sea which plays an important role in the
nature of the broken symmetry phases. We show that it results in two phases
which are absent if this effect is ignored. One is the canted-spin state at
$\sigma_H=0$, which was first discussed in reference [\onlinecite{maxim}].
Another is what we call the valley-spin canted phase which occurs at
$\sigma_H=-1$, which has not been discussed earlier.

The rest of this paper is organized as follows. In Sec.\ref{sec:model} we
start with a realistic lattice model and motivate the effective continuum
interacting model adopted in this paper. Sec.\ref{sec:symmetries} is a
discussion of symmetries of the hamiltonian and the order parameters
corresponding to the $SU(4)$ polarization and the staggered polarization. In
Sec.\ref{sec:variational wftn}, we describe the variational wavefunction used
to compute the variational energy for the interacting model.
Sec.\ref{sec:correlation} gives the details of the technical aspects for
computing the variational energy that takes into account the filled sea of
Landau levels by computing the two point correlators. The two point
correlators are used to describe the nature of the polarization of the Dirac
sea in Sec.\ref{sec:polarisation}. In Sec.\ref{sec:symmetric model}, we
present the results for the symmetric interacting model, i.e. model comprising
the kinetic and the long-ranged Coulomb interaction terms.
Sec.\ref{sec:symmetry breaking} presents the effect of symmetry breaking terms
of our model on the nature of the ground states and the elementary excitations
by specifying the $SU(4)$ polarization. In Sec.\ref{sec:expt}, we compare our
results to recent experiments\cite{kim.nphys,yacoby} and find that there is a
region in our phenomenological parameters, $U$ and $V$,  that is consistent
with the experimental results. We conclude with a summary of the main results
of the paper in Sec.\ref{sec:summary}.

\section{The interacting model} \label{sec:model}
In this section, we describe the motivation for adopting the interacting
continuum model in this article. We start with the lattice model. The
hamiltonian is, 
\begin{equation} \label{latt:model} 
	\mathcal{H}=\mathcal{H}_{0} + \mathcal{H}_{I},
\end{equation}
where $\mathcal{H}_{0} \equiv \mathcal{H}_{t} + \mathcal{H}_{Z}$, is the
non-interacting model for graphene in a magnetic field which includes the
tight binding term, $\mathcal{H}_{t}$ and the Zeeman term, $\mathcal{H}_{Z}$.
$\mathcal{H}_{I} \equiv \mathcal{H}_{\tilde{C}} + \mathcal{H}_{\tilde{V}} +
\mathcal{H}_{U}$, is the interaction term. $\mathcal{H}_{\tilde{C}}$ is the
Coulomb interaction between a pair of point charges on lattice sites.  The
finite extent of the electron wavefunctions modify the interaction at short
distances. This is taken into account by including the repulsive interactions
with the nearest neighbour ($\mathcal{H}_{\tilde{V}}$) and the on-site Hubbard
term ($\mathcal{H}_{U}$) \cite{alicea, herbut, maxim}. The strengths of the
nearest neighbour and the Hubbard interactions are treated as phenomenological
parameters in our analysis. 

We review the tight binding model for graphene and describe the systematic
long wavelength expansion around the Dirac points for the lattice operators.
The leading derivative term obtained after continuum approximation is the
well-studied Dirac hamiltonian of the non-interacting model. We apply
continuum approximation to lattice interaction terms. The leading term is
$SU(4)$ symmetric and long-ranged. The sub-leading terms break the $SU(4)$
symmetry explicitly and the interactions are short-ranged. 

\subsection{The continuum theory}

The tight binding model in the absence of magnetic field for graphene
describes the nearest neighbour hopping of the $\pi$-electrons of carbon
atoms. The lattice hamiltonian is, 
\begin{equation} \label{latt:tight binding}
	\mathcal{H}_{t} = -t \sum_{\bm{n}, \sigma} 
	c_{1, \sigma, \bm{n}}^{\dagger} 
	\Big( \sum_{j=1}^{3} c_{2, \sigma, \bm{n} + \bm{b}_{j}} \Big)
	+ \textrm{h.c.}
\end{equation}
Here $t \approx 3 ~ eV$, is the hopping parameter. $c_{r, \sigma, \bm{n}}$ are
the electron annihilation operators. $\bm{n}$ labels the triangular lattice
sites and $r=1,2$ the sublattices. $\sigma = \uparrow, \downarrow$, is the
spin index. The three vectors, $\bm{b}_{j},~j=1,2,3$, are equal to $0,
\bm{e}_{2}$ and $\bm{e}_{1} + \bm{e}_{2}$, where $\bm{e}_{1}$ and $\bm{e}_{2}$
are the basis vectors for the triangular lattice.  The electron creation and
annihilation operators obey the canonical anti-commutation relation, 
$ \{ c_{r, \sigma, \bm{n}}, c_{\tilde{r}, \tilde{\sigma},
\tilde{\bm{n}}}^{\dagger} \} = \delta_{r, \tilde{r}} 
\delta_{\sigma, \tilde{\sigma}} \delta_{\bm{n}, \tilde{\bm{n}}} $.

At half filling, the Fermi level of the tight binding model for graphene is at
zero. There are two points in the Brillouin zone with zero energy where the
two bands touch: the Dirac points, $ \bm{K}_{\pm} = ({\pm 2\pi/3a,\pm
2\pi/3a}) $.  Near the Dirac points, the quasi-particle energies have a linear
dispersion relation. 

To obtain the low energy effective theory from the lattice hamiltonian, we
project the electron operators to small regions around the Dirac points.
Separating the long wavelength modes from the fast varying modes, we can
write,
\begin{equation} \label{projection} 
	c_{r, \sigma, \bm{n}}  
	\approx a \left(\me^{\mi \bm{K}_{+} \cdot \bm{n}}
	\beta_{r, s} \Psi_{s, +, \sigma}(\bm{n}) +
	\me^{\mi \bm{K}_{-} \cdot \bm{n}} 
	\alpha_{r ,s}^{x} \Psi_{s, - , \sigma}(\bm{n})\right) .
\end{equation}
$a$ is the lattice spacing. $\alpha^{x}, \alpha^{y}, \alpha^{z} \equiv \beta$,
are the Pauli matrices. $\Psi_{r, \eta, \sigma}(\bm{n})$ are the slow varying
modes around the two Dirac points. $\eta =\pm$, is the valley index, labeling
the states around the Dirac points $\bm{K}_{\pm}$.

The leading term from continuum approximation of the lattice kinetic term,
Eq.\eqref{latt:tight binding}, then becomes:
\begin{equation} \label{conti:ham0}
	\mathcal{H}_{t} = \int \dif \bm{x} ~ 
	\Psi^{\dagger}(\bm{x}) \Big( v_{F} \, \bm{\alpha} \cdot 
	\bm{p} \otimes \mathbbm{1}_{4} \Big) \Psi(\bm{x}) .
\end{equation}
This non-interacting hamiltonian describes four species of free massless Dirac
particles. The speed of the massless particle is $v_{F} = \sqrt{3} \, a \, t /
2 \hbar$, the Fermi velocity at the Dirac point. The field operators,
$\Psi(\bm{x})$ are eight component objects and are labeled as $\Psi_{r,A}
(\bm{x})$, where $r$ refers to the Dirac spinor or sublattice index taking
values $r=1, 2$. The spin and valley indices are combined into the $SU(4)$
index, $A$, which takes values $1,\ldots,4$.

For weak magnetic fields, $a/\ell_{c} \ll 1$, the vector potential can be
chosen to be slowly varying on the scale of lattice spacing and the continuum
approximation of the lattice kinetic term is,
\begin{equation} \label{conti:hamt}
	\mathcal{H}_{t} =  \kappa_{t} \int \dif \bm{x} ~ 
	\Psi^{\dagger}(\bm{x}) \Big( \bm{\alpha} \cdot 
	\bm{\pi} \otimes \mathbbm{1}_{4} \Big) \Psi(\bm{x}) .
\end{equation}
Here $\bm{\pi} = \bm{p} -e \bm{A}$ and $\bm{A}$ is the vector potential.  We
have chosen dimensionless variables by the scaling,
\begin{equation*}
	\bm{x} \rightarrow \ell_{c} \bm{x}, \quad 
	\bm{\pi} \rightarrow \frac{\hbar}{\ell_{c}}\bm{\pi}, \quad 
	\Psi(\bm{x}) \rightarrow \frac{1}{\ell_{c}} \Psi(\bm{x}) .
\end{equation*}
The parameter, $\kappa_{t}$ has a square root dependence on the perpendicular
component of the magnetic field, $B_{\perp}$, 
\begin{equation} \label{kappa_t} 
	\kappa_{t} = \frac{\hbar v_{F}}{\ell_{c}} . 
\end{equation}

In the presence of a magnetic field, the lattice Zeeman term is, 
\begin{equation} \label{latt:zeeman}
	\mathcal{H}_{Z} = -g \, \frac{\mu_{B}}{\hbar} \sum_{\bm{n}} \sum_{r} 
	\bm{B} \cdot \bm{S}_{r, \bm{n}} .
\end{equation}
The spin operator in terms of the lattice fermion operators is, 
$S^{j}_{r, \bm{n}} = (\hbar/2) ~ c^{\dagger}_{r, \bm{n}} \sigma^{j} 
c_{r, \bm{n}}$. The continuum approximation results in a term that breaks the 
spin rotation symmetry,
\begin{equation} \label{conti:zeeman}
	\mathcal{H}_{Z} = - \kappa_{Z} \int_{\bm{x}}
	\Psi^{\dagger}(\bm{x}) \, \bm{\sigma} \cdot \mathbf{B} \,
	\Psi(\bm{x}) .
\end{equation}
The parameter, $\kappa_{Z}$, depends on the total magnetic field, $B_{T}(Tesla)$,
\begin{equation} \label{kappa_Z}
	\kappa_{Z} = \frac{1}{2} \, g \, \mu_{B} \, B_{T}
	= 0.67 \, B_{T} \, K .
\end{equation}

\subsection{Interaction terms}

The electron-electron interaction for graphene lattice
is modeled by combining the Coulomb point charge interaction,
$\mathcal{H}_{\tilde{C}}$, the short-ranged nearest neighbour interaction,
$\mathcal{H}_{\tilde{V}}$, and the on-site Hubbard interaction,
$\mathcal{H}_{U}$,
\begin{equation} \label{latt:interaction}
	\mathcal{H}_{I} = \mathcal{H}_{\tilde{C}} + \mathcal{H}_{\tilde{V}} +
	\mathcal{H}_{U} .
\end{equation}
The lattice Coulomb interaction is,
\begin{equation} \label{latt:coulomb}
	\mathcal{H}_{\tilde{C}} = \frac{1}{2} \sum_{\bm{n},\bm{m}} \sum_{r,s} 
	\hat{n}_{r, \sigma, \bm{n}} ~
	V(|\bm{n}_{r} - \bm{m}_{s}|) ~
	\hat{n}_{s, \tilde{\sigma}, \bm{m}} .
\end{equation}
Here $|\bm{n}_{r} - \bm{m}_{s}|$ is the distance between the two sub-lattice
points under consideration, and $V(|\bm{r}|) = e^{2} / (4 \pi \varepsilon \,
|\bm{r}|)$.  $\varepsilon = \varepsilon_{0} \varepsilon_{r}$, where
$\varepsilon_{r}$ is dielectric constant. And there is a constraint on
summation: $r \neq s$ when $\bm{n} = \bm{m}$. 

The interaction effects due to the finite extent of the wavefunctions at short
distances are taken into account by including the nearest neighbour interaction
term,
\begin{equation} \label{latt:nnbr}
	\mathcal{H}_{\tilde{V}} = \tilde{V} \sum_{\bm{n} , \sigma ,\tilde{\sigma}}
	\hat{n}_{1 ,\sigma, \bm{n}} \Big(
	\sum_{j=1}^{3} \hat{n}_{2, \tilde{\sigma}, \bm{n}+\bm{b}_{j}} \Big) .
\end{equation}
and the Hubbard term,
\begin{equation} \label{latt:hubbard}
	\mathcal{H}_{U} = U \sum_{\bm{n}} \sum_{r}
	\hat{n}_{r, \uparrow, \bm{n}} \hat{n}_{r, \downarrow, \bm{n}} .
\end{equation}

The leading $a / \ell_{c}$ term after continuum approximation for the
Coulomb interaction term, Eq.\eqref{latt:coulomb}, is $SU(4)$ symmetric,
\begin{equation} \label{conti:coulomb}
	\mathcal{H}_{C} = \kappa_{C} 
	\int_{\substack{\bm{x}, \bm{y} \\ \bm{x} \neq \bm{y}}} 
	\frac{1}{|\bm{r}|} \, 
	\rho(\bm{x}) \, \rho(\bm{y}) .
\end{equation}
Here $\rho(\bm{x}) = \Psi^{\dagger}(\bm{x}) \Psi(\bm{x})$, is the charge
density operator and $|\bm{r}| = |\bm{x} - \bm{y}|$. The parameter $\kappa_{C}$,
has $\sqrt{B_{\perp}}$ dependence and is inversely proportional
to the dielectric constant,
\begin{equation} \label{kappa_C} 
	\kappa_{C} = \frac{1}{2} \frac{e^{2}}{4 \pi \varepsilon \ell_{c}} .
\end{equation}

There are sub-leading terms which become significant at short distances
because of inverse cube(and higher power) dependence on the distance. They
break the $SU(4)$ symmetry explicitly. This symmetry breaking pattern is
similar to continuum approximation of the nearest neighbour interaction term,
Eq.\eqref{latt:nnbr}. In our continuum model, we combine these symmetry
breaking terms into $\mathcal{H}_{V}$ with a new parameter $V$.
\begin{equation} \label{conti:nnbr}
	\mathcal{H}_{V} = \kappa_{V} 
	\int_{\bm{x}}  \Big( \big(\rho(\bm{x}) \big)^{2} 
	- \big( \Psi^{\dagger}(\bm{x}) \beta \tau^z \Psi(\bm{x})
	\big)^{2} \Big) .
\end{equation}
This approximation forgoes the distance variation of terms with higher order
in $a / \ell_{c}$ for the point charge interaction, Eq.\eqref{latt:coulomb},
but retains the essential physics of $SU(4)$ symmetry breaking. This
approximation simplifies our variational energy computation immensely.

The continuum approximation of the Hubbard term in manifestly spin invariant
form yields the following set of local interaction terms,  
\begin{multline} \label{conti:hubbard}
	\mathcal{H}_{U} = \kappa_{U}
	\int_{\bm{x}} \Big( \big( \rho(\bm{x}) \big)^{2} 
	+ \big( \Psi^{\dagger}(\bm{x}) \beta \tau^{z} \Psi(\bm{x})
	\big)^{2}
	\\
	+ \frac{1}{2} \sum_{j,k=x,y} 
	\big( \Psi^{\dagger}(\bm{x}) \alpha^{j} \tau^{k}
	\Psi(\bm{x}) \big)^{2} \Big) .
\end{multline}

The tight binding term, Eq.\eqref{latt:tight binding}, results in a series of
sub-leading terms after  continuum approximation. These terms involve higher
order derivatives. The first of such sub-leading terms is, 
\begin{equation} \label{conti:kinetic correction}
	\mathcal{H}_{t_{1}} = \kappa_{t_{1}} \int \dif \bm{x} ~ 
	\Psi^{\dagger}(\bm{x}) ~ \mathrm{h}_{t_{1}} \, \Psi(\bm{x}),
\end{equation}
with $\mathrm{h}_{t_{1}} = \big( \alpha^{x} ( \pi_{x}^{2} + 3 \pi_{y}^{2} ) 
- \alpha^{y} 3( \pi_{x} \pi_{y} + \pi_{y} \pi_{x})  \big) \otimes \tau^{z}
$.

The parameters, 
\begin{align} \label{kappa_s}
	\kappa_{V} & = \frac{3 V a^{2}}{4 \ell_{c}^{2}} , &
	\kappa_{U} & = \frac{U a^{2}}{4 \ell_{c}^{2}} , &
	\kappa_{t_{1}} & = \frac{t \, a^{2}}{8 \ell_{c}^{2}} .
\end{align}
have linear dependence on $B_{\perp}$. We treat both $V$ and $U$ as
phenomenological parameters for our interacting continuum model. 

\begin{table}
	\begin{tabular*}{0.4\textwidth}{@{\extracolsep{\fill}} | c | c | c |}
	\hline \hline
	$\kappa_{t}$ & $ \frac{\hbar v_{F}}{\ell_{c}} $ 
	& $0.86 t \big( \frac{a}{\ell_{c}} \big) $\\
	\hline 
	$\kappa_{C}$ & $ \frac{1}{2}\frac{e^{2}}{4 \pi \varepsilon \ell_{c}}$ 
	& $\frac{5.82}{\varepsilon_{r}}  \big( \frac{a}{\ell_{c}} \big) $\\
	\hline 
	$\kappa_{Z}$ & $ \frac{g \mu_{B} B}{2}  $ 
	& $ 0.63 \big( \frac{a}{\ell_{c}} \big)^{2} $\\
	\hline 
	$\kappa_{V}$ & $ \frac{3 V a^{2}}{4 \ell_{c}^{2}}  $ 
	& $ 0.75 V \big( \frac{a}{\ell_{c}} \big)^{2} $\\
	\hline 
	$\kappa_{U}$ & $ \frac{U a^{2}}{4 \ell_{c}^{2}}  $ 
	& $ 0.25 U \big( \frac{a}{\ell_{c}} \big)^{2} $\\
	\hline 
	$\kappa_{t_{1}}$ & $ \frac{t a^{2}}{8 \ell_{c}^{2}}  $ 
	& $ 0.125 t \big( \frac{a}{\ell_{c}} \big)^{2} $\\
	\hline 
	\hline 
\end{tabular*}
\caption{The first two columns list and define all the energy parameters in
	the interacting model. In the third column, we explicitly express the
	parameters as function  of $a/\ell_{c}$ in units of $eV$. For the magnetic
	fields typically achieved in a laboratory, $a/ \ell_{c} \sim 10^{-1}
	\textrm{ to } 10^{-2}$, a small parameter.  The kinetic and Coulomb
	energy parameters are dominant and the remaining ones are smaller by a
	factor of $a/\ell_{c}$.
}
\label{partab}
\end{table}

To summarize, the continuum interacting model hamiltonian that we use in this 
article is,
\begin{align} \label{ham full}
	\mathcal{H} &= \mathcal{H}_{0} + \mathcal{H}_{1} . \\
	\mathcal{H}_{0} &= \mathcal{H}_{t} + \mathcal{H}_{C} . \\
	\mathcal{H}_{1} &= \mathcal{H}_{V} + \mathcal{H}_{U} 
	+ \mathcal{H}_{Z} + \mathcal{H}_{t_{1}} .
\end{align}
$\mathcal{H}_{0}$, defined in Eq.\eqref{conti:hamt} and
Eq.\eqref{conti:coulomb}, is the $SU(4)$ symmetric term.  $\mathcal{H}_{1}$
defined in Eq.\eqref{conti:nnbr}, Eq.\eqref{conti:hubbard},
Eq.\eqref{conti:zeeman} and Eq.\eqref{conti:kinetic correction}, is the
symmetry breaking term. The $a/\ell_{c}$ dependence of the coefficients in
Table \ref{partab} shows that $\mathcal{H}_{0}$ is the leading order term and
$\mathcal{H}_{1}$ is smaller by a factor of $a/\ell_{c}$.
 
Our strategy is to approximately solve $\mathcal{H}_{0}$ using the variational
method. As we will see, this leads to $SU(4)$ symmetry broken ground state
solutions which are degenerate. We then treat $\mathcal{H}_{1}$ in first order
perturbation theory to see which ground state(s) is (are) picked out by the
energetics.

\section{Symmetries and order parameters} \label{sec:symmetries}
In this section, we discuss the details of the $SU(4)$ symmetry displayed by
various terms of our interacting model. We also discuss various order
parameters for the ground states. 

\subsection{The $SU(4)$ symmetry}

Under $SU(4)$ transformations, the fermion field operator transforms as, 
\begin{equation} \label{su4 field}
	\tilde{\Psi}_{r,A}(\bm{x}) = \mathcal{U}_{A,B} \Psi_{r,B}(\bm{x}) ,
\end{equation}
where $\mathcal{U} \in SU(4)$. As mentioned earlier, $\mathcal{H}_{0}$ is
invariant under this transformation.

$\mathcal{H}_{1}$ breaks this $SU(4)$ symmetry. The short-ranged
interactions, $\mathcal{H}_{V}$ and $\mathcal{H}_{U}$, are invariant under
transformations belonging to the subgroup, $U(1) \times SU(2)$, i.e. 
\begin{equation}
	\mathcal{U} = \me^{-\frac{\mi}{2} \theta_{\tau} \tau^{z}} \otimes
	U_{\sigma} .
\end{equation}
Here $U_{\sigma}$ denotes a unitary rotation in $SU(2)$ spin space.  There is
also a discrete symmetry that leaves the short-ranged interactions invariant,
\begin{equation}
	\mathcal{U} = \tau^{x} \otimes U_{\sigma} .
\end{equation}
This transformation amounts to interchanging the valley indices of the field
operators. The group that leaves the short-ranged interaction terms invariant 
is a semi-direct product of $ Z_{2} \rtimes U(1) \otimes SU(2)$.

The sub-leading correction to the kinetic term, $\mathcal{H}_{t_{1}}$, given in
Eq.\eqref{conti:kinetic correction}, is only invariant under $U(1) \otimes
SU(2)$. 

The Zeeman term, $\mathcal{H}_{Z}$, breaks the $SU(2)$ spin symmetry to $U(1)$.
\begin{equation}
	\mathcal{U} = U_{\tau} \otimes \me^{-\frac{\mi}{2}
	\theta_{\sigma} \sigma^{z}} ,
\end{equation}
is the transformation of the field operator that leaves the Zeeman term invariant.
Here $U_{\tau}$ is a unitary rotation in $SU(2)$ valley space. 

The model hamiltonian, Eq.\eqref{ham full}, is invariant under $U(1) \times
U(1)$ rotation. The transformation
\begin{equation}
	\mathcal{U} = \me^{-\frac{\mi}{2} \theta_{\tau} \tau^{z}} \otimes 
	\me^{-\frac{\mi}{2} \theta_{\sigma} \sigma^{z}} ,
\end{equation}
of the field operators leaves $\mathcal{H}$ invariant.

\subsection{Order parameters} \label{subsec-op}

$U(4)$ has sixteen generators. They can be represented by the $4\times 4$
matrices, $T^{\mu \nu} \equiv \tau^{\mu} \otimes \sigma^{\nu}$, $\mu,\nu =
0,x,y,z$, where we define $\tau^{0}$ and $\sigma^{0}$ to be $2 \times 2$
identity matrices. 

The operators representing the $U(4)$ charge densities are $\mathcal{T}^{\mu
\nu}(\bm{x}) \equiv \Psi^{\dagger}(\bm{x}) T^{\mu\nu} \Psi(\bm{x})$. The
ground state expectation value, $ \langle \mathcal{T}^{\mu \nu}(\bm{x})
\rangle $, correspond to the $SU(4)$ polarization of the ground state.

There are also sixteen other order parameters that can be constructed,
$\widetilde{\mathcal{T}}^{\mu \nu}(\bm{x}) \equiv \Psi^{\dagger}(\bm{x}) \beta
\otimes T^{\mu \nu} \Psi(\bm{x})$.  These two types of operators can be
distinguished by their transformation properties under reflection,
\begin{equation}
	\Psi(x,y) \rightarrow \mi \alpha^{y} \Psi(-x,y) .
\end{equation}
Under this transformation, we have
\begin{align*}
	\mathcal{T}^{\mu \nu}(x,y) & \rightarrow \mathcal{T}^{\mu \nu}(-x,y), &
	\widetilde{\mathcal{T}}^{\mu \nu}(x,y) & \rightarrow 
	-\widetilde{\mathcal{T}}^{\mu \nu}(-x,y) .
\end{align*}
The operation of $\alpha^{y}$ switches the Dirac indices, corresponding to a
switch of the sublattice index. We refer to the expectation values of these
operators that are odd under this transformation, as the staggered $SU(4)$
polarization of the ground state. 

In Appendix \ref{appx:dict}, we have given a list of lattice operators
corresponding to each of the 32 order parameters we have discussed above.
Note that many lattice operators can have the same continuum limit.  We have
listed the simplest representative lattice operator.

Some of the operators are linear combinations of the products of electron
operators on the same lattice site. $\mathcal{T}^{00}(\bm{x})$ is the total
charge density in the unit cell and $\mathcal{T}^{0i}(\bm{x})$ the total spin
density. $\widetilde{\mathcal{T}}^{30}(\bm{x})$ is the charge density wave
order parameter corresponding to the difference in charge densities on the
two sublattice points in the unit cell and
$\widetilde{\mathcal{T}}^{3i}(\bm{x})$ is the Neel order parameter
corresponding to the difference in spin densities on the two sublattices.

The operator $\widetilde{\mathcal{T}}^{00}(\bm{x})$ is a time-reversal
symmetry breaking mass term. It is the continuum limit of the operator that
occurs in Haldane's model\cite{haldane} of a Chern insulator on the honeycomb
lattice.  Thus $\widetilde{\mathcal{T}}^{0 \nu}(\bm{x})$ are all bond-order
parameters at the lattice level. 

The operators involving $\tau^{x}$ or $\tau^{y}$ transfer particles from
$\bm{K}_{+} \leftrightarrow \bm{K}_{-}$. The corresponding lattice operators
hence involve a momentum transfer of $\pm(\bm{K}_{+} - \bm{K}_{-})$.

\section{Variational wavefunction} \label{sec:variational wftn}
The main aim of this paper is to study the effects of the Dirac sea of Landau
levels on the $SU(4)$ symmetry breaking induced by the interactions. To this
end, we want to construct variational states such that (i) they allow the
$SU(4)$ polarization of the Dirac sea and (ii) the energy computations are
tractable.

We investigate a simple family of states that satisfy these two conditions,
namely the variational state is constructed using the eigenstates of
one-particle Dirac hamiltonian for a quartet of massive particles in the
presence of a magnetic field, $h_{M}$,
\begin{equation} \label{ham_M}
	h_{M}= \bm{\alpha} \cdot \bm{\pi} + \beta M.
\end{equation} 
Here $M$ is a $4\times 4$ hermitian mass matrix. It is specified by 16
independent real numbers that we treat as variational parameters.

The eigenvectors of $h_{M}$ are of the form, $\Phi^{n,l,q}_{r,A}(\bm{x}) =
\phi^{n,l,q}_{r}(\bm{x}) \chi^{q}_{A}$. $m_{q}$, $q=1,\dots,4$ are the
eigenvalues of $M$ and $\chi^{q}$ the corresponding eigenvectors.
$\phi^{n,l,q}(\bm{x})$ are the Landau levels for a $(2+1)d$ massive Dirac
particle with mass $m_{q}$. We expand the field operators in terms of these
eigenvectors, 
\begin{equation}
\Psi_{r,A}(\bm{x}) = \sum_{n,l,q} \Phi_{r,A}^{n,l,q}(\bm{x}) ~ \psi_{n,l,q} .
\end{equation}
$\psi_{n,l,q}^{\dagger}$ is the creation operator for an electron with Landau
level index $n$, orbital degeneracy index $l$, and $SU(4)$ quantum number $q$. 

The variational ground states are constructed by filling the quartet for
Landau levels with index $n \le -1$, constituting the Dirac sea. The state for
$\sigma_{H}=0$ has half-filled quartet of the $n=0$ Landau level and a
quarter-filled quartet of the $n=0$ Landau level for $\sigma_{H}=-1$. The
variational state for $\sigma_{H}=+1$ is the charge conjugate state of
$\sigma_{H}=-1$, so we are only going to study $\sigma_{H}=-1$.  
\begin{align}
	| \sigma_{H}=0 \rangle & = \Big( \prod_{\substack{l=0, \\ q=1}}^{\infty,2}
	\psi^{\dagger}_{0,l,q} \Big)
	\Big( \prod_{n=-1}^{-N_{c}} \prod_{\substack{l=-|n|, \\ q = 1}}^{\infty, 4} 
	\psi^{\dagger}_{n,l,q} \Big)
	| 0 \rangle . \label{chap4:gs nu0} \\
	| \sigma_{H}=-1 \rangle & = \Big( \prod_{l=0}^{\infty} 
	\psi^{\dagger}_{0,l,1} \Big)
	\Big( \prod_{n=-1}^{-N_{c}} \prod_{\substack{l=-|n|, \\ q = 1}}^{\infty, 4} 
	\psi^{\dagger}_{n,l,q} \Big)
	| 0 \rangle . \label{chap4:gs nu1}
\end{align}
$| 0 \rangle$ is the empty state defined by $\psi_{n,l,q} | 0 \rangle=0$. We
are using the Dirac theory as an effective model of the underlying lattice
model. The lattice model sets an ultraviolet cut-off which is proportional to
the inverse of the lattice constant. We estimate a cut-off for the Landau
level index by equating the number of states of the lattice model to that of
the continuum model, 
\begin{equation} \label{cutoff}
	N_{c} = \frac{2 \pi}{\sqrt{3}} \left( \frac{\ell_{c}}{a} \right)^{2} .
\end{equation}
This amounts to replacing the detailed band structure of the lattice model
with energy levels of the Dirac particle, which is an approximation in our
model calculation. 

We now discuss the transformation properties of the Dirac sea, i.e. the 
fully occupied quartet of the Landau level with $n \le -1$. It is useful to 
write,
\begin{equation}
	\psi^\dagger_{n,l,q} =
	\left( \int_{\bm{x}}\Psi_{r,A}^{\dagger}(\bm{x}) 
	\phi^{n,l,q}_{r}(\bm{x}) \right)
	\chi^{q}_{A} \equiv \tilde{\psi}^{\dagger}_{n,l,q,A} \chi^{q}_{A}
\end{equation}
Now consider the product of the operators that create the quartet at every
occupied $n$, $l$,
\begin{equation} \label{4psiprod}
	\tilde{\psi}_{n,l,1,A_{1}}^{\dagger} \tilde{\psi}_{n,l,2,A_{2}}^{\dagger}
	\tilde{\psi}_{n,l,3,A_{3}}^{\dagger} \tilde{\psi}_{n,l,4,A_{4}}^{\dagger}
	\chi_{A_{1}}^{1} \chi_{A_{2}}^{2} \chi_{A_{3}}^{3} \chi_{A_{4}}^{4}
\end{equation}
If all the eigenvalues, $m_{q}$, are equal, then
$\tilde{\psi}_{n,l,q,A}^{\dagger}$ is independent of $q$. In such a case the
product of the $\tilde{\psi}$ operators is proportional to the totally
antisymmetric tensor $\epsilon_{A_{1},A_{2},A_{3},A_{4}}$. Since the matrix
formed by the eigenvectors, $\chi_{A}^{q}$, is a unitary matrix, the $\chi^{q}$
dependence of the product in Eq.\eqref{4psiprod} is just a phase and thus the
state is invariant under $U(4)$ (and hence $SU(4)$) transformations.

However, if all the $m_{q}$ are not equal to each other, then the product is
not invariant under all the $U(4)$ transformations. The subgroup,
$\mathcal{G}$, under which the product is invariant (up to a phase) is the set
of $U(4)$ matrices that commute with the diagonal matrix, $M_{D}= \{m_{1},
m_{2}, m_{3}, m_{4} \}$. The ground state manifold is then the coset space
$U(4)/{\mathcal G}$.  The smallest $\mathcal{G}$ is $U(1)\times U(1)\times
U(1)\times U(1)$, which is the case where all the four eigenvalues, $m_{q}$,
are unequal. The dimension of the coset space is then 12. 

Thus, our choice of the variational wave function with the mass matrix as the
variational parameter allows us to have an $SU(4)$ polarization both for a
partially filled $n=0$ Landau level quartet and the filled Dirac sea. Several
patterns of symmetry breaking are possible in our ansatz. The specific values
taken by the variational parameters (the elements for the mass matrix, $M$)
and the symmetry breaking pattern will be fixed by the minimization of the
variational ground state energy.

\section{Correlation function} \label{sec:correlation}
The computation of the variational energy requires the evaluation of the
expectation value of the two point field operators. This results from a Wick
decomposition applied to the four fermion terms in the interacting model. We
define a two-point correlation function, 
\begin{equation} 
	\Gamma_{r,A;s,B}(\bm{x}, \bm{y}) = \frac{1}{2}
	\langle GS| [ \Psi^{\dagger}_{s,B}(\bm{y}) , \Psi_{r,A}(\bm{x}) ] |GS
	\rangle .
\end{equation}
$|GS \rangle$ is the ground state under consideration. Expressing the two
point correlation function in terms of wavefunctions of massive Dirac
particles,
\begin{equation} \label{2pt cftn}
	\Gamma(\bm{x}, \bm{y}) = \sum_{q=1}^{4} G_{m_{q}}(\bm{x}, \bm{y}) \, P_{q}
	.
\end{equation}
$P_{q} \equiv \chi^{q} (\chi^{q})^{\dagger}$, are the projection operators
constructed from the eigenvectors of $M$. $G_{m_{q}}(\bm{x}, \bm{y})$ is the
equal time Feynman propagator for a $(2+1)d$ massive Dirac particle with mass 
$m_{q}$ in a magnetic field (Details in Appendix \ref{appx:heat}),
\begin{multline*}
	G_{m_{q}}(\bm{x}, \bm{y}) = \\
	\frac{1}{2} 
	\sum_{\substack{n = -\infty, \\ l = - |n|}}^{\infty, \infty} 
	\Big( \Theta(-\epsilon_{n,l,q}) - \Theta(\epsilon_{n,l,q}) \Big) 
	\phi^{n,l,q}(\bm{x}) \big( \phi^{n,l,q}(\bm{y}) \big)^{\dagger} .
\end{multline*}
$\Theta(x)$ is the Heaviside step function and $\epsilon_{n,l,q}$ are the
eigenvalues for Landau levels for the massive Dirac particle, Eq.\eqref{massive
dirac}. $G_{m_{q}}(\bm{x}, \bm{y})$ can be expressed in the so called `heat
kernel' representation.
\begin{equation} \label{dirac kernal}
	G_{m_{q}}(\bm{x}, \bm{y}) = - \frac{1}{2 \sqrt{\pi}} 
	\int\limits_{\frac{1}{2 N_{c}}}^{\infty} \langle \bm{x} |
	\mathrm{h}_{m_{q}} \me^{- s \, \mathrm{h}^{2}_{m_{q}}} | \bm{y} \rangle .
\end{equation}
Here, $h_{m_{q}} = \bm{\alpha} \cdot \bm{\pi} + \beta m_{q}$, is the 
one-particle hamiltonian for a Dirac particle with mass $m_{q}$ in a magnetic
field.  We make use of the imaginary time propagator for a non-relativistic
$2d$ electron in a magnetic field to evaluate \cite{glasser}
\begin{equation}
	\langle \bm{x} | \me^{-s \mathrm{h}^{2}_{m_{q}}} | \bm{y} \rangle 
	=  \frac{1}{2 \pi} 
	\zeta_{s}(\bm{x}, \bm{y})
	\frac{\me^{-s m_{q}^{2}}}{2 \sinh(s)}
	\left( \begin{array}{cc}
			\me^{-s} & 0 \\
			0 & \me^{s} 
	\end{array} \right) .
\end{equation}
Here, $\zeta_{s}(\bm{x}, \bm{y}) = \me^{- \frac{1}{4}|\bm{x} -
\bm{y}|^{2} \coth (s)} \: \me^{\frac{\mi}{2} ( x_{1} y_{2} - y_{1} x_{2})}$.

By construction, $\Gamma(\bm{x}, \bm{y})$ is an $8 \times 8$ matrix.  The mass
matrix, $M$, for the Dirac particles of graphene encodes the information about
the filling factor and the nature of the ground state,  
\begin{equation}
	M = \sum_{q=1}^{4} m_{q} P_{q} .
\end{equation}
We fix the chemical potential at zero for all the calculations discussed in
this article. The signs of $m_{q}$ then decide which members of the quartet of
the $n=0$ Landau level are occupied. With our convention, the $q^{th}$ member
is occupied if $m_{q}$ is positive. Hence, for $\sigma_{H}=0$, the diagonal
mass matrix takes the form $M_{D} = \{ m_{1}, m_{2}, - m_{3}, - m_{4} \} $,
with $m_{q} > 0$ for $q=1, \ldots ,4$. Similarly, $M_{D} = \{ m_{1}, - m_{2},
- m_{3}, - m_{4} \}$ for $\sigma_{H}=-1$.

The two point correlator, Eq.\eqref{2pt cftn}, enables us to compute the
energy of the long-ranged Coulomb interaction efficiently. It eliminates the
need to compute the matrix elements of individual wavefunctions and the
computation reduces to the evaluation of a few well behaved two dimensional
integrals.

The mean field energy computation for the symmetric breaking terms of the
interacting model is expressed in terms of the coincident correlator, which is
independent of spatial coordinates,
$\Gamma = \sum_{q=1}^{4} G_{m_{q}} (\bm{x}, \bm{x}) \, P_{q} $,
\begin{equation} \label{gamma} 
	\Gamma = \frac{1}{2 \pi} 
	\sum_{q=1}^{4} \frac{1}{2} \Big(
	u_{m_{q}} \, \mathbbmss{1}_{2} 
	- v_{m_{q}} \beta \Big) \otimes P_{q} .
\end{equation}
Here, $u_{m_{q}}$ and $v_{m_{q}}$ are evaluated using the following
integrals,
\begin{align}
	u_{m_{q}} &= \frac{m_{q}}{2 \sqrt{\pi}} \; 
	\int_{\frac{1}{2 N_{C}}}^{\infty} 
	\frac{\dif s}{\sqrt{s}} ~ \me^{-s m_{q}^{2}} . \\
	v_{m_{q}} & = \frac{m_{q}}{2 \sqrt{\pi}} \; 
	\int_{\frac{1}{2 N_{C}}}^{\infty} 
	\frac{\dif s}{\sqrt{s}} ~ \me^{-s m_{q}^{2}} \, \coth(s) .
\end{align}
In the limit of a large cut-off we have $\lim_{N_C\rightarrow\infty}
u_{m}\rightarrow {\rm sgn}(m_{q})/2$. For the magnetic field values relevant
to us, $u_{m}$ is very close to $1/2$. Specifically, $0 \le (0.5 -
|u_{m_{q}}|)  \le 0.001$. This small deviation from $0.5$ is due to our
cut-off procedure. Henceforth we will put $u_{m_{q}}=0.5 \, {\rm sgn}(m_{q})$.
Since $\coth(s) \ge 1$, we always have $|v_{m_{q}}| \ge 0.5$.

The two point correlators, Eq.\eqref{2pt cftn}, and the two point coincident
correlators, Eq.\eqref{gamma}, will be used to evaluate the variational state
energy for the interacting model in the following sections.

\section{Sea Polarization} \label{sec:polarisation}
In section \ref{sec:variational wftn}, we had shown that the filled Dirac sea
of Landau levels ($n=-N_{C},\ldots,-1$), in general, is not an $SU(4)$ singlet
when the parameters $m_{q}$ are unequal. In this section we characterize this
symmetry breaking in terms of the order parameters defined in section
\ref{subsec-op}.

We can use the expression for the coincident correlation function in
Eq.(\ref{gamma}) to evaluate the order parameters,
\begin{align}
	\label{optans}
	\langle\mathcal{T}^{\mu\nu}(\bm{x})\rangle  & =
	\textrm{Tr}[(\mathbbm{1}_{2} \otimes T^{\mu\nu}) \Gamma ] . \\
	\label{optbarans}
	\langle\widetilde{\mathcal{T}}^{\mu\nu}(\bm{x})\rangle & =
	\textrm{Tr}[(\beta \otimes T^{\mu\nu})\Gamma] .
\end{align}
The contribution from the Dirac sea can be obtained by deducting the $n=0$
Landau level contributions from the above equations. Denoting the sea
contribution to the order parameters by $\langle \ldots \rangle_{sea}$, this
gives us,
\begin{align}
	\label{seaoptans}
	\langle\mathcal{T}^{\mu\nu}(\bm{x})\rangle_{sea} &= 
	\frac{1}{2 \pi} 
	\sum_{q=1}^{4} \Big( u_{m_{q}} - \frac{\textrm{sgn}(m_{q})}{2} \Big)
	\textrm{Tr}[T^{\mu\nu} P_{q}] . \\
	\label{seaoptbarans}
	\langle\widetilde{\mathcal{T}}^{\mu\nu}(\bm{x})\rangle_{sea} &=
	-\frac{1}{2\pi} 
	\sum_{q=1}^{4} \Big( v_{m_{q}} - \frac{\textrm{sgn}(m_{q})}{2} \Big)
	\textrm{Tr}[T^{\mu\nu} P_{q}] .
\end{align}

Since $u_{m_{q}}=0.5~{\rm sgn}(m_{q})$, we have
$\langle\mathcal{T}^{\mu\nu}(\bm{x})\rangle_{sea}= 0$.  Since $\vert
v_{m_{q}}\vert >0.5$, the staggered polarization of the sea is, in general,
non-zero. 

Thus the Dirac sea of our variational states breaks the $SU(4)$ symmetry by
developing a staggered polarization. However, the net $SU(4)$ polarization for
the Dirac sea is zero. To visualize this, consider the non-interacting ground
state. This corresponds to $m_{q}=0$ and the sea is a singlet which has all
the charges uniformly distributed over the lattice sites. Now consider making
one of the $m_{q}$ non-zero. The wave function of this species will have
different weights on the two sublattices with the total weight in a unit cell
remaining unchanged. Thus the sea will develop a staggered polarization but
not a net polarization.

\section{Symmetric model} \label{sec:symmetric model}
In this section, we consider the leading order (in $a/\ell_{c}$) part of the
hamiltonian which includes the kinetic term, Eq.\eqref{conti:hamt}, and the
Coulomb term, Eq.\eqref{conti:coulomb}.
\begin{equation}
	\mathcal{H}_{0} = \mathcal{H}_{t} + \mathcal{H}_{C} .
	\label{ham0}
\end{equation}
Both these terms are $SU(4)$ symmetric. The charge densities that occur in
$\mathcal{H}_{C}$ are with respect to the average charge density at half
filling. 

We will evaluate the variational state energy of $\mathcal{H}_{0}$, minimize
it, and show that the $SU(4)$ symmetry is spontaneously broken down to $U(1)
\times SU(3)$ for $\sigma_{H} = -1$ and $SU(2) \times SU(2)$ for $\sigma_{H} =
0$. We also compute the gaps of the quasi-particle(hole) excitations within
the symmetric model.

\subsection{Ground states} \label{subsec:symm}

In this subsection, we use the two point correlator, Eq.\eqref{2pt cftn}, to
compute the expectation value of the kinetic and the Coulomb terms. The
evaluation of these expectation values involves the integration of the $s$
variables of the correlators, Eq.\eqref{dirac kernal}, which is accomplished
numerically. 

The kinetic term has local fermion field operators and the expectation value
can be expressed as, 
\begin{equation} 
	\langle \mathcal{H}_{t} \rangle = \kappa_{t} \int_{\bm{x}} 
	\lim_{\bm{y} \to \bm{x}}
	\langle \Psi^{\dagger}_{r,A}(\bm{x})
	\langle\bm{x}\vert\mathrm{h}_{r,A;s,B}\vert\bm{y}\rangle 
	\Psi_{s,B}(\bm{y}) \rangle .
\end{equation}
Here $\mathrm{h}_{r,A;s,B} = \big( \bm{\alpha} \cdot \bm{\pi} \big)_{r,s}
\big( \mathbbmss{1}_{4} \big)_{A,B}$, Eq.\eqref{conti:hamt}.   Using the two
point correlator, Eq.\eqref{2pt cftn},  the kinetic energy density,
$\mathcal{E}_{t} \equiv \langle \mathcal{H}_{t} \rangle$, of the variational
state is evaluated to give 
\begin{equation} \label{avg kinetic}
	\mathcal{E}_{t} 
	= \frac{\kappa_{t}}{2 \pi} \sum_{q=1}^{4} \eta_{t}(m_{q}^{2}) .
\end{equation}
Here the quantity $\eta_{t}$ depends on the square of the mass parameter and
is a positive quantity which we evaluate numerically (Details in Appendix
\ref{appx kinetic}).

\begin{figure}
	\begin{center}
		\includegraphics[width=0.9\columnwidth]{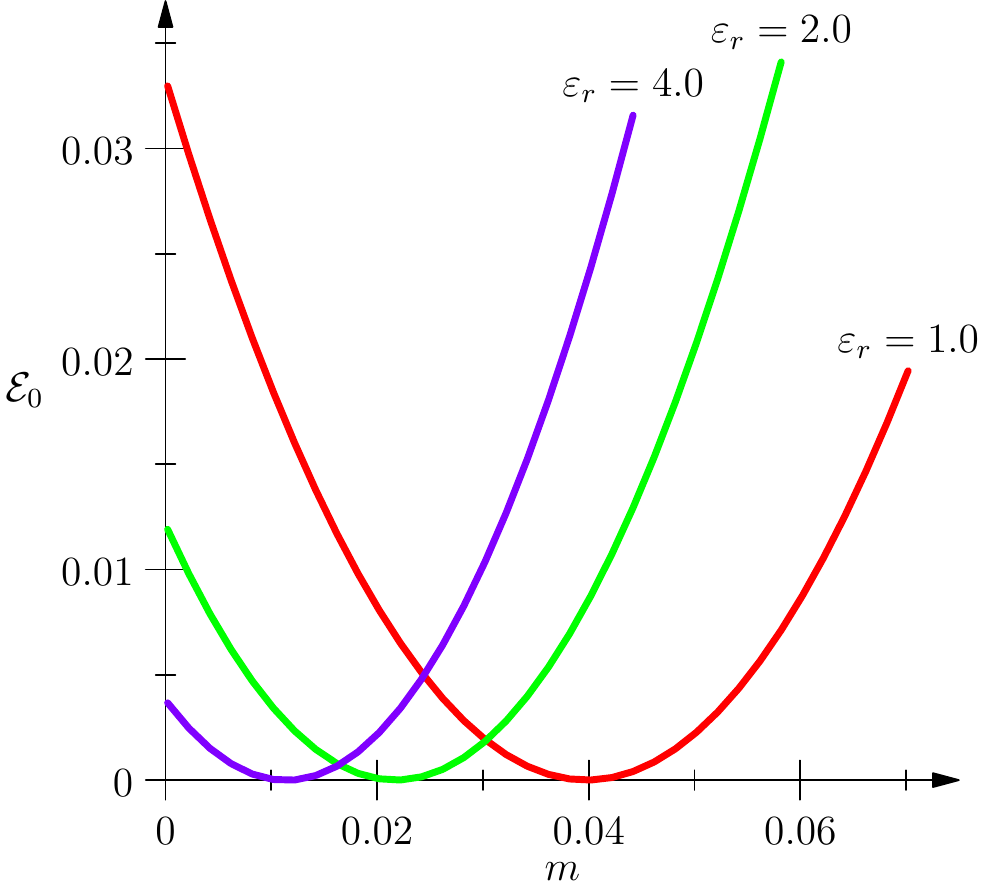}
	\end{center}
	\caption{This figure shows the ground state energy density as a function
		of the variational parameter $m$ for $\varepsilon_{r}=1,2,4$. The
		minima of the energy density function decreases with increasing
		$\varepsilon_{r}$. The curves have been adjusted to align the minima
		along the $m$-axis.
	} 
	\label{energy Vs mass} 
\end{figure}

The expectation value of the Coulomb term, Eq.\eqref{conti:coulomb}, can be
expressed in terms of the two point correlator,
\begin{equation*}
	\langle \mathcal{H}_{C} \rangle = 
	- \kappa_{C}
	\iint\limits_{\bm{x},\bm{y}} 
	\frac{1}{|\bm{x} - \bm{y}|} 
	\textrm{Tr}[ \Gamma(\bm{x}, \bm{y}) \,
	\Gamma(\bm{y}, \bm{x}) ].
\end{equation*}
The Coulomb energy density, $\mathcal{E}_{C} \equiv \langle
\mathcal{H}_{C} \rangle$, is evaluated using the two-point correlators,
\begin{equation} \label{avg coulomb}
	\mathcal{E}_{C} = - \frac{\kappa_{C}}{2 \pi} \sum_{q=1}^{4} 
	\eta_{C_{2}}(m_{q}^{2}). 
\end{equation}
The quantity $\eta_{C_{2}}$ also depends on the square of the mass parameter
and its numerical evaluation involves double integration of the variable $s$
of the correlator (Details in Appendix \ref{appx coulomb}).

The total energy of the leading order hamiltonian, $\mathcal{H}_{0}$, is
therefore a sum of functions of $m_{q}^{2}$,
\begin{equation} \label{symm energy}
	\mathcal{E}_{0} = \mathcal{E}_{t} + \mathcal{E}_{C} = 
	\sum_{q=1}^{4} \Big( \kappa_{t} \, \eta_{t}(m^{2}_{q}) 
	- \kappa_{C} \eta_{C_{2}}(m^{2}_{q}) \Big) .
\end{equation}
Thus the energy minimization condition, $\partial\mathcal{E}_{0}/\partial
m_{q}=0$, implies that the magnitudes of all the masses, $m_{q}$, are equal.
Fig.\ref{energy Vs mass} shows the energy as a function of $m_{q}$. The
minimum of energy density at a non-zero value of $m_{q}$ decreases with an
increasing dielectric constant (decreasing interaction strength). 

The filling factor(Hall conductivity) determines the signs of the masses.  For
$\sigma_{H}=-1$, three of the masses are negative and one positive. For
$\sigma_{H}=0$, two are positive and two negative. The non-zero values of
$m_{q}$ imply that the $SU(4)$ symmetry is spontaneously broken. As we discuss
below, the signs of the masses determine the patterns of the symmetry breaking
for the two cases.

\subsubsection{Ground state at manifold $\sigma_{H} = -1$} \label{gs nu1}

The many body ground state at $\sigma_{H} = -1$ has a quarter-filled quartet
of the $n=0$ Landau level. Thus after minimization, the diagonal mass matrix
for the ground state at $\sigma_{H} = -1$ takes the form,
\begin{equation} \label{massd nu1}
	M_{D} = \{m, - m, - m, - m \} .
\end{equation}
The ground state at Hall conductivity $\sigma_{H}=-1$ is therefore invariant
under a subgroup, $U(1) \times U(3)$, of $U(4)$. The ground state manifold is
a coset space $U(4)/(U(1) \times U(3)) = CP^{3}$. The dimension of this coset
space is six and is completely specified by a normalized four component
complex vector (up to a phase). This complex vector corresponds to the $SU(4)$
component of the wavefunction of the occupied member of the $n=0$ Landau level
quartet.  Without loss of generality, we choose the $SU(4)$ component to be
$\chi^{1}$ and parameterize it as,
\begin{equation} \label{nu1 ket1}
	| \chi^{1} \rangle = \cos( \frac{\gamma_{1}}{2} ) 
	| + \rangle | \bm{n}_{1} \rangle 
	+ \me^{\mi \Omega_{1}} \sin ( \frac{\gamma_{1}}{2} )
	| - \rangle | - \bm{n}_{2} \rangle .
\end{equation}
Here $| \pm \rangle | \bm{n}_{i} \rangle \equiv | \pm \rangle \otimes |
\bm{n}_{i} \rangle$. And $| \pm \rangle$ are the eigenvectors of the valley
operator $\tau^{z}$. $| \bm{n}_{i} \rangle$ is a vector pointing in an
arbitrary direction in the spin space. 
\begin{equation}
	| \bm{n}_{i} \rangle = \cos ( \frac{\theta_{i}}{2} ) | \! \uparrow \rangle 
	+ \me^{\mi \varphi_{i}} \sin ( \frac{\theta_{i}}{2} ) | \! \downarrow 
	\rangle .
\end{equation}
Here $| \! \uparrow \rangle$ and $| \! \downarrow \rangle$ are the
eigenvectors of $\sigma^{z}$. 

The mass matrix for the ground state  at $\sigma_{H} = -1$ is 
\begin{equation} \label{mass nu1}
	M = m \Big( 2 | \chi^{1} \rangle \langle \chi^{1} | 
	-  \mathbbmss{1}_{4} \Big) \equiv m Q^{(-1)} ,
\end{equation}
where $Q^{(-1)}(\gamma_{1}, \Omega_{1}, \theta_{1}, \phi_{1}, \theta_{2},
\phi_{2})$ is a $4\times 4$ matrix valued function of the six angle
parameters. It is in one-to-one correspondence with the elements of the coset
space $U(4)/\left(U(3)\times U(1)\right)$. It satisfies the properties,
$(Q^{(-1)})^2 = \mathbbmss{1}_{4}$ and $\textrm{Tr}[Q^{(-1)}] = -2$.
 
The mass parameter $m$ and $Q$, which is parameterized by the six angle
parameters, make it seven variational parameters to describe the degenerate
ground states of $\mathcal{H}_{0}$ at $\sigma_H=-1$.

\subsubsection{Ground state manifold at $\sigma_{H} = 0$} \label{gs nu0}

For the ground state at $\sigma_{H} = 0$, the diagonal mass matrix is,
\begin{equation} \label{massd nu0}
	M_{D} = \{m, m, -m, -m \} .
\end{equation}
The manifold for the ground state in this case is a coset space  $U(4)/(U(2)
\times U(2))$. The dimension of this coset space is eight. 

For this case, we need to construct two four component vectors that specify
the $SU(4)$ components of the two occupied members of the $n=0$ Landau level
quartet. Eq.\eqref{nu1 ket1} describes the first vector. The second vector is
orthonormal to Eq.\eqref{nu1 ket1} and can be constructed explicitly,
requiring two more angle parameters, 
\begin{equation} \label{nu0 ket2} 
	| \chi^{2}\rangle = \cos( \frac{\gamma_{2}}{2} ) 
	| + \rangle | - \bm{n}_{1} \rangle 
	+ \me^{\mi \Omega_{2}} \sin ( \frac{\gamma_{2}}{2} )
	| - \rangle | \bm{n}_{2} \rangle .
\end{equation}
The mass matrix is of the form, 
\begin{equation} \label{mass nu0}
	M = m \left( 2 \left(| \chi^{1} \rangle \langle \chi^{1} | 
	+ | \chi^{2} \rangle \langle \chi^{2} |\right) - \mathbbmss{1}_{4} \right)
	\equiv mQ^{(0)} , 
\end{equation}
where $Q^{(0)}(\gamma_{1}, \Omega_{1}, \gamma_{2}, \Omega_{2}, \theta_{1},
\phi_{1}, \theta_{2}, \phi_{2})$ is a $4\times 4$ matrix valued function of
the eight angle parameters. It is in one-to-one correspondence with the
elements of the coset space $U(4)/\left(U(2)\times U(2)\right)$. It satisfies
the properties, $(Q^{(0)})^{2} = \mathbbmss{1}_{4}$ and
$\textrm{Tr}[Q^{(0)}]=0$.

The mass parameter $m$ and $Q$, which is parameterized by the eight angle
parameters, make it nine variational parameters to describe the degenerate
ground states of $\mathcal{H}_{0}$ at $\sigma_{H}=0$.

\subsection{Excitation states and gaps} \label{sec:symm gaps}

In this subsection we will compute the activation gaps for $\mathcal{H}_{0}$,
which is the excitation energy of a well separated quasi-particle and
quasi-hole pair. In our case, the excited states at Hall conductivity,
$\sigma_{H} = 0, - 1$, the particle and the hole quantum numbers belong to
different members of the $n=0$ Landau level quartet.  The excited state is
defined as,
\begin{equation} \label{excited state}
| ES \rangle = \psi^{\dagger}_{0,l_{p},q_{p}} \psi_{0,l_{h},q_{h}}
| GS \rangle .
\end{equation}
The activation gap is evaluated from the expectation value of the hamiltonian
for the excited state and for the ground state, defined as the following, 
\begin{equation} \label{gaps}
	\Delta_{gap} = \frac{1}{2} \big( \langle ES | \mathcal{H} | ES \rangle - 
	\langle GS | \mathcal{H} | GS \rangle \big) .
\end{equation}
The ground state expectation value of the hamiltonian is expressed in terms of
two point correlators, used for variational state energy calculation. In a
similar fashion, we can define the two point correlator for excited states,
$\Upsilon(\bm{r}_{1}, \bm{r}_{2})$ as, 
\begin{equation}
	\Upsilon_{r,A;s,B} (\bm{r}_{1}, \bm{r}_{2})  = 
	\langle ES | \Psi^{\dagger}_{s,B}(\bm{r}_{2}) \Psi_{r,A}(\bm{r}_{1}) 
	| ES \rangle .
\end{equation}
This correlator is an $8 \times 8$ matrix and can be expressed as following,
\begin{equation} \label{2pt es}
	\Upsilon (\bm{r}_{1}, \bm{r}_{2}) = 
	\Gamma^{(p)}(\bm{r}_{1}, \bm{r}_{2})
	- \Gamma^{(h)}(\bm{r}_{1}, \bm{r}_{2})
	+ \Gamma(\bm{r}_{1}, \bm{r}_{2}) .
\end{equation}
Here $\Gamma^{(p)}(\bm{r}_{1}, \bm{r}_{2})$ and $\Gamma^{(h)}(\bm{r}_{1},
\bm{r}_{2})$ are the two point correlators for the particle and the hole
states respectively and $\Gamma(\bm{r}_{1}, \bm{r}_{2})$ is the two point
correlator for the ground state.

The correlator for the particle or hole, (${\tt x} = p,h$)
\begin{multline} \label{correlator x}
	\Gamma^{({\tt x})}(\bm{r}_{1}, \bm{r}_{2}) 
	= \Phi^{0,l_{{\tt x}},q_{{\tt x}}}(\bm{r}_{1}) 
	{\Phi^{0,l_{{\tt x}},q_{{\tt x}}}}^{\dagger}(\bm{r}_{2}) \\
	= \frac{1}{4 \pi}
	( \mathbbmss{1}_{2} - \beta ) \frac{1}{l_{{\tt x}}!} 
	\Big( \frac{\bar{z}_{1} z_{2}}{2} \Big)^{l_{{\tt x}}} \\
	\me^{- \frac{1}{4}(\bar{z}_{1} z_{2} + z_{1} \bar{z}_{2})}
	\me^{- \frac{1}{4}|z_{1} - z_{2}|^{2}} 
	P_{q_{{\tt x}}} .
\end{multline}

Within the symmetric model ($\mathcal{H} = \mathcal{H}_{0}$ in
Eq.\eqref{gaps}), we will show that the $SU(4)$ component of the correlator
gets traced out and there is no angle dependence for the activation gaps.
Moreover, we consider the case where the particle and the hole are separated
by a large distance, hence excluding the possibility of particle-hole bound
states.

The contribution to the activation gap from the kinetic term is evaluated as, 
\begin{multline}
	\Delta_{t} = \frac{\kappa_{t}}{2} 
	\Big( \int_{\bm{r}} \lim_{\bm{r}_{0} \to \bm{r}} 
	\textrm{Tr}[\mathrm{h} \, \Upsilon(\bm{r} , \bm{r}_{0})] \\
	- \int_{\bm{r}} \lim_{\bm{r}_{0} \to \bm{r}} 
	\textrm{Tr}[\mathrm{h} \, \Gamma(\bm{r} , \bm{r}_{0})] \Big) .
\end{multline}
We find no contribution to the activation gap from the kinetic term. This is
not surprising because of the fact that the $n=0$ Landau level wavefunctions
are the same for both the massless and the massive Dirac particle. And the
massless Dirac particle has zero eigenvalue for the $n=0$ Landau level.

\begin{figure}
	\begin{center}
		\includegraphics[width=0.45\textwidth]{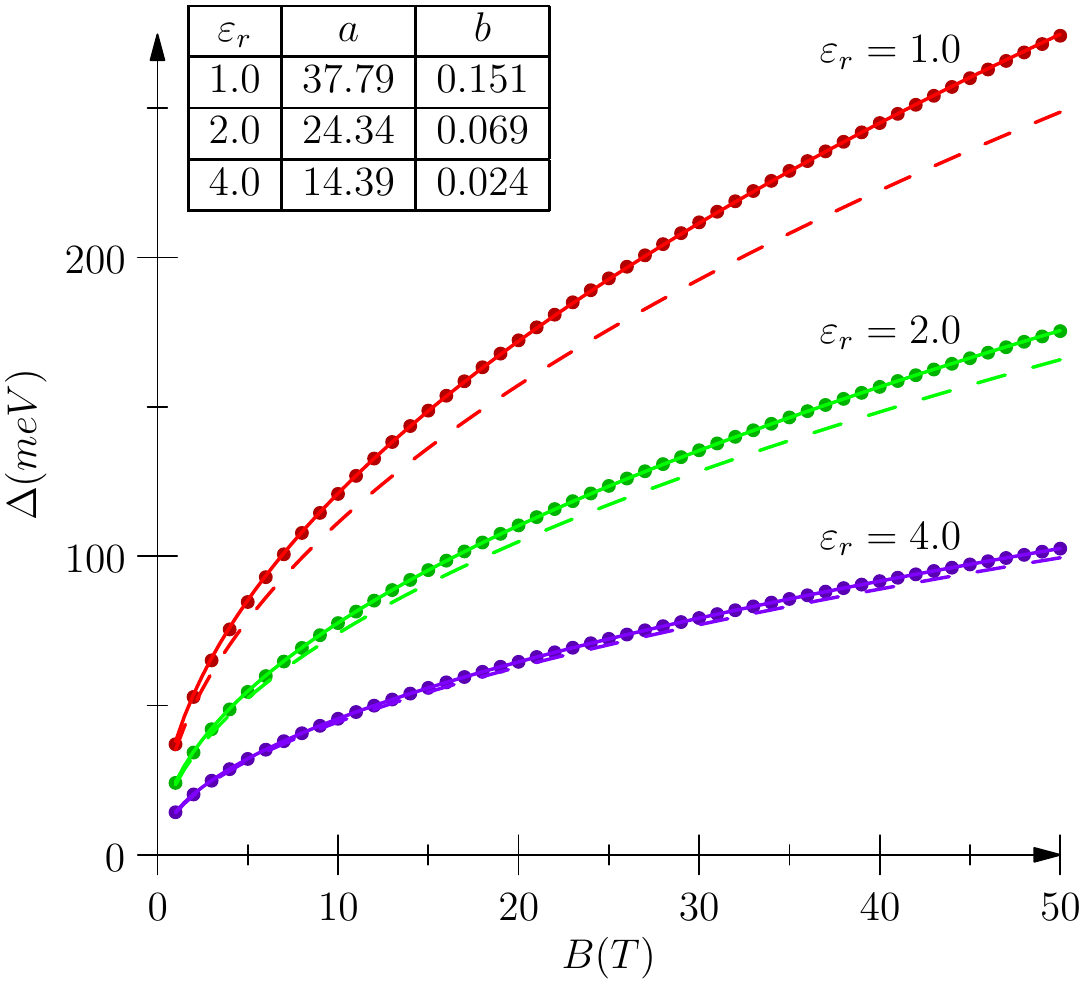}
	\end{center}
		\caption{This figure shows the variation of activation gap with
			magnetic field. The gaps for the Hall conductivity at
			$\sigma_{H}=0$ and $\sigma_{H}=-1$ are same. The solid points are
			values for the gap obtained from our calculations. The solid line
			shows the best fit for the gaps as function of magnetic field of
			the form $a \sqrt{B} + b \, B$.  $a,b$ are best fit parameters
			shown in the inset for three dielectric constants. We have shown
			the variation of gaps with change in the dielectric constant of
			the substrate. The dielectric constant $\varepsilon_{r}=1$
			corresponds to the suspended graphene. The best fit showed the
			dominant $\sqrt{B}$ contribution and linear $B$ decreases as the
			dielectric constant of substrate increased. The dashed lines shows
			the gaps when the effects of the filled Dirac sea are ignored. 
		}
\label{gap Vs B}
\end{figure}

The large distance separating the particle and hole results in Coulomb term
contributions to the activation gap which is given by,
\begin{multline} 
	\Delta_{C} = \kappa_{C}
	\iint\limits_{\bm{r}_{1}, \bm{r}_{2}} 
	\frac{1}{|\bm{r}_{1} - \bm{r}_{2}|} 
	\Big( \textrm{Tr}[\Gamma^{(h)}(\bm{r}_{1}, \bm{r}_{2}) 
	\Gamma(\bm{r}_{2},\bm{r}_{1})] \\
	- \textrm{Tr}[\Gamma^{(p)}(\bm{r}_{1},\bm{r}_{2})
	\Gamma(\bm{r}_{2},\bm{r}_{1})] \Big) .
\end{multline}
The gaps turn out to be the same for both the cases, $\sigma_{H}=0$ and
$\sigma_{H}=-1$. The ground state correlators are specified by their
respective mass matrix and we also use the fact from energy minimization,
$m_{h} = m = - m_{p} $.

The activation gap within the symmetric model for the Hall conductivity at
$\sigma_{H}=0,-1$, is  
\begin{equation} \label{gap coulomb}
	\Delta_{gap} = \kappa_{C} \, \eta_{C_{1}}(m) .
\end{equation}
The coefficient $\eta_{C_{1}}(m)$ is evaluated numerically,
\begin{equation}
	\eta_{C_{1}} = \frac{m}{2} 
	\int_{\frac{1}{2 N_{C}}}^{\infty}
	\dif s \frac{\me^{-m^{2} (s - \frac{1}{2})}}{\sqrt{s \, \sinh(s)}} .
\end{equation}
The variation of the particle-hole gap with the magnetic field is shown in
Fig.\ref{gap Vs B}. We observe that the gaps have dominant $\sqrt{B}$
contribution and a few percent contribution from the linear term.

We have compared the results of our analysis with the case when the effects of
the filled Dirac sea are ignored. In Fig.\ref{gap Vs B}, we have also
presented the variation of gaps with dielectric constant of the substrate. At
low magnetic fields, there is negligible contribution from the filled Dirac
sea.  Gaps shows systematic decrease in the contribution from the filled Dirac
sea with the increase of dielectric constant.

\section{Symmetry breaking terms} \label{sec:symmetry breaking}
In this section, we investigate the effects of the symmetry breaking terms of
our interacting model, Eq.\eqref{ham full}, on the ground state and
excitations at $\sigma_{H}=0,-1$. The symmetry breaking terms are a factor of
$a / \ell_{c}$ smaller when compared with the $SU(4)$ symmetric terms. Hence,
a change to the minimum value of the mass parameters obtained from the
symmetric model analysis will be smaller by a factor of  $a / \ell_{c}$.  The
range of parameter ($U$ and $V$) values of our interest and $a / \ell_{c} \ll
1$, allow us to conclude that the effect of the symmetry breaking terms on the
minimum value of the mass parameter is negligibly small. We fix the mass
parameter values from the symmetric model analysis, Eq.\eqref{mass nu0} and
Eq.\eqref{mass nu1} at $\sigma_{H}=0, -1$ respectively. This amounts to
treating the symmetry breaking terms as perturbations about the symmetric
model solutions. 

The variational state energy can be expressed in terms of various traces of
the coincident correlator, Eq.\eqref{gamma}, which takes the form, 
\begin{equation} \label{2pt coincident}
	\Gamma = \frac{1}{4 \pi} 
	(u_{m} \mathbbmss{1}_{2} - v_{m} \beta) \otimes
	Q^{(\sigma_{H})} .
\end{equation}
Here $Q^{(\sigma_{H})}$ are $4 \times 4$ matrices specified at $\sigma_{H} =
0, -1$ discussed in Sec.\ref{gs nu0} and Sec.\ref{gs nu1}. 
\begin{equation}
	Q^{(\sigma_{H})} = 
	\sum_{j \in \textrm{occ}} P_{j} - \sum_{j \in \textrm{unocc}} P_{j}
	= 2 \sum_{j \in \textrm{occ}} P_{j} - \mathbbmss{1}_{4} ,
\end{equation}
is constructed from the $SU(4)$ components of the occupied members of the $n=0$
Landau level quartet. The coincident correlator is position independent,
the spatial integration is trivial and results in volume of the system. We
derive the general expression for the expectation value of symmetry
breaking terms in terms of various traces of the coincident correlator and
then apply to the specific cases of $\sigma_{H}=0$ and $\sigma_{H}=-1$. 

The variational energy density for the nearest neighbour term, $
\mathcal{E}_{V} \equiv \langle \mathcal{H}_{V} \rangle $, is expressed in
terms of the traces of the coincident correlator as following,
\begin{multline} \label{E_V}
	\mathcal{E}_{V} = \frac{\kappa_{V}}{4 \pi^{2}} 
	\Big( - \textrm{Tr}[\Gamma \, \Gamma]
	- \big( \textrm{Tr}[\beta \, \tau^{z} \, \Gamma] \big)^{2} \\
	+ \textrm{Tr}[\beta \, \tau^{z} \, \Gamma \, 
	\beta \, \tau^{z} \, \Gamma]
	\, \Big) .
\end{multline}
Similarly, $\mathcal{E}_{U} \equiv \langle \mathcal{H}_{U} \rangle $,
the variational state energy density for the Hubbard term,
\begin{multline} \label{E_H}
	\mathcal{E}_{U} = \frac{\kappa_{U}}{4 \pi^{2}} 
	\bigg( - \textrm{Tr}[\Gamma \Gamma]
	+ \big( \textrm{Tr}[\beta \tau^{z} \Gamma] \big)^{2}
	- \textrm{Tr}[\beta \tau^{z} \Gamma  
	\beta \tau^{z} \Gamma] \\
	+ \frac{1}{2} \sum_{j,k=x,y} \Big( 
	\big( \textrm{Tr}[\alpha^{j} \tau^{k} \Gamma] \big)^{2}
	- \textrm{Tr}[\alpha^{j} \tau^{k} \Gamma 
	\alpha^{j} \tau^{k} \Gamma] \Big) \bigg) .
\end{multline}
The Zeeman energy density, $\mathcal{E}_{Z} \equiv \langle
\mathcal{H}_{Z} \rangle $, in terms of the coincident correlator,
\begin{equation} \label{E_Z}
	\mathcal{E}_{Z} = - 
	\frac{\kappa_{Z}}{2 \pi} \, \textrm{Tr}[\sigma^{z} \, \Gamma] .
\end{equation}
The expectation value of $\mathcal{H}_{t_{1}}$ is obtained by a procedure
similar to the one discussed in Sec.\ref{subsec:symm} for the kinetic term. 
\begin{equation} \label{avg t1}
	\langle \mathcal{H}_{t_{1}} \rangle = 
	\kappa_{t_{1}} \,
	\int_{\bm{r}} \lim_{\bm{r} \to \bm{r}_{0}} 
	\textrm{Tr}[\mathrm{h}_{t_{1}} \, \Gamma(\bm{r}, \bm{r}_{0})] .
\end{equation}
$ {\displaystyle \lim_{\bm{r} \to \bm{r}_{0}} }
	\mathrm{Tr}[\mathrm{h}_{t_{1}} \Gamma(\bm{r}, \bm{r}_{0})] 
= \sum_{q=1}^{4}f_{1}(m_{q}^{2}) \textrm{Tr}[\tau^{z}] $,
here $f_{1}(m_{q}^{2})$ results from the action of the off-diagonal elements
of $\mathrm{h}_{t_{1}}$ on  $b_{m}(\bm{r}, \bm{r}_{0})$, Eq.\eqref{b_m}, and
$d_{m}(\bm{r}, \bm{r}_{0})$, Eq.\eqref{d_m}. The trace over the $SU(4)$
indices vanishes, hence $ \langle \mathcal{H}_{t_{1}} \rangle = 0 $.
Henceforth we drop the sub-leading kinetic term. 

The net variational state energy, $\mathcal{E}_{1} \equiv \langle
\mathcal{H}_{1} \rangle$, from the symmetry breaking terms,
\begin{equation} \label{E_1}
	\mathcal{E}_{1} = \mathcal{E}_{V} + \mathcal{E}_{U} +\mathcal{E}_{Z} .
\end{equation}
is numerically minimized for the angle parameters and hence determines the
$SU(4)$ polarization of the variational ground state.

\subsection{Ground states at $\sigma_{H} = 0$} \label{sec:gnd0}

The mass matrix for the $\sigma_{H}=0$, $M_{D} = \{ m, m , -m, -m \}$ and
$Q^{(0)}$ parameterization scheme discussed in Sec.\ref{gs nu0} are included
in the two point coincident correlator Eq.\eqref{2pt coincident}. Various
traces are evaluated with this two point coincident correlator to obtain the
contribution from the symmetry breaking terms as functions of the variational
angle parameters.

\begin{figure}
	\begin{center}
		\includegraphics[width=0.42\textwidth]{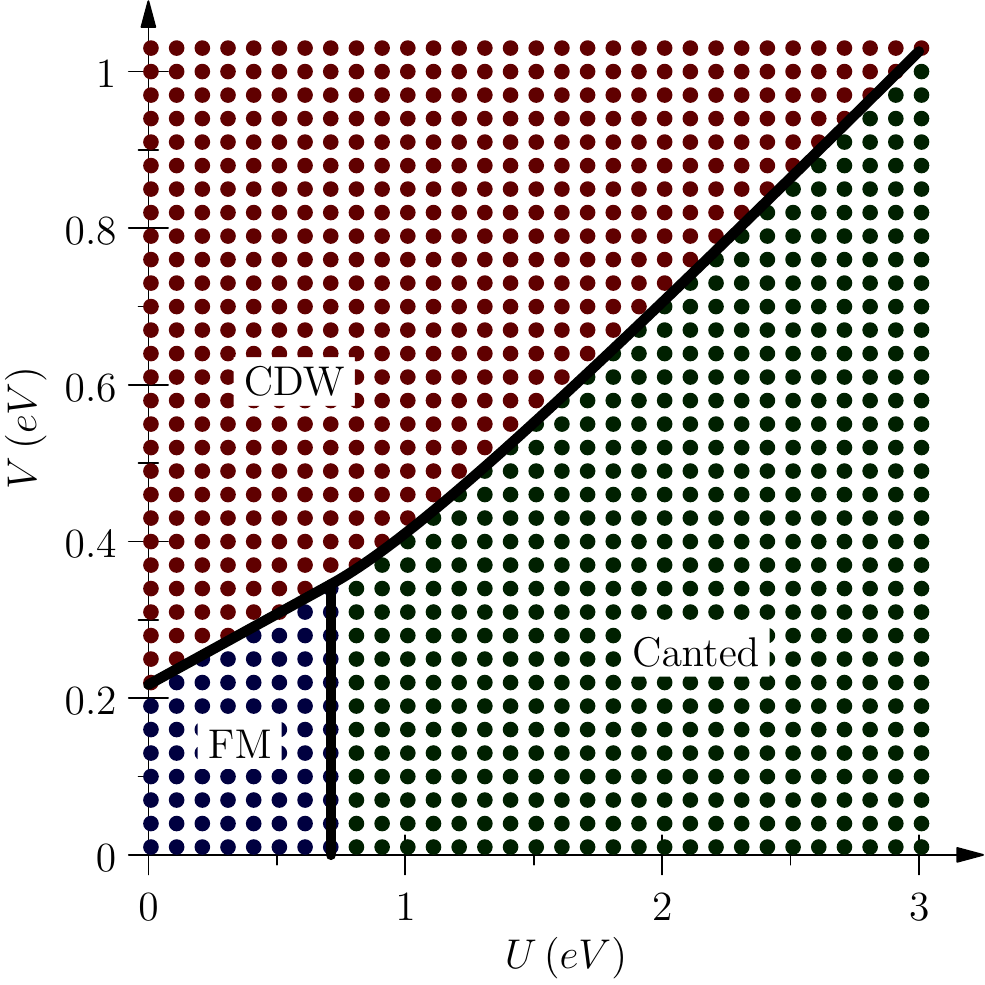}
	\end{center}
	\caption{The figure shows the phase diagram for possible ground states for
		the Hall conductivity at $\sigma_{H}=0$ in $U$-$V$ parameter space.
		This phase diagram is for the magnetic field $B = 30 \, T$ and 
		$ \varepsilon_{r} = 1.0$. The region
		marked `FM' is the ferromagnetic ordered state, `CDW' is the charge
		ordered state and `Canted' is the canted spin ordered state.  The three
		regions marked with dots represents solutions obtained with numerical
		minimization corresponding to three phases. The bold line separating
		the three phases were obtained analytically after equating each pair of
		analytical expressions for the variational energies.  Equations for
		these lines are given in Eq.\eqref{nu0 line1}, Eq.\eqref{nu0 line2} and
		Eq.\eqref{nu0 line3}. Phase transition across the `FM' and `Canted'
		region is continuous whereas others are first order transition.
	}
	\label{fig pd nu0}
\end{figure}
The nearest neighbour interaction energy density,
\begin{multline} \label{nu0 nnbr}
	\mathcal{E}_{V} = - \frac{\kappa_{V}}{4 \pi^{2}} 
	\bigg( 
	8 v_{m}^{2} 
	\cos(\gamma_{1}) \cos(\gamma_{2}) \\
	+ 2  \big( v_{m}^{2} - u_{m}^{2} \big) 
	\big( \cos^{2}(\gamma_{1}) + \cos^{2}(\gamma_{2}) \big) 
	\bigg) ,
\end{multline}
is a function of only two angle parameters $\gamma_{1} \textrm{ or }
\gamma_{2}$. Since $v_{m} > u_{m}$, the minimization of the nearest neighbour
energy is solely decided by the first term in Eq.\eqref{nu0 nnbr}. It
minimizes for the values $\gamma_{1} = \gamma_{2} = 0 \textrm{ or } \pi$,
which results in a charge ordered state.

The variational state energy density for the Hubbard term,
\begin{multline} \label{nu0 hubbard}
	\mathcal{E}_{U} = \frac{\kappa_{U}}{4 \pi^{2}} 
	\bigg( 8 v_{m}^{2} \cos(\gamma_{1}) \cos(\gamma_{2}) \\
	+ 2 \big( v_{m}^{2} - u_{m}^{2} \big) 
	\big( \cos(\gamma_{1}) - \cos(\gamma_{2}) \big)^{2} 
	|\langle \bm{n}_{1} | \bm{n}_{2} \rangle|^{2} \bigg) ,
\end{multline}
minimizes for the values $\gamma_{1} = 0 \textrm{ or } \pi$, $\gamma_{2} = \pi
- \gamma_{1}$ and $|\langle \bm{n}_{1} | \bm{n}_{2} \rangle|^{2} = 0$. These
values indicate that the Hubbard term prefers a Neel ordered ground state. 

The Zeeman energy contribution,
\begin{equation} \label{nu0 zeeman}
	\mathcal{E}_{Z} = -  \frac{\kappa_{Z}}{2 \pi} u_{m} 
	\big( \cos(\gamma_{1}) - \cos(\gamma_{2}) \big)
	\big( \cos(\theta_{1}) + \cos(\theta_{2}) \big) ,
\end{equation}
minimizes for the variational angle parameters that correspond to a
ferromagnetic ordered ground state, for example: $\gamma_{1} = 0$, $\gamma_{2}
= \pi$, $\theta_{1} = 0$ and $\theta_{2} = \pi$. 

The angle parameters that minimize the total energy $\mathcal{E}_{1}$ for
$\sigma_{H}=0$ were numerically obtained. Numerical solutions correspond to
three phases in the $U$-$V$ parameter space shown in Fig.\ref{fig pd nu0} and
described below.

\subsubsection{Charge ordered state}  \label{CDW}

The region marked `CDW' shown in Fig.\ref{fig pd nu0} is the charge
ordered ground state. The angle parameters $\gamma_{1} = \gamma_{2} = 0
(\textrm{or } \pi)$, minimize $\mathcal{E}_{1}$, which result in doubly
degenerate ground states. This is clearly indicated by the two $Q$ matrices,
which take the form,  
\begin{equation}
	Q^{(0)}_{\textrm{CDW}} = \pm \, \tau^{z} \otimes \mathbbmss{1}_{2} .
\end{equation}

\begin{figure}
	\centering
	\subfloat[]{
		\includegraphics[width=0.44\columnwidth]{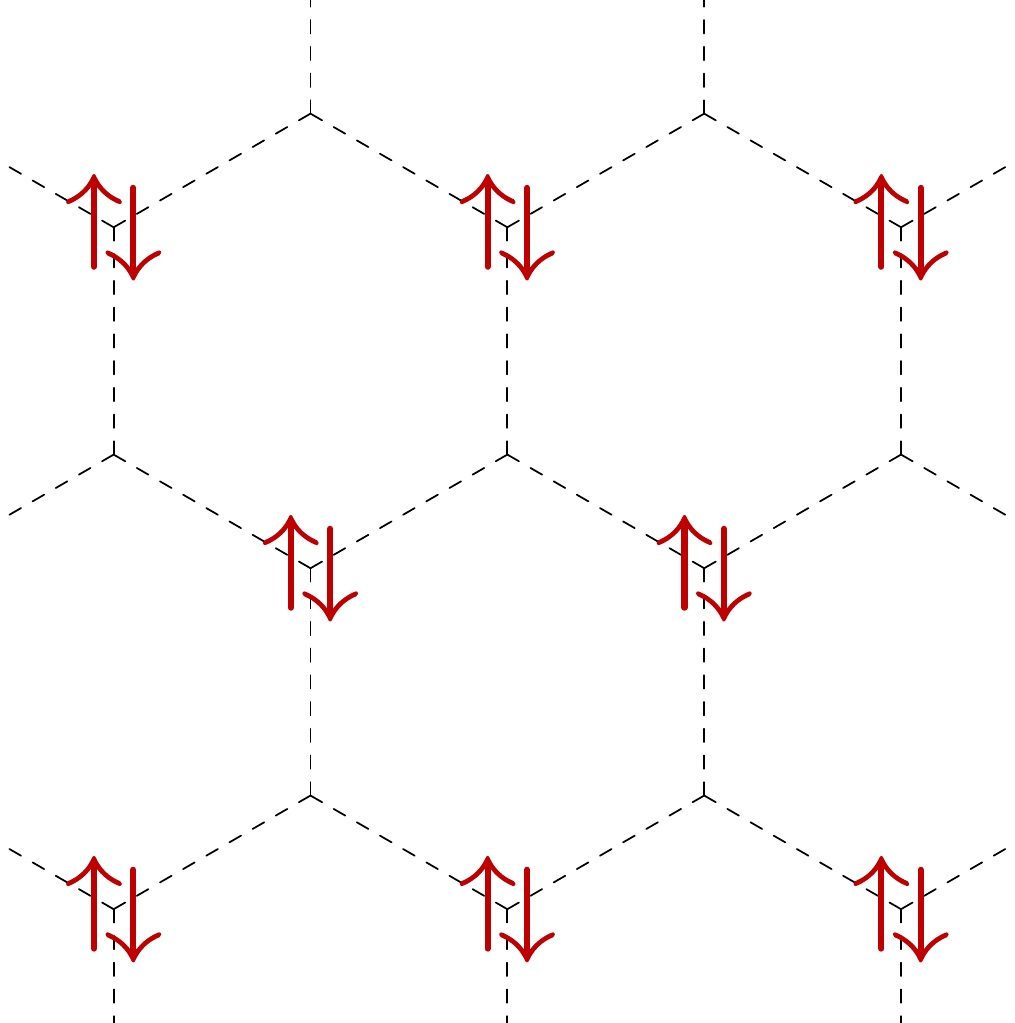}
			\label{fig cdw:latt}
		}
		\hfill
		\subfloat[]{
			\includegraphics[width=0.44\columnwidth]{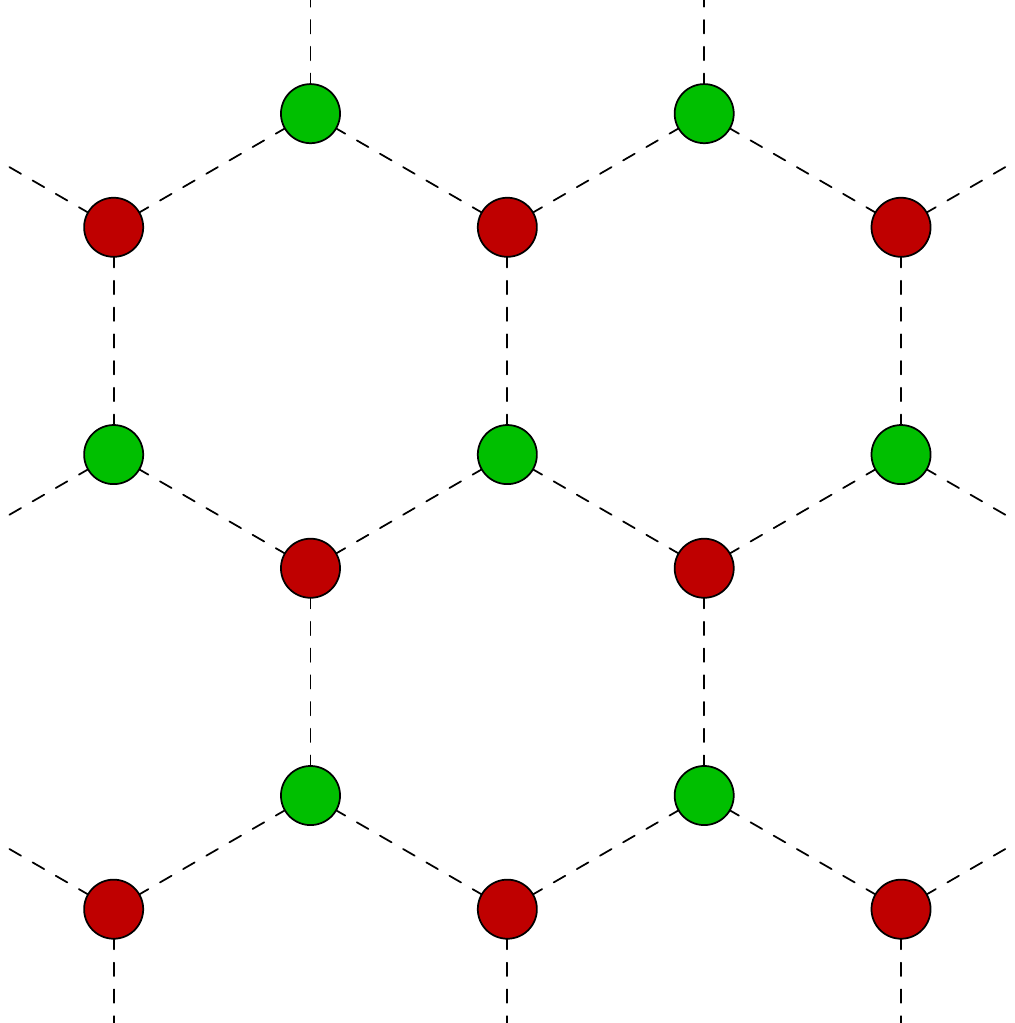}
			\label{fig cdw:op}
		}
		\caption{One of the arrangement of $SU(4)$ components of the $n=0$
			Landau levels for the doubly degenerate charge ordered ground
			state shown in (a), anti-parallel spins located on same
			sub-lattice. The site order parameter for charge ordering shown in
			(b). Two colors represent the staggered charge densities on two
			sublattices. 
		}
\end{figure}

The $SU(4)$ components of the two occupied quartets of the $n=0$ Landau levels
can be either $| + \rangle | \bm{n} \rangle, | + \rangle | - \bm{n} \rangle $
or $| - \rangle | \bm{n} \rangle, | - \rangle | - \bm{n} \rangle$. This
results in an additional charge localized on one of the sublattice points.
One of the arrangements on the graphene lattice is shown in Fig.\ref{fig
cdw:latt}.  There are two order parameters for the charge ordered ground
state, $\langle \mathcal{T}^{30}(\bm{x}) \rangle $ and $\langle
\widetilde{\mathcal{T}}^{30}(\bm{x}) \rangle$. The $SU(4)$ polarization of
operator $\tau^{z}$ is the order parameter $\langle \mathcal{T}^{30}(\bm{x})
\rangle $ which results in equal bond order between the same sublattice on the
graphene lattice. The staggered $SU(4)$ polarization for the $\tau^{z}$
operator, $\langle \widetilde{\mathcal{T}}^{30}(\bm{x}) \rangle$ corresponds
to a staggered charge distribution on the two sublattices for graphene as
shown in Fig.\ref{fig cdw:op}. This site order parameter has dominant filled
Dirac sea contributions.

\subsubsection{Ferromagnetic ordered state} \label{FM}

\begin{figure}
	\centering
	\subfloat[]{
		\includegraphics[width=0.44\columnwidth]{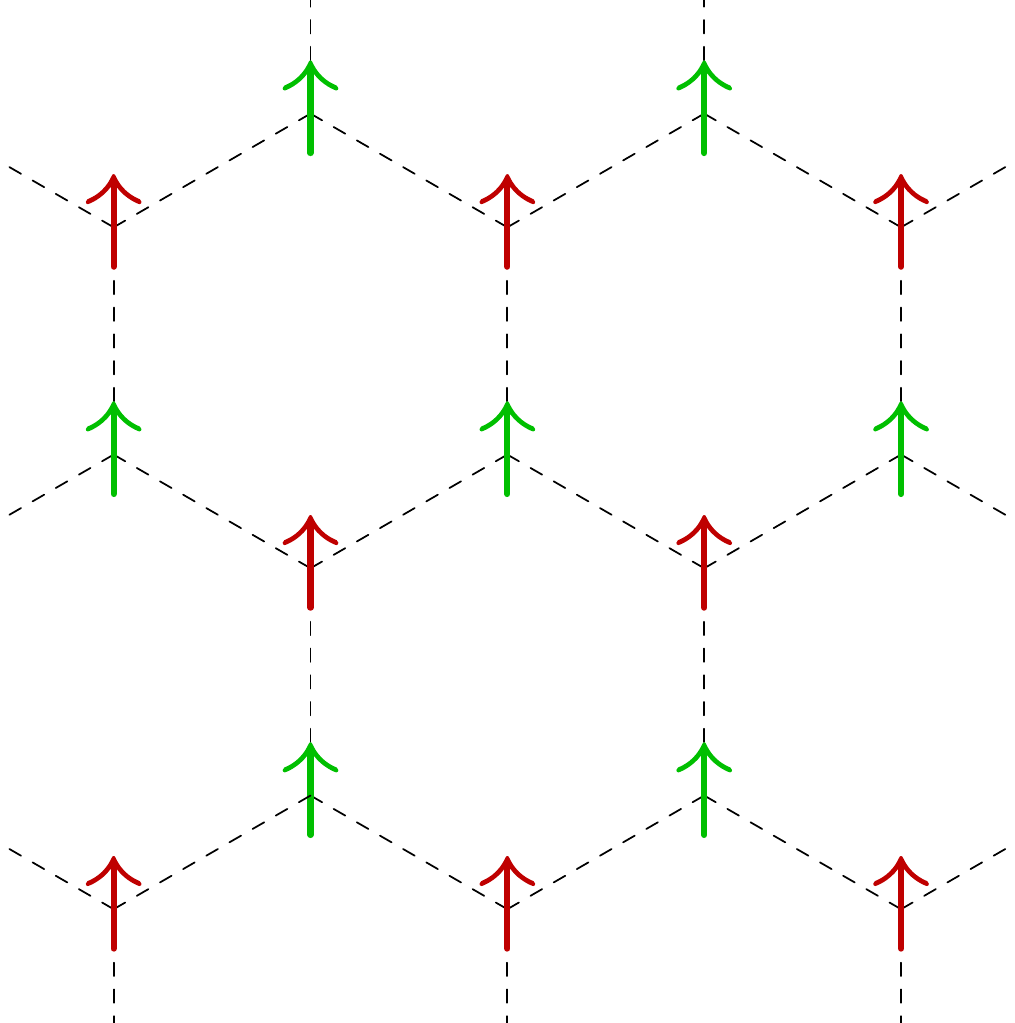}
			\label{fig fm:latt}
		}
		\hfill
		\subfloat[]{
			\includegraphics[width=0.44\columnwidth]{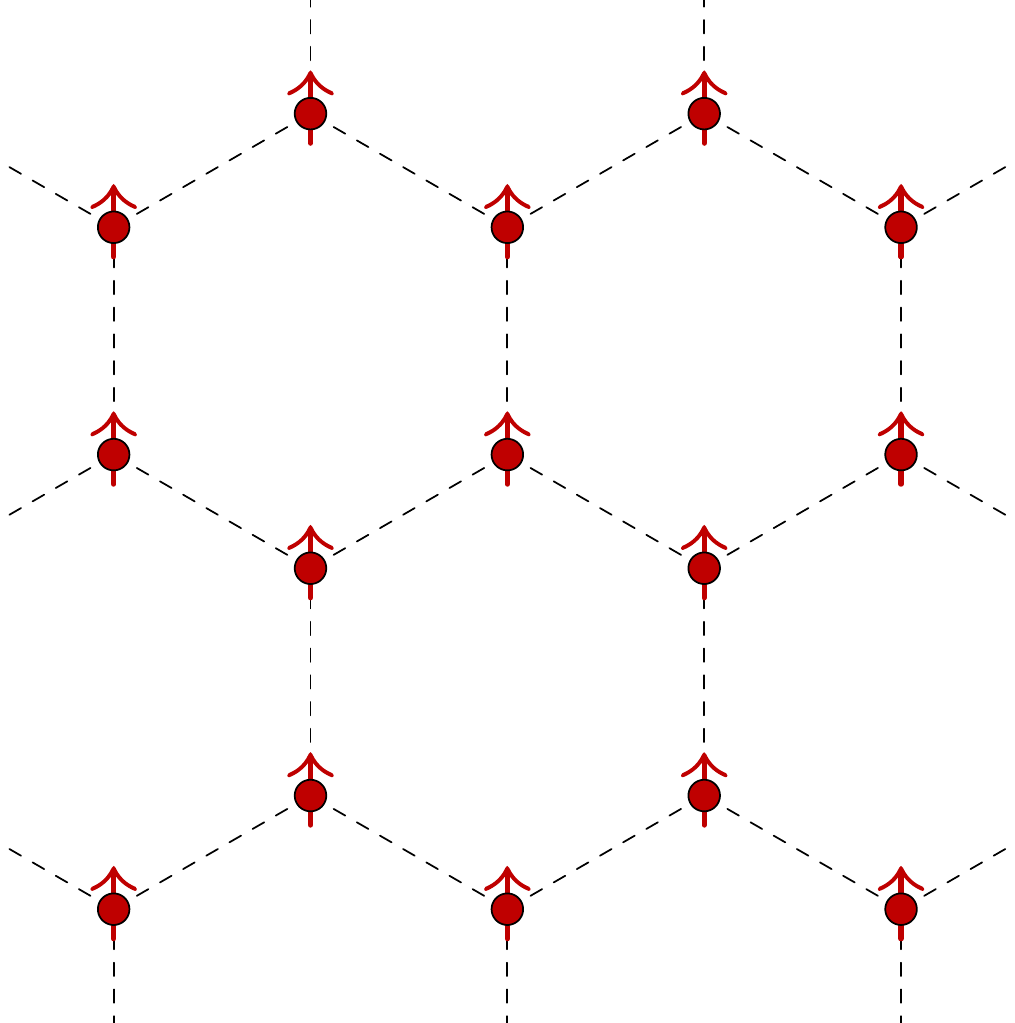}
			\label{fig fm:op}
		}
		\caption{The $SU(4)$ components of the $n=0$ Landau level for the
			ferromagnetic ground states is shown in (a). The spin pointing along
			the direction of the magnetic field on each sub-lattice shown with
			different colors. The site order parameter for the ferromagnetic
			state, the total spin shown in (b). The small circle size
			indicates that this site order parameter magnitude comes from the
			half-filled $n=0$ Landau level quartet.
	}
\end{figure}

The region marked `FM' for small values of $U$ and $V$ in Fig.\ref{fig pd nu0}
is the ferromagnetic ordered ground state. The minimization of
$\mathcal{E}_{1}$ results in the angle parameter taking values,  $\gamma_{1} =
0 (\pi)$, $\gamma_{2} = \pi (0)$, $\varphi_{1} - \varphi_{2} = \pi$,
$\theta_{1} = \theta_{2} = 0 (\pi)$. For these values of the angle parameters,
the $SU(4)$ components of the two occupied quartet of the $n=0$ Landau levels
are $| + \rangle | \! \uparrow \rangle$ and $| - \rangle | \! \uparrow
\rangle$, shown in Fig.\ref{fig fm:latt}. The ferromagnetic ordered state
completely lifts the $SU(4)$ degeneracy and this is indicated by the $Q$
matrix, which takes the form, 
\begin{equation}
	Q_{\textrm{FM}}^{(0)} = \mathbbmss{1}_{2} \otimes \sigma^{z} .
\end{equation}

The order parameters for the ferromagnetic ground state are $\langle
\mathcal{T}^{03}(\bm{x}) \rangle$ and  $\langle
\widetilde{\mathcal{T}}^{03}(\bm{x}) \rangle $. $SU(4)$ polarization of
$\sigma^{z}$ is  $\langle \mathcal{T}^{03}(\bm{x}) \rangle$, which at lattice
level corresponds to total spin polarization of the system along the direction
of the magnetic field, shown in Fig.\ref{fig fm:op}. The staggered $SU(4)$
polarization of $\sigma^{z}$ is $\langle \widetilde{\mathcal{T}}^{03}(\bm{x})
\rangle $, which at lattice level results in bond order parameters for the two
sublattices with equal magnitude but opposite signs. This is analogous to the
Haldane mass term\cite{haldane}.

\subsubsection{Canted-spin ordered state} \label{Canted}
	
The canted-spin ordered ground state in $U$-$V$ parameter space is shown
in Fig.\ref{fig pd nu0} with the region marked `Canted'. The angle
parameters that minimize $\mathcal{E}_{1}$ are $\gamma_{1} = 0 (\pi)$,
$\gamma_{2} = \pi (0)$, $\varphi_{1} - \varphi_{2} = \pi$ and $\theta_{1} =
\theta_{2} = \theta_{0}$, with 
\begin{equation} \label{cant angle}
	\cos(\theta_{0}) = +(-) \frac{1}{4}
	\frac{u_{m}}{v_{m}^{2} - u_{m}^{2}} \frac{2 \pi \kappa_{Z}}{\kappa_{U}} .
\end{equation}

\begin{figure}
	\centering
	\subfloat[]{
		\includegraphics[width=0.65\columnwidth]{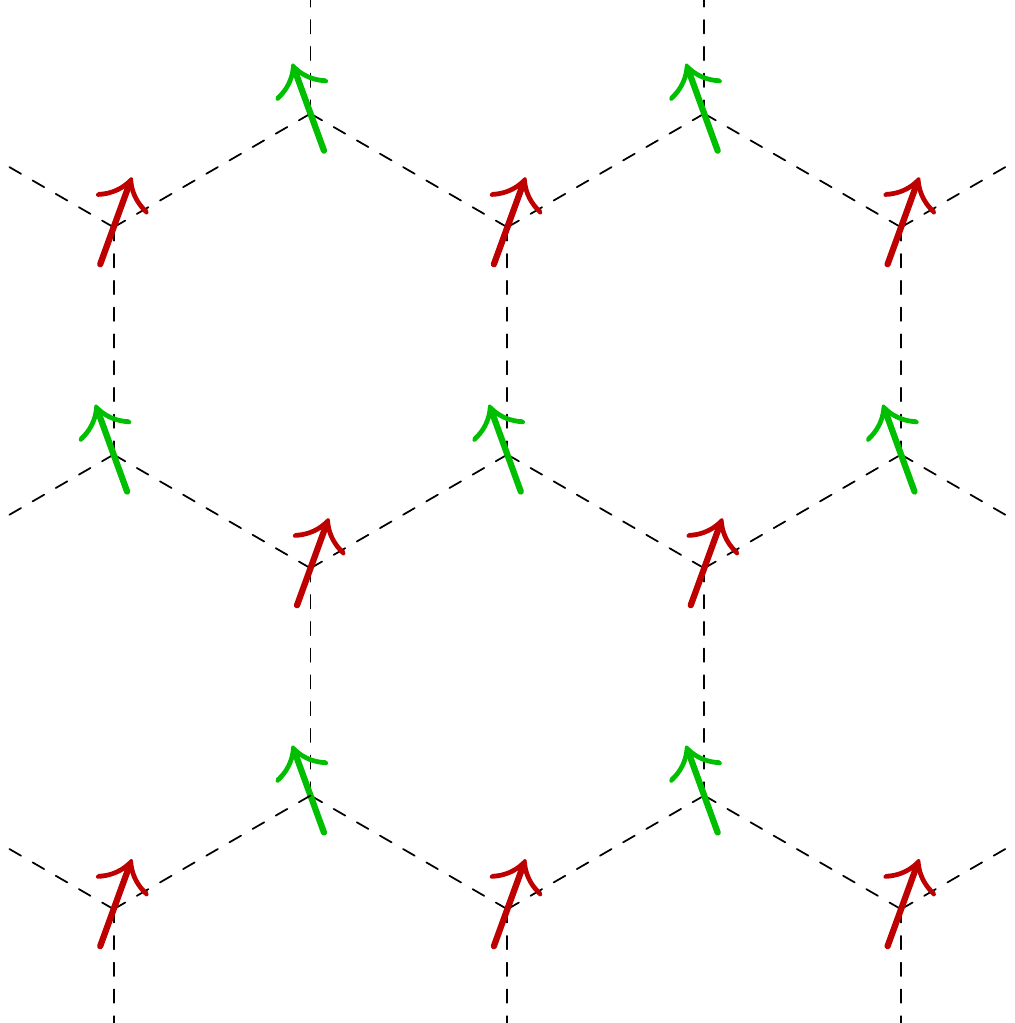}
			\label{fig canted:latt}
		} \\
		\subfloat[]{
			\includegraphics[width=0.44\columnwidth]{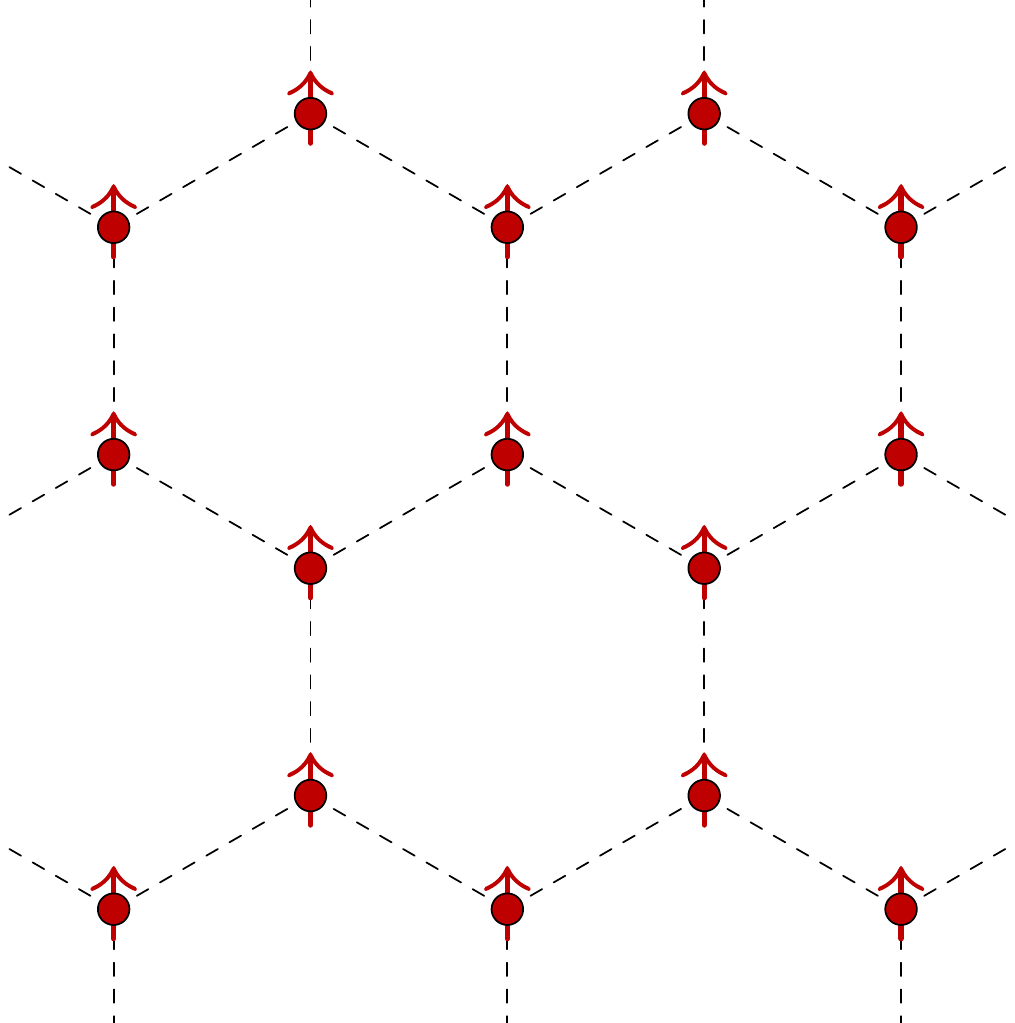}
			\label{fig canted:op1}
		} 
		\hfill
		\subfloat[]{
			\includegraphics[width=0.44\columnwidth]{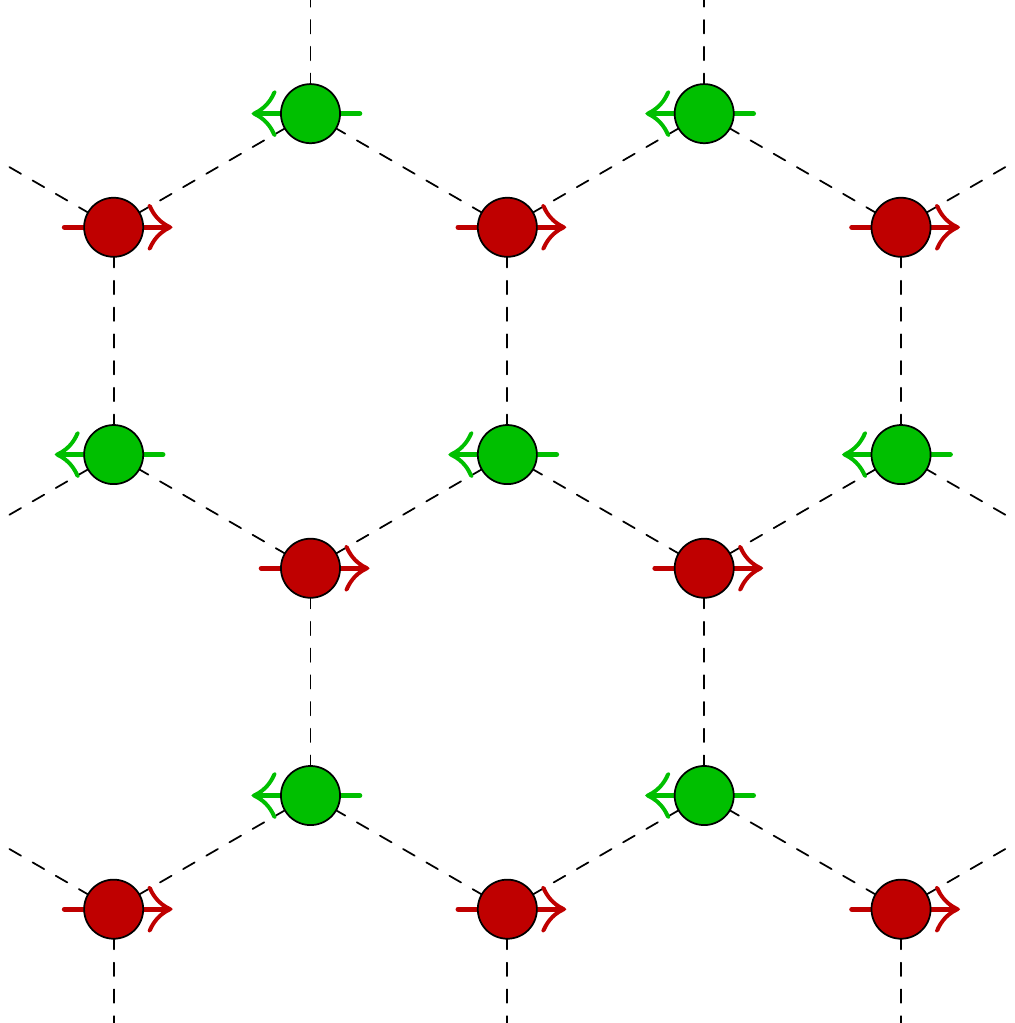}
			\label{fig canted:op2}
		}
		\caption{The $SU(4)$ components of the $n=0$ Landau level for the canted
			ground state shown in (a). The spins on sub-lattice shown in
			different colors with cant angle $2 \theta_{0}$ between them. The
			site order parameter in (b), the total spin along the direction of
			the magnetic field on each sub-lattice which has contributions on
			from the occupied quartets of the $n=0$ Landau levels shown with
			small circles. (c) shows the order parameter corresponding to Neel
			ordering in the direction perpendicular to the magnetic field.
			Larger circle size compared to order parameter in (b) is to
			indicate contributions from filled Dirac sea to this site order
			parameter.
		}
\end{figure}

The $SU(4)$ components of the $n=0$ Landau level for the canted ground state
are $| + \rangle | \bm{n}_{1} \rangle$, $| - \rangle | \bm{n}_{2} \rangle$ and
is shown in Fig.\ref{fig canted:latt}. $SU(4)$ degeneracy is not completely
lifted, there is a remnant $U(1)$ rotation as only $\varphi_{1} - \varphi_{2}$
is fixed by the minimization. There is a family of $Q$ matrices for the canted
ground state parameterized by the parameter $\varphi$ shown below, 
\begin{multline}
	Q_{\textrm{Canted}}^{(0)}(\varphi) = \cos(\theta_{0}) 
	\mathbbmss{1}_{2} \otimes \sigma^{z} \\
	+ \sin(\theta_{0}) \tau^{z} \otimes 
	\big( \cos(\varphi) \sigma^{x} - \sin(\varphi) \sigma^{y} \big) .
\end{multline}

In general, there are six order parameters for the canted ground state, 
$ \langle \mathcal{T}^{03}(\bm{x}) \rangle $,
$ \langle \mathcal{T}^{31}(\bm{x}) \rangle $,
$ \langle \mathcal{T}^{32}(\bm{x}) \rangle $,
$ \langle \widetilde{\mathcal{T}}^{03}(\bm{x}) \rangle $,
$ \langle \widetilde{\mathcal{T}}^{31}(\bm{x}) \rangle $ and 
$ \langle \widetilde{\mathcal{T}}^{32}(\bm{x}) \rangle $.
To draw the connection between these order parameters with the lattice
ordering, we consider $Q_{\textrm{Canted}}^{(0)}$ at $\varphi=0$.  The order
parameters $ \langle \mathcal{T}^{03}(\bm{x}) \rangle $ and $ \langle
\tilde{\mathcal{T}}^{03}(\bm{x}) \rangle $ indicate ferromagnetic ordering
component, described earlier in Sec.\ref{FM}. The staggered order parameter
for the operator $\tau^{z} \sigma^{x}$ corresponds to Neel ordering along
a direction perpendicular to the magnetic field. Here in our case the Neel
ordering is along the $x$-axis, shown in Fig.\ref{fig canted:op2}. 

The three phases separated by bold lines shown in Fig.\ref{fig pd nu0}, can be
obtained by comparing the energies for the corresponding phases.  The equation
of the straight line separating the first order transition between the
ferromagnetic and the charge ordered phases is given by
\begin{equation} \label{nu0 line1}
	3V = \frac{1}{2} \Big( 1 + \frac{u_{m}^{2}}{v_{m}^{2}} \Big) U 
	+ \frac{1}{4} \frac{u_{m}}{v_{m}^{2}} \tilde{Z} .
\end{equation}
Here, 
\begin{equation}
	\tilde{Z} = 
	\frac{4 \pi g \mu_{B} \hbar}{a^{2} e^{2}} \frac{B_{T}}{B_{\perp}} ~ (eV) ,
\end{equation}
is the parameter that depends on the tilt angle and for $B_{T} = B_{\perp}$,
$\tilde{Z} \approx 15.85 ~ eV$.  The first order transition between the charge
ordered and the canted spin ordered state is separated by, 
\begin{equation} \label{nu0 line3}
	3V = U + \frac{1}{32} \frac{u_{m}^{2}}{v_{m}^{2}(v_{m}^{2} - u_{m}^{2})} 
	\frac{\tilde{Z}^{2}}{U} .
\end{equation}
This equation in the large $U$ and $V$ limit becomes a straight line, $3V = U$. 

\begin{figure}
	\centering
	\subfloat[]{
		\includegraphics[width=0.46\columnwidth]{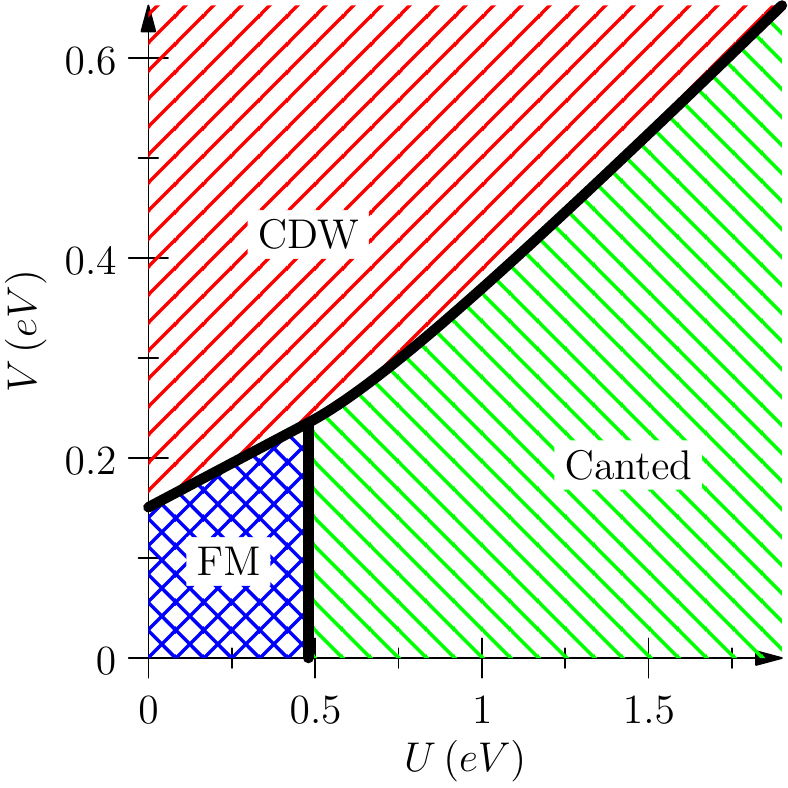}
			\label{fig pd sea}
		}
		\hfill
		\subfloat[]{
			\includegraphics[width=0.46\columnwidth]{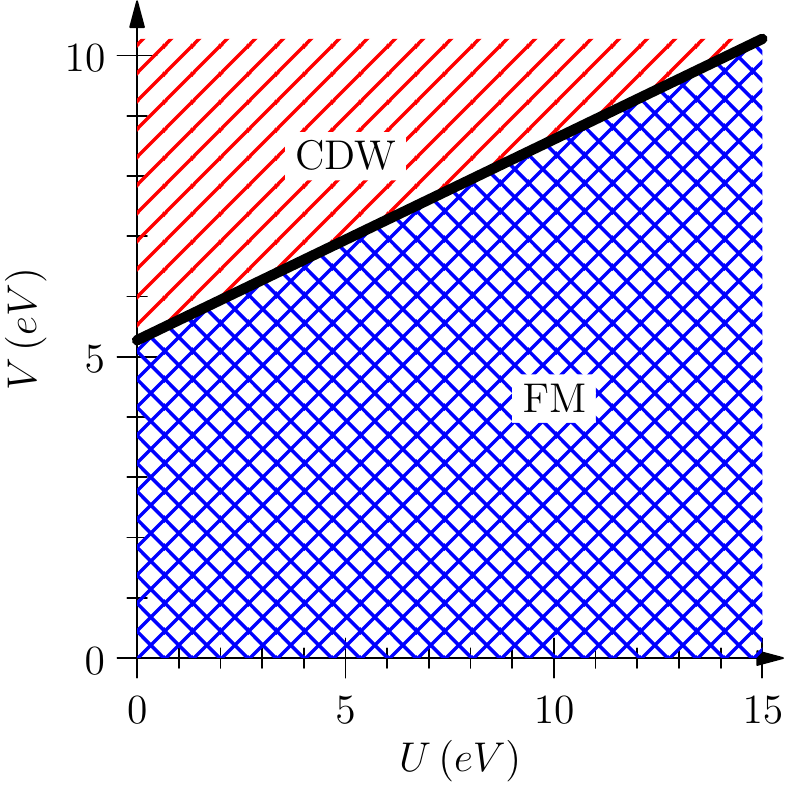}
			\label{fig pd ll}
		}
		\caption{(a) shows the phase diagram for $\sigma_{H}=0$ at $B=10T$ for
			our interacting model. (b) shows phase diagram when the effects
			of filled Dirac sea are ignored at any magnetic field.
	}
\end{figure}

The transition between the ferromagnetic and the canted-spin ordered
state is continuous and the straight line separating them is parallel to the
$V$-axis given by, 
\begin{equation} \label{nu0 line2}
	U = \frac{1}{4} \frac{u_{m}}{v_{m}^{2} - u_{m}^{2}} \tilde{Z} .
\end{equation}

We compare our results of the variational state analysis with the calculations
restricted to $n=0$ Landau level i.e. ignoring the filled Dirac sea. The
filled Dirac sea contributions to the symmetry breaking terms can be easily
neglected by substituting $v_{m} = u_{m} = \frac{1}{2}$ in equations:
\eqref{nu0 nnbr}, \eqref{nu0 hubbard} and \eqref{nu0 zeeman}. The minimization
of this energy results in charge ordered state and ferromagnetic ordered state
in $U$-$V$ parameter space and the straight line separating the first order
transition between the two phases is given by,  $3V = U + \tilde{Z}$, shown in
Fig.\ref{fig pd ll}. 

The inclusion of the filled Dirac sea in our variational state analysis has
resulted in a canted-spin ordered state in $U$-$V$ parameter space. Moreover
our variational state analysis predicts all three phases for moderate values
$( < 2 ~ eV)$ for $U$ and $V$ in contrast to only ferromagnetic state when
calculation restricted to the $n=0$ Landau level.

\subsection{Ground state at $\sigma_{H} = -1$}

The diagonal mass matrix for the $\sigma_{H} = -1$, $M_{D} = \{ m, - m , -m, -m
\}$ and $Q^{(-1)}$ parameterization scheme discussed in Sec. \ref{gs nu1} are
included in the two point coincident correlator, Eq.\eqref{2pt coincident}.
The variational state energy for the symmetry breaking terms is evaluated with
this two point coincident correlator.  

The nearest neighbour interaction energy is a function of only one angle
parameter, 
\begin{equation} \label{nu1 nnbr}
	\mathcal{E}_{V} = - \frac{\kappa_{V}}{4 \pi^{2}} 
	\bigg( 2 \big( v_{m}^{2} - u_{m}^{2} \big) 
	\cos^{2}(\gamma_{1}) \bigg)
\end{equation}
and this term individually minimizes for $\gamma_{1} = 0 \textrm{ or } \pi$,
which results in a valley polarized ground state with the spin pointing in an 
arbitrary direction.  

The expectation value for the Hubbard term, 
\begin{equation} \label{nu1 hubbard}
	\mathcal{E}_{U} = - \frac{\kappa_{U}}{4 \pi^{2}} 
	\bigg( 2 \big( v_{m}^{2} - u_{m}^{2} \big) \sin^{2}(\gamma_{1}) 
	|\langle \bm{n}_{1} | \bm{n}_{2} \rangle|^{2} \bigg) ,
\end{equation}
minimizes when $\gamma_{1} = \pi/2$ and $|\langle \bm{n}_{1} | \bm{n}_{2}
\rangle|^{2} = 1$. This indicates that the Hubbard term prefers a Neel ordered
spin ground state.

The Zeeman energy density depends on three angle variables, 
\begin{multline} \label{nu1 zeeman}
	\mathcal{E}_{Z}  = - \frac{\kappa_{Z}}{2 \pi}
	\bigg( u_{m} \Big(
	\big( \cos(\theta_{1}) - \cos(\theta_{2}) \big) \\
	+ \cos(\gamma_{1}) 
	\big( \cos(\theta_{1}) + \cos(\theta_{2}) \big) \Big)
	\bigg)
\end{multline}
and it minimizes for angle parameters, $\gamma_{1} = 0(\pi), \theta_{1} = 0
(\theta_{2} = \pi)$. Hence the Zeeman term prefers a valley-spin polarized
ground state with spins pointing along the magnetic field.

\begin{figure}
	\begin{center}
		\includegraphics[width=0.85\columnwidth]{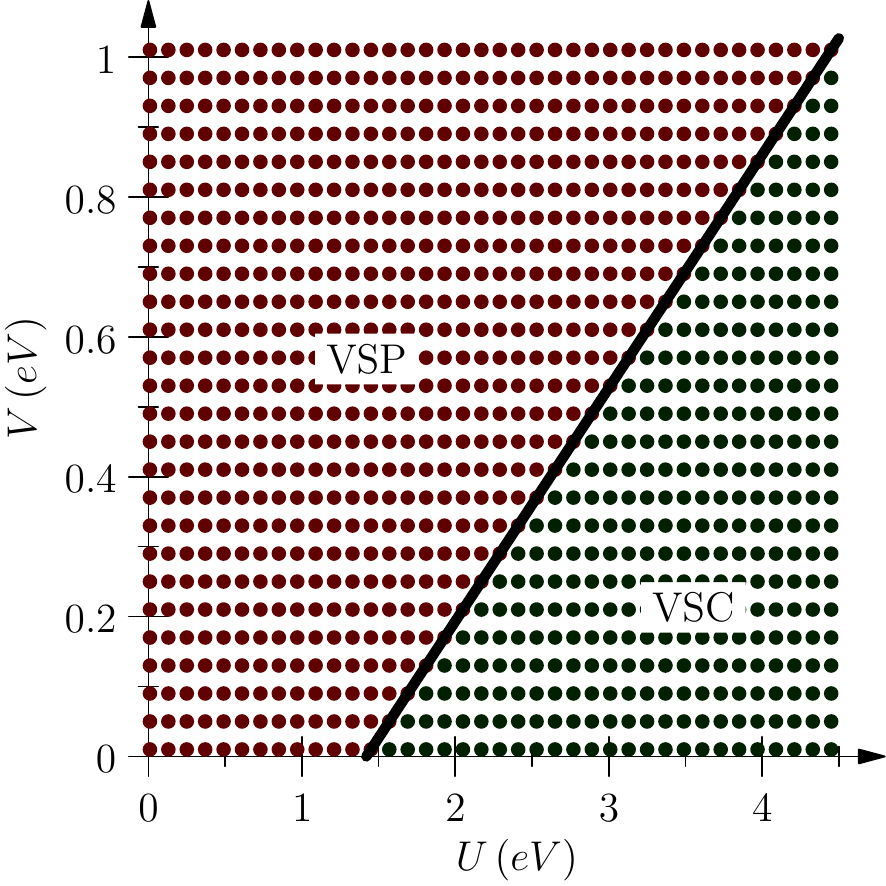}
	\end{center}
	\caption{The figure shows the phase diagram for the ground state at
		$\sigma_{H}=-1$ in $U$-$V$ parameter space. The phase diagram shown
		here is for magnetic field, $B = 30 \,T$ and $\varepsilon_{r} = 1.0$. 
		The red dots mark the region
		`VSP' corresponds to the area in parameter space where the numerical
		minimization yielded angle parameters for a valley-spin polarized
		state. The region marked `VSC' corresponds where the angle parameters
		result in valley-spin canted state. The bold line separating the two
		phases is obtained by equating the energies of these two phases and
		equation for this line is given in Eq.\eqref{nu1 line}. The order
		transition across this line is continuous.
	}
	\label{fig pd nu1}
\end{figure}

The angle parameters obtained from the  numerical minimization of the
variational state energy, $\mathcal{E}_{1}$ shown in Fig.\ref{fig pd nu1} and
the solutions correspond to two phases which we have enumerated below.

\subsubsection{Valley-spin polarized state}

\begin{figure}
	\centering
	\subfloat[]{
		\includegraphics[width=0.44\columnwidth]{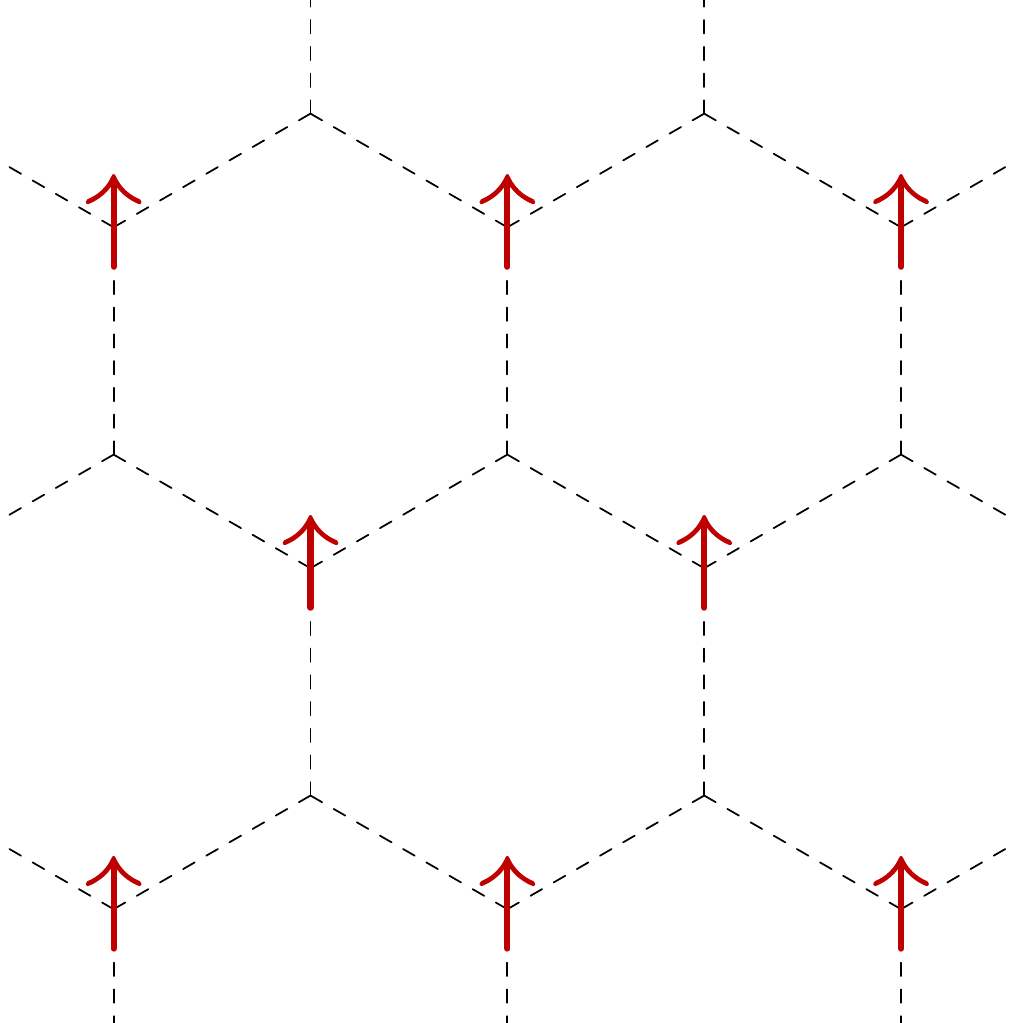}
			\label{fig vsp:latt}
		}
		\hfill
		\subfloat[]{
			\includegraphics[width=0.44\columnwidth]{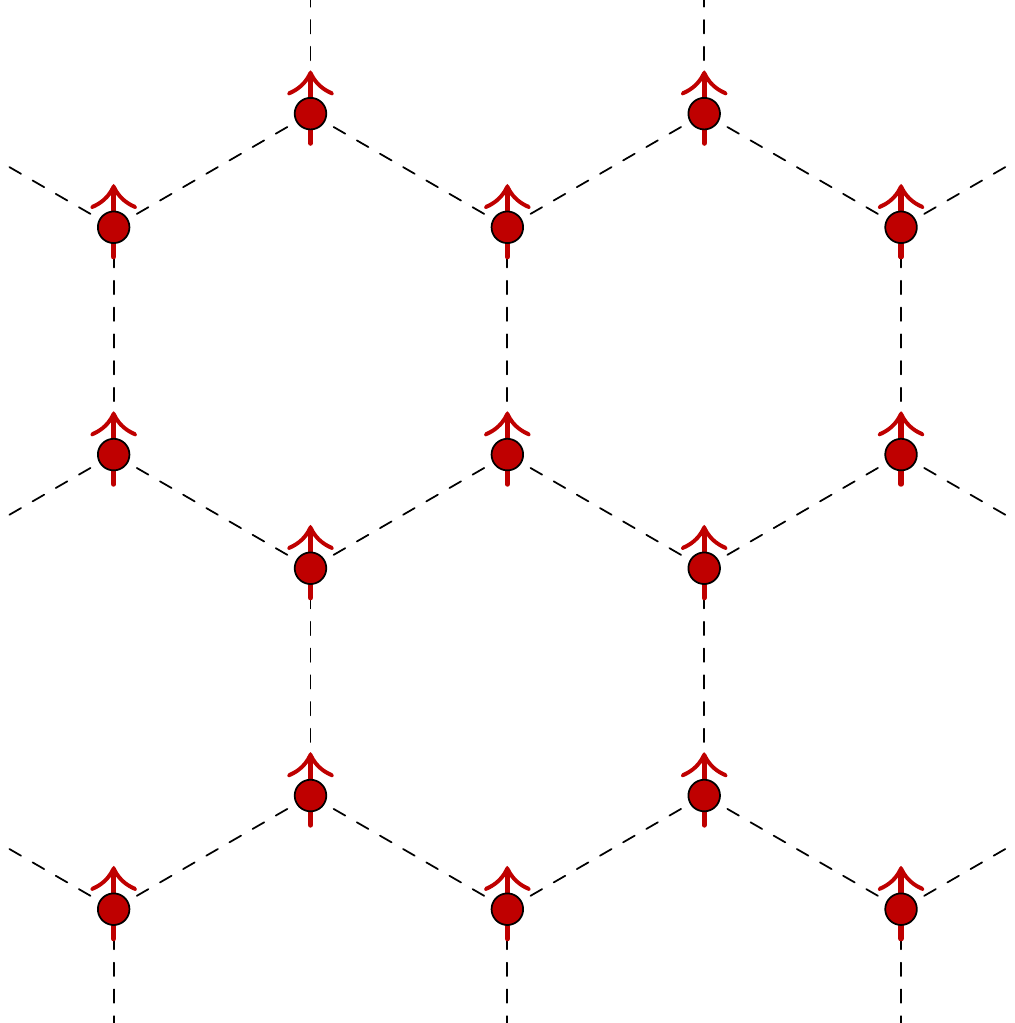}
			\label{fig vsp:op1}
		} \\
		\subfloat[]{
		\includegraphics[width=0.44\columnwidth]{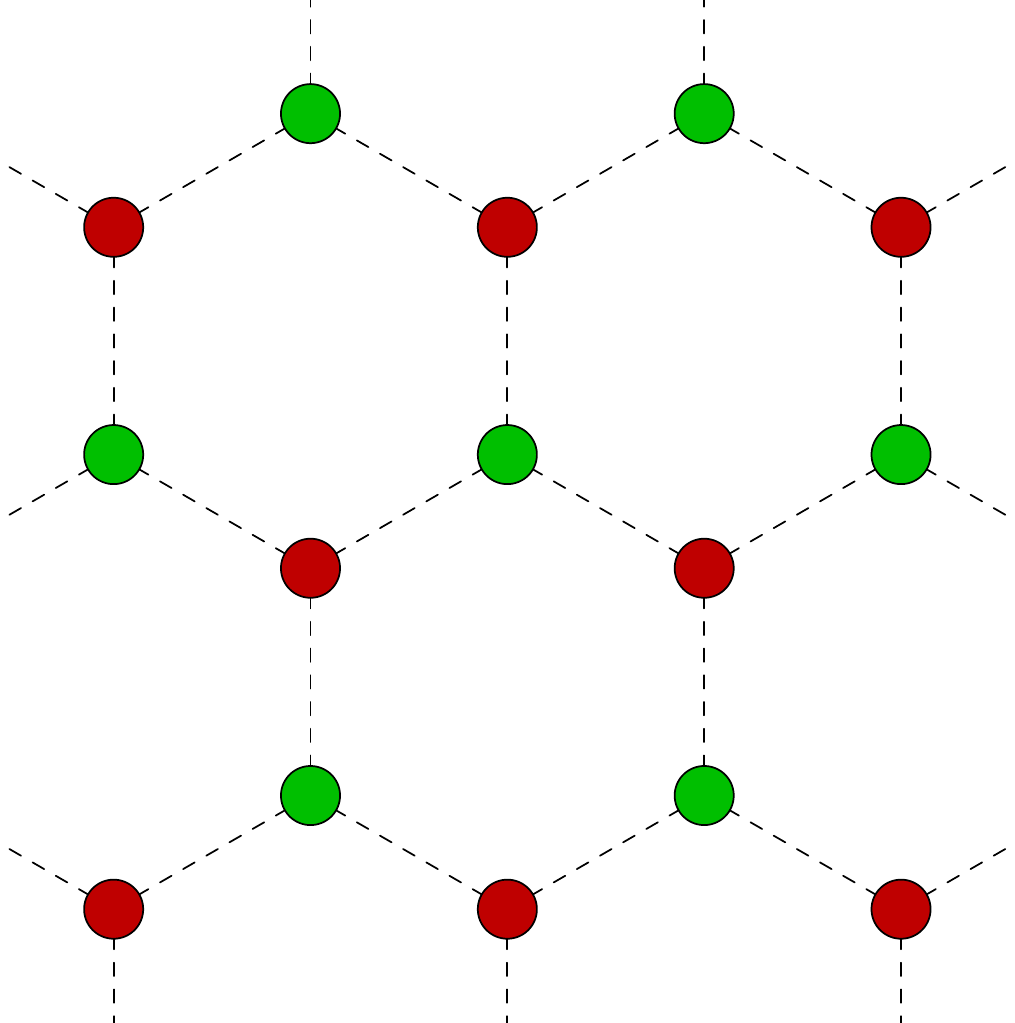}
			\label{fig vsp:op2}
		}
		\hfill
		\subfloat[]{
			\includegraphics[width=0.44\columnwidth]{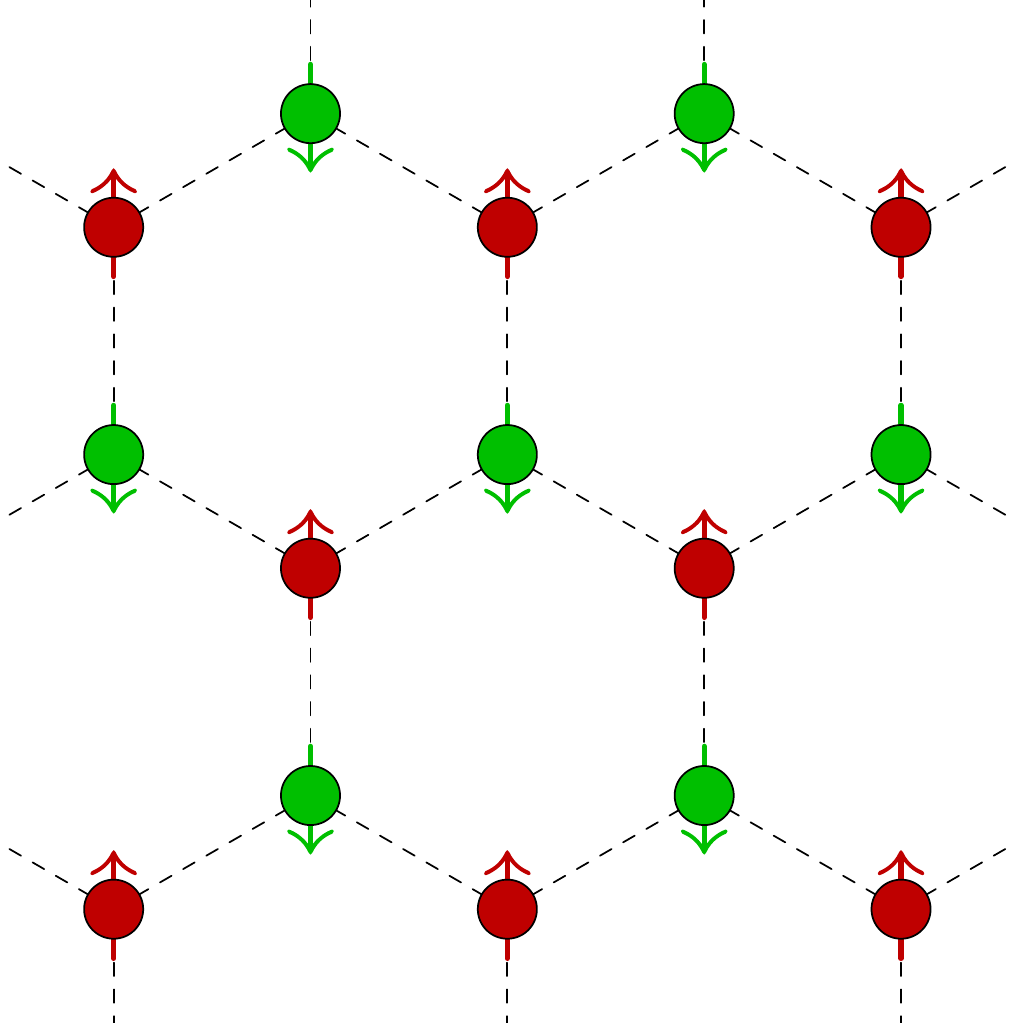}
			\label{fig vsp:op3}
		}
		\caption{(a) shows the $SU(4)$ component of $n=0$ Landau level for one
			of the two degenerate ground state for valley-spin polarized
			ground state. (b) is the lattice manifestation of the site order
			parameter for the total spin along the direction of the magnetic
			field. (c) represents the site order parameter which is the
			staggered charge density on the sub-lattice points. (d) shows the
			Neel ordering of the spins along the direction of the magnetic
			field for the valley-spin polarized ground state. 
}
\end{figure}

The region marked `VSP' in Fig.\ref{fig pd nu1} is the valley-spin polarized
state. The angle parameters that minimize $\mathcal{E}_{1}$ are,  $ \gamma_{1}
= 0, \theta_{1} = 0$ or $ \gamma_{1} = \pi, \theta_{2} = 0 $. The $SU(4)$
components of the $n=0$ Landau level for valley-spin polarized state are $| +
\rangle | \! \uparrow \rangle $ or $ | - \rangle | \! \uparrow \rangle$ and
the former is shown in Fig.\ref{fig vsp:latt}.  This ground state is doubly
degenerate and the $Q$ matrices for this ground state are 
\begin{equation}
	Q^{(-1)}_{\textrm{VSP}} = \frac{1}{2} \big(
	\pm (\tau^{z} + \tau^{z} \sigma^{z} )
	+ \sigma^{z} - \mathbbmss{1}_{4} \big)
\end{equation}

There are eight order parameters in total. Four $SU(4)$ polarization, 
$\langle \mathcal{T}^{00}(\bm{x}) \rangle$,
$\langle \mathcal{T}^{03}(\bm{x}) \rangle$,
$\langle \mathcal{T}^{30}(\bm{x}) \rangle$ and
$\langle \mathcal{T}^{33}(\bm{x}) \rangle$,
and four staggered polarization, 
$\langle \widetilde{\mathcal{T}}^{00}(\bm{x}) \rangle$,
$\langle \widetilde{\mathcal{T}}^{03}(\bm{x}) \rangle$,
$\langle \widetilde{\mathcal{T}}^{30}(\bm{x}) \rangle$ and
$\langle \widetilde{\mathcal{T}}^{33}(\bm{x}) \rangle$.
The lattice manifestation of order parameters that are site order parameters:
$\langle \mathcal{T}^{00}(\bm{x}) \rangle$, is the total charge density
indicating that the Fermi surface is away from the half filling for $\sigma_{H} =
-1$. $\langle \mathcal{T}^{03}(\bm{x}) \rangle$ is the total spin along the
direction of the magnetic field, similar to what we had seen for the
ferromagnetic ordered state in Sec.\ref{FM}, shown in Fig.\ref{fig vsp:op1}.
$\langle \widetilde{\mathcal{T}}^{30}(\bm{x}) \rangle$ gives the measure of
the staggered charge order shown in Fig.\ref{fig vsp:op2}.  $\langle
\widetilde{\mathcal{T}}^{33}(\bm{x}) \rangle$ is the Neel ordering of the
spins along the direction of the magnetic field and Fig.\ref{fig vsp:op3}
shows a caricature on graphene.

$\langle \mathcal{T}^{30}(\bm{x}) \rangle$ and
$\langle \mathcal{T}^{33}(\bm{x}) \rangle$ 
are equal bond order parameters for both sub-lattice points. $\langle
\widetilde{\mathcal{T}}^{00}(\bm{x}) \rangle$ is the bond parameter for same
sublattice points with equal magnitude and opposite sign. $\langle
\widetilde{\mathcal{T}}^{03}(\bm{x}) \rangle$ is also the bond order between same
sublattice points with equal magnitude and relative sign. It also has relative
sign between the spin up and spin down.

\subsubsection{Valley-spin canted state: }

In Fig.\ref{fig pd nu1}, the region marked `VSC' in the parameter space is
where we find angle parameters that correspond to the valley-spin canted state
after numerical minimization. The angle parameters take values, 
$ \theta_{1} = 0 (\pi)$,  $\theta_{2} = \pi (0) $ and 
\begin{equation} \label{vsc theta}
	\cos(\gamma_{0}) = +(-) \frac{1}{2}
	\frac{u_{m}}{v_{m}^{2} - u_{m}^{2}} 
	\frac{2 \pi \kappa_{Z}}{\kappa_{U} - \kappa_{V}} .
\end{equation}
The $SU(4)$ component of the occupied $n=0$ Landau quartet, 
\begin{equation}
	| \chi^{1} \rangle = \cos(\frac{\gamma_{0}}{2}) | + \rangle | \uparrow
	\rangle + \me^{\mi \Omega} \sin(\frac{\gamma_{0}}{2}) 
	| - \rangle | \downarrow \rangle .
\end{equation}
Here $\Omega$ is a free parameter. The ground state is doubly degenerate
and is parameterized by one parameter, $\Omega$, a $U(1)$ symmetry remnant of
$SU(4)$. The $Q$ matrix that characterizes the family of ground states for the
valley-spin canted ground state is, 
\begin{multline}
	Q^{(-1)}_{\textrm{VSC}} = \frac{1}{2} \Big( 
	\pm \tau^{z} \sigma^{z} - \mathbbmss{1}_{4} 
	+ \cos(\gamma_{0}) (\sigma^{z} \pm \tau^{z}) \\
	+ \sin(\gamma_{0}) \big( \cos(\Omega) ( \tau^{x} \sigma^{x} 
	\mp \tau^{y} \sigma^{y} ) \\
	+ \sin(\Omega) ( \tau^{x} \sigma^{y} \pm \tau^{x} \sigma^{y} ) \big)
	\Big) .
\end{multline}

There are 16 order parameters for the valley-spin polarized ground
state. To keep things simple, we set the parameter $\Omega=0$ for the $Q$
matrix, to analyze the manifestation of order parameters on the lattice. This 
leaves us with twelve order parameter. Eight of these parameters we have
encountered in when we described the valley-spin polarized ground state.
The remaining four are,
$\langle \mathcal{T}^{11}(\bm{x}) \rangle$,
$\langle \mathcal{T}^{22}(\bm{x}) \rangle$,
$\langle \widetilde{\mathcal{T}}^{11}(\bm{x}) \rangle$ and
$\langle \widetilde{\mathcal{T}}^{22}(\bm{x}) \rangle$. 
These correspond to bond order parameter between the nearest neighbour
sublattice points which also involves a momentum transfer of $\pm (
\bm{K}_{+} - \bm{K}_{-} )$. 

\begin{figure}
	\centering
		\includegraphics[width=0.33\textwidth]{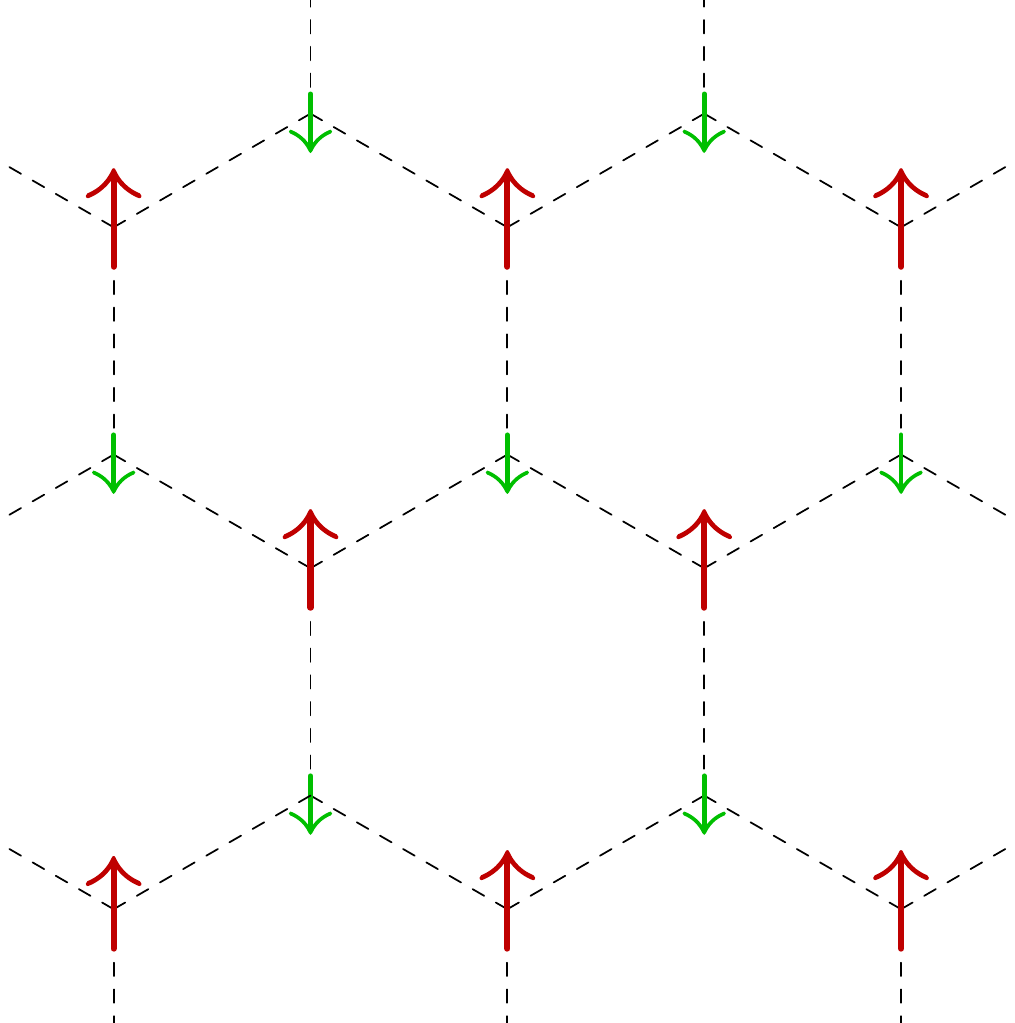}
		\caption{The figure shows the arrangement of $SU(4)$ components of the
			occupied quartet of the $n=0$ Landau level for the valley-spin
			canted ground state. The colors of arrows indicate spins localized
			on the two sub-lattice points and different lengths represents
			unequal weights on the sub-lattice points. The site order
			parameter on the lattice for the valley-spin canted ground state
			are same as for the valley-spin polarized ground state, shown in
			figures \ref{fig vsp:op1}, \ref{fig vsp:op2} and \ref{fig
			vsp:op3}.
		}
		\label{fig vsc:latt}
\end{figure}

The equation of the straight line that separates the two phases in $U$-$V$
parameter space in Fig.\ref{fig pd nu1} is,
\begin{equation} \label{nu1 line}
	3V = U - \frac{1}{2} \frac{u_{m}}{v_{m}^{2} - u_{m}^{2}} \tilde{Z} ,
\end{equation}
which we obtained by comparing the energies of the two phases. The order of
the phase transition across this straight line is continuous. 

Once again we compare the results of energy minimization at $\sigma_{H}=-1$
with the case when the effects of the filled Dirac sea are ignored. We find no
contribution to the variational state energy from the symmetry breaking
interaction terms. This is not surprising as $\sigma_{H} = -1$ has
quarter-filled $n=0$ Landau level. The only contribution comes from the Zeeman
term which results in the valley-spin polarized ground state in the entire
$U$-$V$ parameter space. The valley-spin canted ground state is the
consequence of the filled Dirac sea effects within our variational analysis.

\subsection{Excitation states and gaps} 

The quasi-particle(hole) excitations from the symmetric model provided the
magnitude of the gaps. The exact nature of these excitations, i.e. $SU(4)$
components of the excitations were left unspecified. In this section we
compute the contributions to the particle-hole excitations from the $SU(4)$
symmetry breaking terms of our model. The ground state has been fixed in the
$U$-$V$ phase space by local terms as we had seen in the previous section.
The gaps from the local interaction terms are lower by a factor of $(a/
\ell_{c})$ when compared to the magnitude from long-ranged Coulomb
interaction. But the angle parameter dependence that results from symmetry
breaking terms will fix the $SU(4)$ components of the excitations.

To compute the contribution to the particle-hole gap from the short ranged
interaction terms, we need to compute
\begin{equation*}
	\langle ES | \big(\Psi^{\dagger}(\bm{r}) \, \mathbbmss{G} \, 
	\Psi(\bm{r}) \big)^{2} | ES\rangle
	- \langle GS | \big(\Psi^{\dagger}(\bm{r}) \, \mathbbmss{G} \, 
	\Psi(\bm{r}) \big)^{2} | GS\rangle ,
\end{equation*}
where the matrix $\mathbbmss{G} = \mathbbmss{1}_{4}, \beta \tau^{z},
\alpha^{j} \tau^{k}$ and $j,k = x,y$. A similar procedure that we followed for
particle-hole excitations for symmetric model in section \ref{sec:symm gaps},
we define the two point coincident correlator for the excited state, 
\begin{equation} \label{2pt coincdent es}
	\Upsilon (\bm{r}, \bm{r}) = 
	\Gamma^{(p)}(\bm{r}, \bm{r}) - \Gamma^{(h)}(\bm{r}, \bm{r})
	+ \Gamma(\bm{r}, \bm{r}) .
\end{equation}
$\Gamma^{(p)}$ and $\Gamma^{(h)}$ are coincident correlators for the particle and
the hole state using the $n=0$ Landau level wavefunctions.  $\Gamma(\bm{r},
\bm{r})$ is the coincident correlator for the ground state and has no
coordinate dependence.  The correlator for the particle and the hole,
Eq.\eqref{correlator x}, after the coincident limit, the coordinate
integration is accomplished using gamma functions, $\Gamma^{(\mathtt{x})}  =
\int_{\bm{r}} \Gamma^{(\mathtt{x})}(\bm{r}) = \frac{1}{2} ( \mathbbmss{1}_{2}
- \beta ) P_{\mathtt{x}} $, here $\mathtt{x} = p, h$.

The particle-hole gap contribution from each of the local interaction terms for
the case where the separation is large between the particle and the hole 
states is evaluated using the following expression, 
\begin{multline} \label{es terms}
	\Delta_{\mathbbmss{G}} \propto \frac{1}{2}
	\Big( \, \big(\textrm{Tr}[ \mathbbmss{G} \, \Gamma^{(p)}]
	- \textrm{Tr}[\mathbbmss{G} \, \Gamma^{(h)}] \big)
	\textrm{Tr}[ \mathbbmss{G} \, \Gamma] \\
	- \textrm{Tr}[\mathbbmss{G} \, \Gamma^{(p)}
	\, \mathbbmss{G} \, \Gamma] 
	+ \textrm{Tr}[\mathbbmss{G} \, \Gamma^{(h)}
	\, \mathbbmss{G} \, \Gamma] \Big) .
\end{multline}
The Zeeman term contribution to the gap is, 
\begin{equation}
	\Delta_{Z} = \frac{1}{2} \kappa_{Z} \big( 
	\textrm{Tr}[\sigma^{z} \Gamma^{(p)}] - \textrm{Tr}[\sigma^{z} \Gamma^{(h)}]
	\big) .
\end{equation}

\subsubsection{Excitations for ground states at $\sigma_{H} = 0$}

We construct the quasi-hole and the quasi-particle state by taking a linear
combination of the occupied and the unoccupied members of the $n=0$ Landau
level quartet respectively for each ground state at $\sigma_{H}=0$. By doing
this we span all possible $SU(4)$ polarizations for the quasi-particle and the
quasi-hole states for each ground state in $U$-$V$ parameter space.

\begin{figure}[htb]
	\centering
	\subfloat[]{
		\includegraphics[width=0.75\columnwidth]{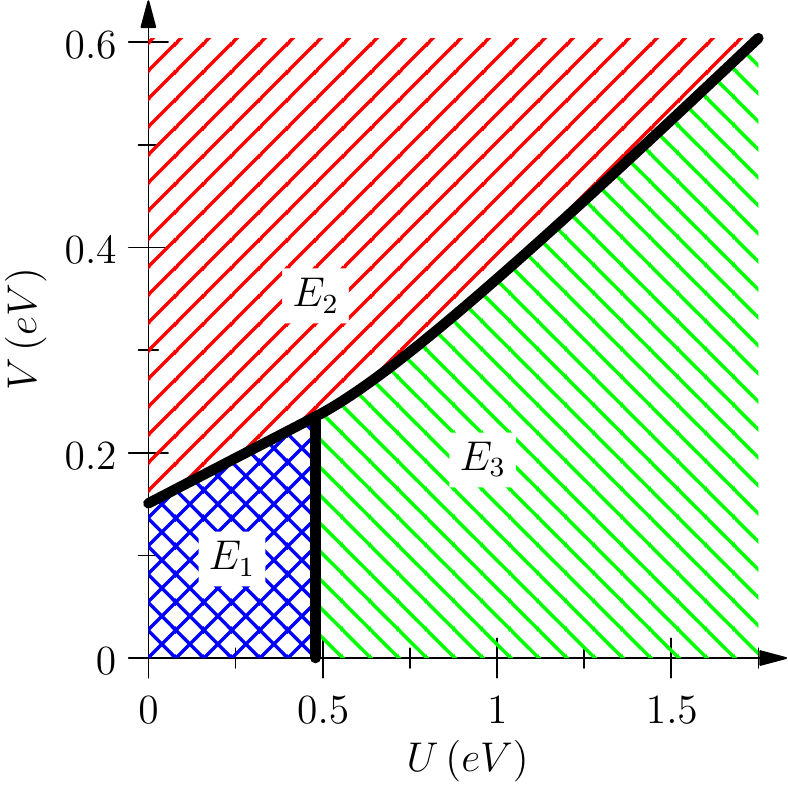}
			\label{fig extn0}
		} \\
		\subfloat[]{
			\includegraphics[width=0.60\columnwidth]{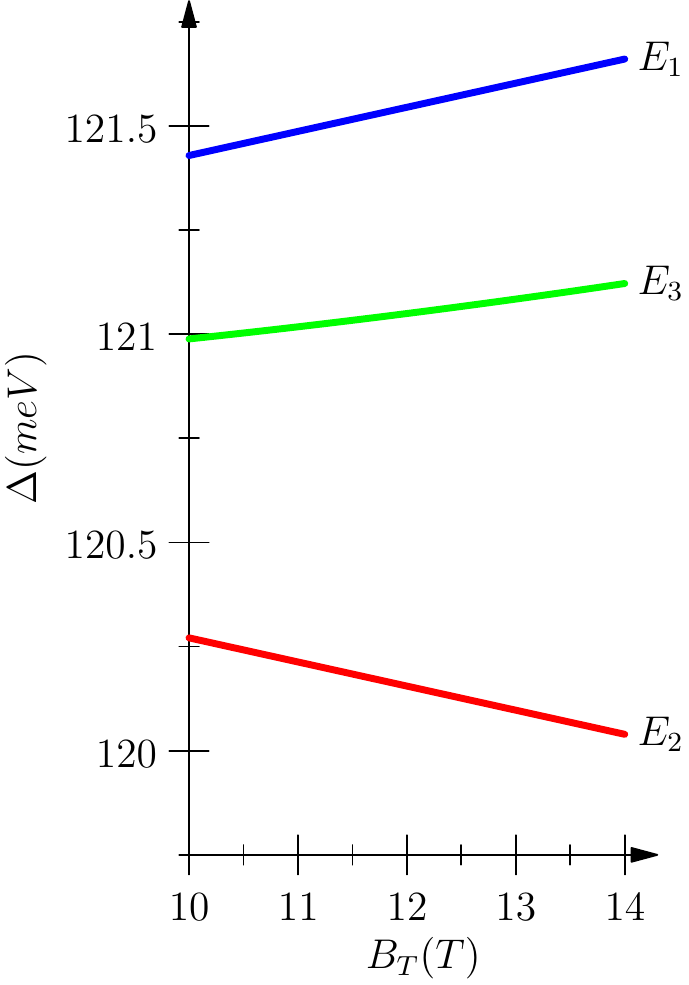}
			\label{fig gapt0}
		}
		\caption{(a) shows the region for all excitations from the $\sigma_{H}
			= 0$ ground state at $B_{\perp} = 10 T$ and $\epsilon_{r} =1.0$.
			(b) shows the variation of particle-hole gaps with respect to
			$B_{T}$ (tilt angle $< 45^{\circ}$) for all the three phases for
			the Hall state at $\sigma_{H}=0$. The gap for ferromagnetic spin
			ordered state labeled $E_{1}$, increases linearly with $B_{T}$,
			whereas gap for the charge ordered state, labeled $E_{2}$,
			decreases linearly. The gap for canted-spin phase, labeled
			$E_{3}$, shows an increase with $B_{T}$ with a quadratic
			dependence. In the canted phase, gap is inversely proportional to
			the Hubbard interaction strength $U$. In the figure we have
			chosen $U= 2 \, eV$ for the shown curve. 
		}
	\label{extn nu0}
\end{figure}

\paragraph{Ferromagnetic ordered state:}

The $SU(4)$ degeneracy is completely lifted for the ferromagnetic ground
state. A general hole state is a linear combination of up-spin states,  
$| h \rangle  = | \gamma_{h} \rangle | \! \uparrow \rangle$ and 
particle state of down-spin states, 
$| p \rangle  = | \gamma_{p} \rangle | \! \downarrow \rangle $
Here $|\gamma_{\mathtt{x}} \rangle = \cos(\frac{\gamma_{\mathtt{x}}}{2}) 
| + \rangle + \me^{\mi \Omega_{\mathtt{x}}}
\sin(\frac{\gamma_{\mathtt{x}}}{2}) | - \rangle $,  with $\mathtt{x} = p, h$.
The net gap from the symmetry breaking terms for the ferromagnetic ground
state is independent of angle parameters. The excitations of the ferromagnetic
ground state are, 
\begin{equation}
	E_{1}: ~ |h \rangle = | \gamma_{h} \rangle | \! \uparrow \rangle, 
	~ |p \rangle | = | \gamma_{p} \rangle | \! \downarrow \rangle .
\end{equation}

The quasi-particle and the quasi-holes states can be linear combinations of
unoccupied and occupied members of the $n=0$ Landau level quartet
respectively. The region for the excitations, $E_{1}$, for the ferromagnetic
ground state is marked in Fig.\ref{fig extn0}. The ferromagnetic ordered state
is fully spin polarized hence the excitations will involve a spin flip, which
costs a Zeeman energy, $\Delta_{Z} = \kappa_{Z}$.  This brings in a linear
dependence on $B_{T}$ to the excitation gap shown in Fig.\ref{fig gapt0}.

\paragraph{Charge ordered state:}

The charge ordered state is doubly degenerate as it can be localized on either
valley. We choose one of them and construct the quasi-hole state by choosing a
spin in an arbitrary direction localized in the valley, $|+\rangle$:   $| h
\rangle  = |+ \rangle | \bm{n}_{h} \rangle $ and $| - \rangle $ for the
quasi-particle state:   $| p \rangle  = |- \rangle | \bm{n}_{p} \rangle $.
Here the spin vector, $|\bm{n}_{\mathtt{x}} \rangle =
\cos(\frac{\theta_{\mathtt{x}}}{2}) | \! \uparrow \rangle + \me^{\mi
\varphi_{\mathtt{x}}} \sin(\frac{\theta_{\mathtt{x}}}{2}) | \! \downarrow
\rangle $,  with $\mathtt{x} = p, h$.  The only angle dependence comes from
the Zeeman term, $\Delta_{Z} = - \frac{1}{2} \kappa_{Z} ( \cos(\theta_{p}) -
\cos(\theta_{h}) )$. The net gap minimizes for $\theta_{p} = 0$ and $\theta_{h}
= \pi$. The $SU(4)$ components of the excitations for the charge ordered
ground state are,
\begin{equation}
	E_{2} : ~ | h \rangle = | + \rangle | \downarrow \rangle, ~
	| p \rangle = | - \rangle | \uparrow \rangle .
\end{equation}
The region for the excitations, $E_{2}$, in $U$-$V$ parameter space is shown
in Fig.\ref{fig extn0}. The particle-hole excitations involve flipping both
the valley and the spin quantum numbers and hence the lower the Zeeman energy,
$\Delta_{Z} = - \kappa_{Z} $.  The gap decreases linearly with $B_{T}$, shown
in Fig.\ref{fig gapt0}. 

\paragraph{Canted spin state:}

For the canted spin ground state, the $SU(4)$ degeneracy is not completely
lifted. We construct a family of quasi-hole states by taking a linear
combination of the occupied members of the $n=0$ Landau level quartet, 
$| h \rangle = 
\cos(\frac{\theta_{h}}{2}) |+ \rangle | \bm{n}_{1} \rangle 
+ \me^{\mi \varphi_{h}} \sin(\frac{\theta_{h}}{2}) 
|- \rangle | \bm{n}_{2} \rangle $,
where $\varphi_{1} - \varphi_{2} = \pi$ and $\theta_{1} = \theta_{2} =
\theta_{0}$. The cant angle is given by the Eq.\eqref{cant angle}.
A family of quasi-particle states is constructed by taking the linear 
superposition of unoccupied members of the $n=0$ Landau level quartet, 
$| p \rangle = 
\cos(\frac{\theta_{p}}{2}) |+ \rangle | - \bm{n}_{1} \rangle 
+ \me^{\mi \varphi_{p}} \sin(\frac{\theta_{p}}{2}) 
|- \rangle | - \bm{n}_{2} \rangle $.
Here, like the case for the excitations of the ferromagnetic ordered
ground state, the net gap from the symmetry breaking terms turns out to be
independent of the angle parameters of the particle and the hole states. The
excitations from the canted-spin ground state are linear combinations of the
occupied and unoccupied members of the $n=0$ Landau level quartet,  
\begin{multline}
	E_{3} : ~ 
	| h \rangle = 
	\cos(\frac{\theta_{h}}{2}) |+ \rangle | \bm{n}_{1} \rangle 
	+ \me^{\mi \varphi_{h}} \sin(\frac{\theta_{h}}{2}) 
	|- \rangle | \bm{n}_{2} \rangle \\
	| p \rangle = 
	\cos(\frac{\theta_{p}}{2}) |+ \rangle | - \bm{n}_{1} \rangle 
	+ \me^{\mi \varphi_{p}} \sin(\frac{\theta_{p}}{2}) 
	|- \rangle | - \bm{n}_{2} \rangle .
\end{multline}
The region for these excitations in $U$-$V$ parameter space is shown in
Fig.\ref{fig extn0}.  The canted-spin state is not fully spin polarized, the
excitations cost a Zeeman energy, $\Delta_{Z}  =  \kappa_{Z} \cos(\theta_{0})$.
The gaps in tilted magnetic have an increasing quadratic dependence on 
$B_{T}$ and is shown in Fig.\ref{fig gapt0}.

The tilted magnetic field measurements for the activation gaps in quantum Hall
experiments are good to decipher the spin of the excitations. In tilted field
experiment, $B_{\perp}$, the perpendicular component of the magnetic field is
kept fixed and $B_{T}$, the total magnetic field is varied by rotating the
sample. In the Zeeman term, the spin sees only the total magnetic field.
Whereas rest of the terms of our interacting model depends on the magnetic
field perpendicular the plane. The particle-hole excitations from the
interaction terms have a dominant contribution from the long-ranged Coulomb
term and the short-ranged interactions are suppressed by a factor of $a /
\ell_{c}$, hence we ignore them in the following analysis. The dependence of
total gap can be written as
\begin{equation}
	\Delta = \Delta_{C}(B_{\perp}) + \Delta_{Z}(B_{T}) .
\end{equation} 
For fixed $B_{\perp}$, the Coulomb contribution is constant, whereas the
Zeeman contribution varies with $B_{T}$. The variation of gaps with $B_{T}$
for the three ground states for the Hall conductivity at $\sigma_{H}=0$ are
shown in Fig.\ref{fig gapt0}. 

\subsubsection{Excitations for ground states at $\sigma_{H} = -1$}

The ground state at $\sigma_{H}=-1$ has quarter-filled $n=0$ Landau level
quartet and its $SU(4)$ component is used to construct the quasi-hole state.
We construct a quasi-particle state by taking a linear combination of the
three of the unoccupied members of the $n=0$ Landau level quartet. 

\begin{figure}
	\centering
	\subfloat[]{
		\includegraphics[width=0.75\columnwidth]{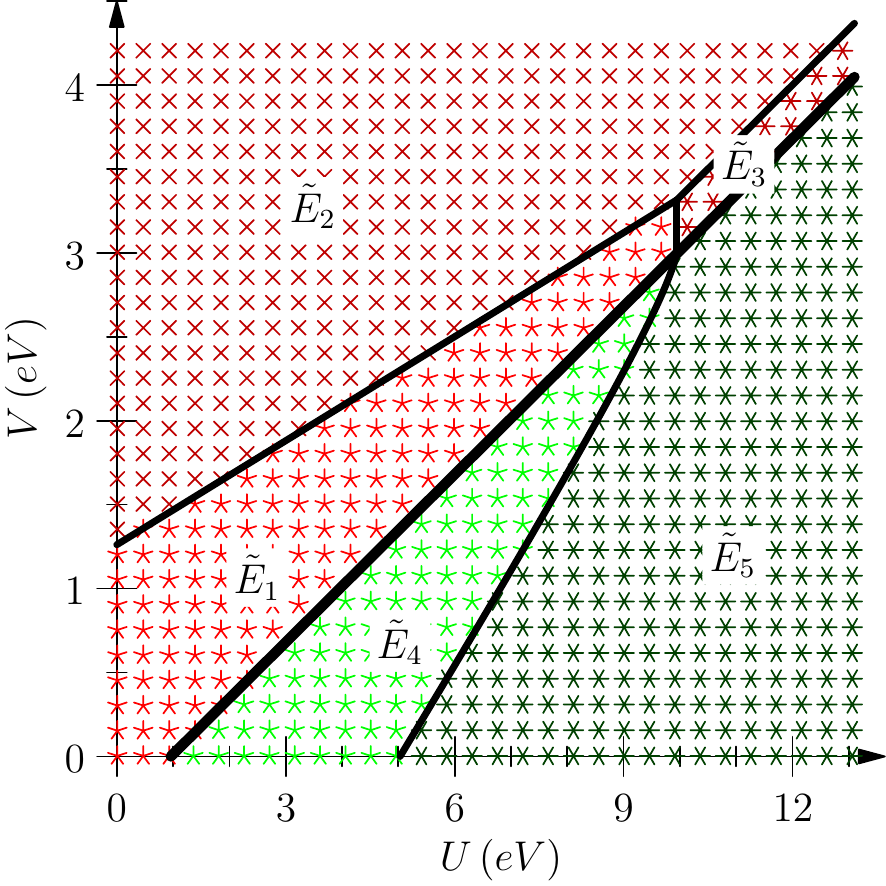}
			\label{fig extn1}
		} \\
		\subfloat[]{
			\includegraphics[width=0.60\columnwidth]{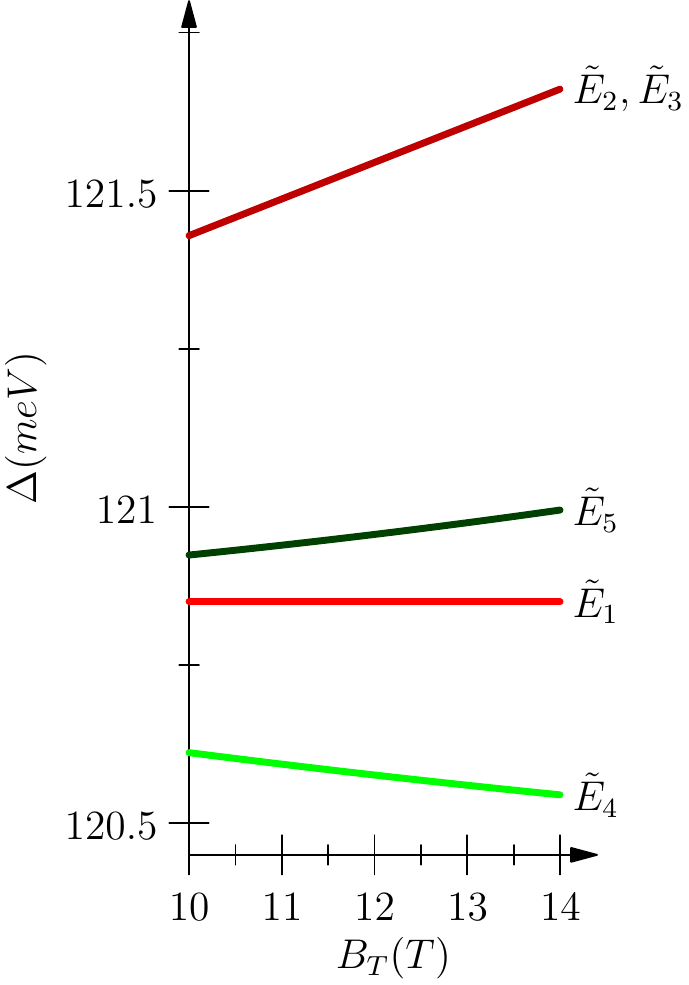}
			\label{fig gapt1}
		}
		\caption{The figures show the excitations described in the main text
			for the ground states at $\sigma_{H} =
			-1$ in the $U$-$V$ parameter space, (a), and their variation with
			$B_{T}$ (tilt angle $< 45^{\circ}$) in (b). The excitations
			$\tilde{E}_{1}$, $\tilde{E}_{2}$ and $\tilde{E}_{3}$ are for the
			valley-spin polarized state and $\tilde{E}_{4}$ and
			$\tilde{E}_{5}$ for the valley-spin canted state. The excitations
			$\tilde{E}_{2}$ and $\tilde{E}_{3}$ of the valley-spin polarized
			state have a linear variation with $B_{T}$ whereas $\tilde{E}_{1}$
			has none. The excitation $\tilde{E}_{5}$ of the valley-spin canted
			state has a quadratic dependence on $B_{T}$ whereas
			$\tilde{E}_{4}$ has competing increasing quadratic and decreasing
			linear $B_{T}$ dependence. For small tilt, angles the decreasing
			linear $B_{T}$ dependence dominate as shown in (b).
	}
\label{nu1 pd excit}
\end{figure}

\paragraph{Valley-spin polarized state:}

The valley-spin polarized ground state is doubly degenerate, we choose the
$SU(4)$ component for the quasi-hole state, $|h \rangle = |+ \rangle | \! \uparrow
\rangle$.  The $SU(4)$ components for the quasi-particle can be constructed as
the following linear combination of states, 
\begin{equation}
	|p \rangle = \cos(\frac{\theta_{p}}{2}) | - \rangle | \! \uparrow \rangle 
	+ \me^{\mi \varphi_{p}} \sin(\frac{\theta_{p}}{2})
	| \gamma_{p} \rangle | \! \downarrow \rangle .
\end{equation}

The minimization of the net gap from the symmetry breaking terms was
numerically obtained and we find three excitations for the valley-spin polarized
state and they are, 
\begin{align}
	\tilde{E}_{1} &:& 
	\theta_{p} &= 0, &  
	|p \rangle &= | - \rangle | \uparrow \rangle, &
	\Delta_{Z} &= 0 . \\
	\tilde{E}_{2} &:& 
	\theta_{p} &= \pi, \gamma_{p} = 0, & 
	|p \rangle &= | + \rangle | \downarrow \rangle, &
	\Delta_{Z} &= \kappa_{Z} .\\
	\tilde{E}_{3} &:&
	\theta_{p} &= \pi, \gamma_{p} = \pi, & 
	|p \rangle &= | - \rangle | \downarrow \rangle, &
	\Delta_{Z} &= \kappa_{Z} .
\end{align}

The region in $U$-$V$ parameter space for these excitations is shown in
Fig.\ref{fig extn1}.  The variation of excitation gaps with $B_{T}$ are shown
in Fig.\ref{fig gapt1}. The excitation $\tilde{E}_{1}$ has no variation with
$B_{T}$ whereas both $\tilde{E}_{2}$ and $\tilde{E}_{3}$ have linear and
increasing dependence on $B_{T}$.

\paragraph{Valley-spin canted state:}

The valley-spin canted ground state is doubly degenerate and has a free
parameter $\Omega$. Here we choose the $SU(4)$ components of the quasi-hole
state, 
$| h \rangle = 
\cos(\frac{\gamma_{0}}{2}) |+ \rangle | \! \uparrow \rangle 
+ \me^{\mi \Omega} 
\sin(\frac{\gamma_{0}}{2}) |- \rangle | \! \downarrow \rangle $.
$\gamma_{0}$ is given by Eq.\eqref{vsc theta}.
The $SU(4)$ components of the quasi-particle state are obtained by taking the
following linear combination,  
\begin{multline}
	|p \rangle = \cos(\frac{\theta_{p}}{2}) \Big(\sin(\frac{\gamma_{0}}{2}) 
	|+ \rangle | \! \uparrow \rangle 
	- \me^{\mi \Omega}  \cos(\frac{\gamma_{0}}{2}) 
	|- \rangle | \! \downarrow \rangle  \Big) \\
	+ \me^{\mi \varphi_{p}} 
	\sin(\frac{\theta_{p}}{2}) \Big( \cos(\frac{\gamma_{p}}{2}) 
	|+ \rangle | \! \downarrow \rangle  
	+ \me^{\mi \Omega_{p}} \sin(\frac{\gamma_{p}}{2}) 
	|- \rangle | \! \uparrow \rangle  \Big) .
\end{multline}

Minimization of the net gap from the symmetry breaking terms with respect to the
variational angle parameters was performed numerically and we find two
possible excitations enumerated below, 
\begin{multline}
	\tilde{E}_{5} : ~ \theta_{p} = 0,   
	|p \rangle = \sin(\frac{\gamma_{0}}{2}) |+ \rangle | \! \uparrow \rangle 
	- \me^{\mi \Omega}  \cos(\frac{\gamma_{0}}{2}) 
	|- \rangle | \! \downarrow \rangle, \\
	\Delta_{Z} = \kappa_{Z} \cos(\gamma_{0}) .
\end{multline}
\begin{multline}
	\tilde{E}_{4} : ~ \theta_{p} = \pi, \gamma_{p} = \pi, 
	|p \rangle = |+ \rangle | \! \uparrow \rangle, \\
	\Delta_{Z} = -\frac{1}{2} \kappa_{Z} ( 1 - \cos(\gamma_{0}) ) .
\end{multline}

The regions for these two excitations are marked $\tilde{E}_{4}$ and
$\tilde{E}_{5}$, in $U$-$V$ parameter space shown in Fig.\ref{fig extn1}. The
excitation $\tilde{E}_{5}$ has quadratic dependence on $B_{T}$ as shown in
Fig.\ref{fig gapt1}. The excitation, $\tilde{E}_{4}$, has a decreasing linear
and a increasing quadratic dependence on $B_{T}$ which compete with each
other. At low tilt angles the decreasing linear $B_{T}$ dependence dominates
as shown in Fig.\ref{fig gapt1}.

\section{Comparison with experiments} \label{sec:expt}
In this section, we compare the behaviour of gaps w.r.t. $B_{\perp}$ and
$B_{T}$ with experiments\cite{kim.nphys, yacoby}. This analysis will demarcate
a region in $U$-$V$ parameter space where our model results are consistent with
experiments.  

In the reference [\onlinecite{yacoby}], it was reported that the gap at
$\sigma_H={0}$ for suspended graphene was good fit to $\Delta_{\sigma_{H}=0} =
(23.1 \sqrt{B} - 11.1) meV$.  In Sec.\ref{sec:symm gaps}, we had evaluated the
quasi-particle(hole) gap at $\sigma_{H}=0$ and found that it was a good fit
to, $\Delta(B_{\perp}) = (37.8 \sqrt{B_{\perp}} + 0.15 B_{\perp}) meV$
Fig.\ref{gap Vs B}. Here we have taken the bare value $\varepsilon_{r} = 1$
for suspended graphene. For the range of magnetic fields ($B \in (1,12)T$) our
model gaps are 30-50\% larger than the experimentally observed ones.  This is
a reasonable agreement considering our analysis does not include the effects
of disorder and ignores the screening effects of the long-ranged Coulomb
interactions, which is known to reduce the theoretical values of gaps of clean
samples. 

Reference [\onlinecite{kim.nphys}] fits a straight line to the gap at
$\sigma_{H}=0$ for graphene on boron nitrite substrates. While the fit to our
model results does have a linear term, it is small and arises from our cut-off
procedure. This experimental observation \cite{kim.nphys} is hence not
consistent with our model.

The tilted field measurements by Young et.al.\cite{kim.nphys} show a decrease
in the gaps with increasing tilt angle. The gaps vary linearly as a function
of $B_{T}$ (for fixed $B_{\perp}$).  In Sec.\ref{sec:gnd0} we showed that
there are three ground states in $U$-$V$ parameter space: charge ordered,
ferromagnetic and canted-spin state. The variation of particle-hole gaps w.r.t
$B_{T}$ for each ground state is shown in Fig.\ref{fig gapt0}. The gaps for
the ferromagnetic and canted-spin states increase with increasing $B_{T}$.
Only the charge ordered state shows a linear decrease.  Thus we conclude that
the region labeled $E_{2}$ in Fig.\ref{fig extn0} demarcates the region in
$U$-$V$ consistent with experiments.

\begin{figure}
	\begin{center}
		\includegraphics[width=0.80\columnwidth]{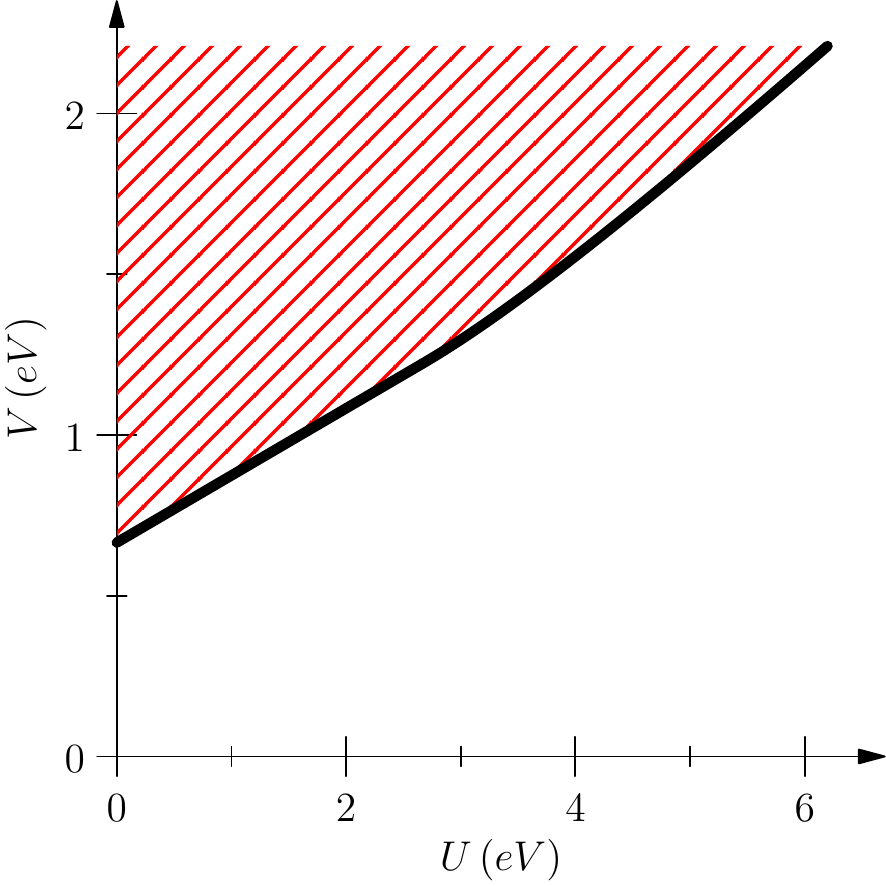}
	\end{center}
	\caption{The figure shows the delineated region in $U$-$V$ parameter space
		that is consistent with excitations observed\cite{kim.nphys} for the
		Hall conductivity at $\sigma_{H}=0$ for graphene sample on boron
		nitrite. The dashed area shown is the region where the gaps decrease
		with increasing $B_{T}$. The plot is shown for $B_{\perp}= 5\, T$ and
		$\varepsilon_{r} = 4.0$ for boron nitrite substrate. 
	}
	\label{fig delineate nu0}
\end{figure}

The gaps\cite{yacoby} for $\sigma_{H}=-1$ were reported to be
$\Delta_{\sigma_{H}=-1} = (8.5 \sqrt{B} - 1.8) meV$. The large difference
between the gaps at $\sigma_{H}=0$ and $\sigma_{H}=-11$ strongly indicate
skyrmionic excitations\cite{kim.nphys, yacoby} about the ground state at
$\sigma_{H}=-1$.  The values of our model gaps are the same at $\sigma_{H}=0$
and $\sigma_{H}=-1$ since we compute the gaps for single quasi-particles and
do not consider skyrmionic excitations. For this reason we do not compare our
model results for the tilt angle dependence of the gaps with experimental
observations.

To conclude, the behaviour of the quasi-particle gaps as a function of the tilt
angle at $\sigma_{H}=0$ is consistent with the experimental results in reference
[\onlinecite{kim.nphys}] in the region shown in Fig.\ref{fig delineate nu0}.
This region has been computed for graphene on boron nitrite,
$\varepsilon_{r}=4.0$.

\section{Summary} \label{sec:summary}
In this paper, we have investigated the role of the Dirac Sea and long-ranged
Coulomb interactions in $SU(4)$ symmetry breaking in graphene. We have
concentrated on the integer quantum Hall states in the vicinity of the charge
neutral point. 

We started with a realistic interacting lattice model for graphene. The model
includes long-ranged Coulomb point-charge interactions and short ranged
corrections to it. The short ranged corrections are parameterized by two
phenomenological parameters $U$ and $V$. We have presented a systematic
derivation of  the continuum limit of this lattice model.  All our
computations are done using this effective continuum theory.

We take a variational approach to the problem.  We have developed a technique,
using the heat kernel representation of the Dirac propagator in the presence
of a magnetic field to make the computations of Coulomb energy of the Dirac
sea tractable.  Our variational wavefunctions allow for a staggered $SU(4)$
polarization of the Dirac sea. We show that this indeed happens and results in
phases that are absent if the Dirac sea were ignored.  The canted phase for
$\sigma_{H}=0$ and the valley-spin canted phase for $\sigma_{H}=-1$ were shown
to be direct consequences of the inclusion of the filled Dirac sea in our
calculations. 

The ground state manifolds for $\sigma_H=0$ and $\sigma_H=-1$ are the coset
spaces $U(4)/(U(2)\times U(2))$ and $U(4)/(U(3)\times U(1))$ respectively.
Thus there will be 8 gapless collective modes for $\sigma_H=0$ and 6 for
$\sigma_H=-1$.  The explicit $SU(4)$ symmetry breaking terms, the short-ranged
interactions and the Zeeman term, pick out specific ground states which depend
on $U$ and $V$.  We have provided explicit parameterization of the ground
state manifolds and have done a complete search to compute the phase diagram
in $U$-$V$ space. We find three phases at $\sigma_H=0$ and two at
$\sigma_H=-1$. The canted-spin state at $\sigma_H=0$ and the valley-spin
canted phase at $\sigma_H=-1$ have a remnant $U(1)$ degeneracy. In the other
three cases the degeneracy is at most a discrete set. Thus we expect one
Goldstone boson to remain gapless in the first two cases and all to get
gapped in the other three phases.

A comparison of our model with tilted field experiments \cite{kim.nphys} at
$\sigma_H=0$ shows that only the charge ordered phase is consistent with the
observed behaviour of the gaps. This enables us to demarcate a region in
$U$-$V$ space consistent with experiments on graphene. This region is shown in
Fig.\ref{fig delineate nu0}.

While we have only analyzed the symmetry breaking of the $n=0$ integer quantum
Hall states, our method is applicable to $n\neq 0$ also. We will be reporting
on these cases in the near future.

\begin{acknowledgments}
	We acknowledge useful discussions with J.K. Jain and S.R. Hassan.
\end{acknowledgments}

\appendix
\section{Heat kernel representation of two point correlation function}
\label{appx:heat}
In this section, we evaluate the two point correlator using the time ordered
Feynman propagator for a massive Dirac particle in the presence of a magnetic
field. Consider a massive Dirac particle in $(2+1)d$ subjected to a magnetic
field along the $z$-axis. The hamiltonian, 
\begin{equation} \label{ham massive}
	\mathrm{h} =  \frac{\hbar v_{F}}{\ell_{c}} 
	\left( 
		\begin{array}{cc}
			m & \pi_{x} - \mi \pi_{y} \\
			\pi_{x} + \mi \pi_{y} & -m 
		\end{array}
	\right) .
\end{equation}
Here $m>0$ is expressed in units of $\hbar v_{F} / \ell_{c} $, and the
eigenvalues are,
\begin{equation} \label{massive dirac}
	\begin{split}
		\epsilon_{n,l} & = \textrm{sgn}(n) 
		\big(\frac{\hbar \, v_{F}}{\ell_{c}}\big)
		\sqrt{ 2 |n| + m^{2}} . \\
		\epsilon_{0,l} & = - m .
	\end{split}
\end{equation}
The completeness relation is,
\begin{equation*}
	\mathbbmss{1}_{2} = \sum_{n = - \infty}^{\infty} \sum_{l = -|n|}^{\infty}
	\Big( \Theta(- \epsilon_{n}) |n,l \rangle \langle n,l |  
	+ \Theta(\epsilon_{n}) \|n,l \rangle \langle n,l | \Big) .
\end{equation*}
Here $\mathbbmss{1}_{2}$ is $2 \times 2$ identity matrix, $\epsilon_{n} > 0$
and the first term is a summation over all negative energy eigenvalue states
which include all the Landau levels with negative indices and the $n=0$
Landau level since the eigenvalue is $-m$ with $m>0$. The second term is for
states with positive energies i.e. Landau levels with indices $n > 0$.

We are interested in constructing the two point correlator, 
\begin{equation}
	G_{s,s^{'}}(\bm{r}, \bm{r}_{0}) =
	\frac{1}{2} \langle 0 | [ \Psi^{\dagger}_{s}(\bm{r}) , 
	\Psi_{s^{'}}(\bm{r}_{0}) ] | 0 \rangle .
\end{equation}
Here $|0 \rangle$, the vacuum is constructed by occupying Landau levels with
index $n=0$ and $n<0$ i.e. all the Landau levels with negative energy
eigenvalues. This definition takes into account the subtraction of background
charge from the positive charged ions. 
\begin{equation} \label{G}
	G  = \frac{1}{2} 
	\sum_{n = - \infty}^{\infty} \sum_{l = -|n|}^{\infty}
	\big( \Theta(- \epsilon_{n}) - \Theta(\epsilon_{n}) \big) 
	|n,l \rangle \langle n,l | .
\end{equation}
Let  
\begin{equation} \label{K}
	K = \Theta(- \epsilon_{n}) - \Theta(\epsilon_{n}) .
\end{equation}
$K$ can be defined as following by considering equal time for the time ordered
Feynman propagator, 
\begin{equation} \label{K lim}
	K = \lim_{\tau \to 0^{-}} K(\tau) + \lim_{\tau \to 0^{+}} K(\tau) .
\end{equation}
Here
\begin{align}
	K(\tau) & = \me^{-\tau \epsilon_{n}} 
	\Theta(- \tau)  \Theta(- \epsilon_{n}) 
	- \me^{-\tau \epsilon_{n}} 
	\Theta( \tau) \Theta(\epsilon_{n}) \\
	& = - \int\limits_{-\infty}^{+ \infty} \frac{\dif \omega}{2 \pi} 
	\frac{\me^{\mi \omega \tau}}{\epsilon_{n} + \mi \omega} .
\end{align}
We make use of the integral representation, 
\begin{equation}
	\frac{1}{\epsilon_{n}^{2} + \omega^{2}} = \int\limits_{0}^{\infty}
	\dif s ~ \me^{-s (\epsilon_{n}^{2} + \omega^{2})} .
\end{equation}
and evaluate the integrals and obtain the correlator for massive Dirac 
particle, 
\begin{equation} \label{Gs}
	G = - \frac{1}{2 \sqrt{\pi}}
	\int\limits_{0}^{\infty} 
	\frac{\dif s}{\sqrt{s}} ~ \mathrm{h}  \, \me^{-s \mathrm{h}^{2}} .
\end{equation}
This is the so called `heat kernel' representation. 

To express the heat kernel operator in configuration space, we use the fact
that $\mathrm{h}^{2}$ is diagonal. and from the knowledge of imaginary time,
i.e. $t = -\mi \hbar \beta$, the propagator for the hamiltonian $( \pi_{x}^{2}
+ \pi_{y}^{2} ) (\hbar \omega_{c}/2)$ in configuration space \cite{glasser}, 
\begin{multline*}
	\sum_{n,l} \varphi_{n,l}^{*}(\bm{r}_{0})
	\me^{-\beta (\frac{\hbar \omega_{c}}{2}) (\pi_{x}^{2} + \pi_{y}^{2})} 
	\varphi_{n,l}(\bm{r}) \\
	= \frac{1}{2 \pi \ell_{c}^{2}} 
	\frac{ \me^{- \frac{1}{4 \ell_{c}^{2}}|\bm{r} - \bm{r}_{0}|^{2} 
	\coth (\frac{\beta \hbar \omega_{c}}{2})} }
	{ 2 \sinh(\frac{\beta \hbar \omega_{c}}{2}) }
	\me^{\frac{\mi}{2 \ell_{c}^{2}} ( x y_{0} - y x_{0})} .
\end{multline*}
This enables us to write $\me^{-s \mathrm{h}^{2}}$ in configuration space, 
\begin{equation} 
	\langle \bm{r_{0}} | \me^{-s \mathrm{h}^{2}} | \bm{r} \rangle 
	=  \frac{1}{2 \pi \ell_{c}^{2}} 
	\zeta_{s}(\bm{r}, \bm{r}_{0})
	\frac{\me^{-s m^{2}}}{2 \sinh(s)}
	\left( \begin{array}{cc}
			\me^{-s} & 0 \\
			0 & \me^{s} 
	\end{array} \right) .
\end{equation}
here, 
\begin{equation} \label{zeta_s}
	\zeta_{s}(\bm{r}, \bm{r}_{0}) =
	\me^{- \frac{1}{4}|\bm{r} - \bm{r}_{0}|^{2} \coth (s)} \:
	\me^{\frac{\mi}{2} ( x y_{0} - y x_{0})} .
\end{equation}
Since $s$ has dimension of the inverse of the square energy and $\bm{r}$ has
the dimension of length,. 
\[ \frac{s}{\epsilon^{2}} \to s, \qquad \qquad 
\frac{\bm{r}}{\ell_{c}} \to \bm{r} .\]

To obtain the propagator, we need to evaluate $\mathrm{h} \, \me^{-s
\mathrm{h}^{2}}$. This is achieved by finding the action of operators
$\pi_{+}$ and $\pi_{-}$ on $\zeta_{s}(\bm{r}, \bm{r}_{0})$. To accomplish this
we express the operators $\pi_{\pm}$ in terms of the complex variable $z = x +
\mi y$ and its conjugate $\bar{z}$. We have to take into account the
ultra-violet cut-off because of the underlying lattice structure of graphene.
This is taken into account by taking the limits of integration from $1/2N_{C}$
to infinity for $s$ variable in Eq.\eqref{Gs}. In configuration space the
equal time two point correlator, $\langle \bm{r}_{2} | G | \bm{r}_{1} \rangle
= G_{m}(\bm{r}_{1}, \bm{r}_{2}) $
\begin{equation} \label{green ftn}
	G_{m}(\bm{r}_{1}, \bm{r}_{2}) 
	= \frac{1}{2} \frac{1}{2 \pi \ell_{c}^{2}} 
	\left(
		\begin{array}{cc}
			- f_{m}(\bm{r}_{1}, \bm{r}_{2})
			& d_{m}(\bm{r}_{1}, \bm{r}_{2})  \\
			b_{m}(\bm{r}_{1}, \bm{r}_{2})  &
			g_{m}(\bm{r}_{1}, \bm{r}_{2})
		\end{array}	
	\right) .
\end{equation}
Here, 
\begin{equation} \label{f_m}
	f_{m}(\bm{r}_{1}, \bm{r}_{2}) = \frac{m}{2\sqrt{\pi}} \;
	\int\limits_{\frac{1}{2 N_{C}}}^{\infty} 
	\frac{\dif s}{\sqrt{s}} ~ \frac{\me^{-s(m^{2} + 1)}}{\sinh(s)} \;
	\zeta_{s}(\bm{r}_{1}, \bm{r}_{2}) .
\end{equation}
\begin{equation} \label{g_m}
	g_{m}(\bm{r}_{1}, \bm{r}_{2}) = \frac{m}{2\sqrt{\pi}} \;
	\int\limits_{\frac{1}{2 N_{C}}}^{\infty} 
	\frac{\dif s}{\sqrt{s}} ~ \frac{\me^{-s(m^{2} - 1)}}{\sinh(s)} \;
	\zeta_{s}(\bm{r}_{1}, \bm{r}_{2}) .
\end{equation}
\begin{equation} \label{b_m}
	b_{m}(\bm{r}_{1}, \bm{r}_{2}) = 
	- \mi \,\frac{(z_{1}-z_{2})}{4\sqrt{\pi}} \;
	\int\limits_{\frac{1}{2 N_{C}}}^{\infty} 
	\frac{\dif s}{\sqrt{s}} ~ \frac{\me^{-s m^{2}}}{\sinh^{2}(s)} 
	\zeta_{s}(\bm{r}_{1}, \bm{r}_{2}) .
\end{equation}
\begin{equation} \label{d_m}
	d_{m}(\bm{r}_{1}, \bm{r}_{2}) = 
	- \mi \,\frac{(\bar{z}_{1}-\bar{z}_{2})}{4\sqrt{\pi}} \;
	\int\limits_{\frac{1}{2 N_{C}}}^{\infty} 
	\frac{\dif s}{\sqrt{s}} ~ \frac{\me^{-s m^{2}}}{\sinh^{2}(s)} 
	\zeta_{s}(\bm{r}_{1}, \bm{r}_{2}) .
\end{equation}

\section{Variational state energy for symmetric model}
\label{appx:symmetric}
\subsection{Kinetic term} \label{appx kinetic}

The kinetic term has local fermion field operators and the expectation value
can be expressed as, 
\begin{equation} \label{H_t}
	\langle \mathcal{H}_{t} \rangle =  \kappa_{t} \int_{\bm{r}} 
	\mathrm{h}_{r,A;s,B} \langle \Psi^{\dagger}_{r,A}(\bm{r}) 
	\Psi_{s,B}(\bm{r}) \rangle .
\end{equation}

Here $\mathrm{h}_{r,A;s,B} =  \big( \bm{\alpha} \cdot \bm{\pi} \big)_{r,s}
\big( \mathbbmss{1}_{4} \big)_{A,B}$. To compute the average of kinetic term we
need to take into account the action of the operator $\mathrm{h}$ which makes
it non-local because of the action of conjugate momentum operator. We cannot
apply the coincident correlator here, instead we compute the action of the
operator $\mathrm{h}$ on the two point correlator and take the coincident
limit. 
\begin{equation} 
	\langle \mathcal{H}_{t} \rangle = \kappa_{t} \int_{\bm{r}} 
	\lim_{\bm{r} \to \bm{r}_{0}}
	\mathrm{Tr}[ \mathrm{h} \, \Gamma(\bm{r}, \bm{r}_{0})] .
\end{equation}
The operator $\bm{\alpha} \cdot \bm{\pi}$ has only off-diagonal elements, hence 
\begin{multline}
	\mathrm{Tr}[ \mathrm{h} \, \Gamma(\bm{r}, \bm{r}_{0})] 
	= \frac{1}{2 \pi} 
	\frac{\kappa_{t}}{2}
	\sum_{q=1}^{4} \big( \pi_{-} b_{m_{q}}(\bm{r}, \bm{r}_{0}) \\
	+ \pi_{+} d_{m_{q}}(\bm{r}, \bm{r}_{0}) \big) .
\end{multline}
Here $\pi_{\pm} = \pi_{x} \pm \mi \pi_{y}$, we obtain
\begin{multline}
	\lim_{\bm{r} \to \bm{r}_{0}} \pi_{-} b_{m_{q}}(\bm{r}, \bm{r}_{0}) 
	= \lim_{\bm{r} \to \bm{r}_{0}} \pi_{+} d_{m_{q}}(\bm{r}, \bm{r}_{0}) \\
	= - \frac{1}{2 \sqrt{\pi}} \int_{s} 
	\frac{\me^{-s m_{q}^{2}}}{\sqrt{s} \sinh^{2}(s)} .
\end{multline}
The spatial integration in Eq.\eqref{H_t} is trivial and results in the 
volume of the system. The kinetic energy density, 
\begin{equation} \label{chap5:kinetic tmp1}
	\mathcal{E}_{t} 
	= -  \frac{\kappa_{t}}{2 \pi}
	\frac{1}{2 \sqrt{\pi}} \sum_{q=1}^{4}
	\int_{\frac{1}{2N_{C}}}^{\infty} \dif s \, 
	\frac{\me^{-sm_{q}^{2}}}{\sqrt{s} \sinh^{2}(s)} .
\end{equation}
The integrand in Eq.\eqref{chap5:kinetic tmp1}, is a diverging function of $s$
near zero. We note that leading contribution of this integration is independent
of variational parameters, which is just a constant from the minimization point
of view. So we drop this constant in the process of computing the coefficients
with variational parameter dependence. The kinetic term density that has
variational parameter dependence can be expressed as, 
\begin{equation}
	\mathcal{E}_{t} 
	= \kappa_{t}  \sum_{q=1}^{4} \eta_{t}(m_{q}^{2}) .
\end{equation}
The quantity
\begin{equation}
	\eta_{t}(m_{q}^{2}) = 
		\frac{1}{2 \sqrt{\pi}} 
		\int_{\frac{1}{2N_{C}}}^{\infty} \dif s \, 
		\frac{(1 - \me^{-sm_{q}^{2}})}{\sqrt{s} \sinh^{2}(s)} .
\end{equation}
The coefficient $\eta_{t}(m_{q}^{2})$ always yields positive values for the
range of integration and parameter values that are of our interest.

\subsection{Coulomb interaction term} \label{appx coulomb}

The variational state energy contributions from the Coulomb term come from
the exchange term and can be expressed in terms of the two point correlator, 
\begin{equation}
	\langle \mathcal{H}_{C} \rangle = 
	- \kappa_{C}
	\iint_{\bm{r}_{1},\bm{r}_{2}} 
	\frac{1}{|\bm{r}|} 
	\textrm{Tr}[ \Gamma(\bm{r}_{1}, \bm{r}_{2}) \,
	\Gamma(\bm{r}_{2}, \bm{r}_{1}) ] .
\end{equation}
Here $|\bm{r}| = |\bm{r}_{1} - \bm{r}_{2}|$. We use the fact,
$\textrm{Tr}[P_{q} P_{\tilde{q}}] = \delta_{q, \tilde{q}}$, to evaluate the
trace of the correlator, This indicated that the Coulomb expectation value is
independent of the angle parameters. 

Now consider the integral
\begin{multline}
	\mathcal{I}_{1} = 
	\frac{1}{(2 \pi)^{2}} 
	\frac{1}{4}
	\iint_{\bm{r}_{1}, \bm{r}_{2}} \frac{1}{|\bm{r}|}
	\Big( f_{m_{q}}(\bm{r}_{1}, \bm{r}_{2}) f_{m_{q}}(\bm{r}_{2}, \bm{r}_{1})
	\\
	+ g_{m_{q}}(\bm{r}_{2}, \bm{r}_{1}) g_{m_{q}}(\bm{r}_{1}, \bm{r}_{2})
	\Big) .
\end{multline}
Once again we plug in $f_{m_{q}}(\bm{r}_{1}, \bm{r}_{2})$ from
Eq.\eqref{f_m} and $g_{m_{q}}(\bm{r}_{1}, \bm{r}_{2})$ from
Eq.\eqref{g_m} in the above equation,
\begin{multline}
	\mathcal{I}_{1} = 
	\frac{1}{(2 \pi)^{2}} 
	\frac{m_{q}^{2}}{8 \pi} \iint_{s_{1},s_{2}}
	\frac{\me^{-(s_{1} + s_{2}) m_{q}^{2}} \, \cosh(s_{1} + s_{2})}
	{\sqrt{s_{1} \, s_{2}} \, \sinh(s_{1}) \sinh(s_{2})} \\
	\iint_{\bm{r}_{1}, \bm{r}_{2}} 
	\frac{\me^{-\frac{1}{4}|\bm{r}|^{2} 
	(\coth(s_{1}) + \coth(s_{2}))}}{|\bm{r}|} .
\end{multline}
The spatial integration involves only the magnitude of the relative position
coordinates, hence we transform the spatial integration from two position
coordinates to the center of mass and the relative coordinates. The center of
mass coordinate is trivial and yields the volume of the system and the relative
coordinate is a gaussian integral.
\begin{equation}
	\mathcal{I}_{1} = 
	\frac{\mathrm{V}}{2 \pi} \, \eta_{fg}(m_{q}^{2}) .
\end{equation}
\begin{multline}
	\eta_{fg}(m_{q}^{2}) = 
	\frac{m_{q}^{2}}{8 \sqrt{\pi}} \iint_{s_{1},s_{2}}
	\frac{\me^{-(s_{1} + s_{2}) m_{q}^{2}}}{\sqrt{s_{1} \, s_{2}}} \\
	\frac{\cosh(s_{1} + s_{2})}
	{\sqrt{\sinh(s_{1}) \sinh(s_{2}) \sinh(s_{1} + s_{2})}} .
\end{multline}

Now consider the integral,
\begin{multline}
	\mathcal{I}_{2} = 
	\frac{1}{(2 \pi)^{2}} 
	\frac{1}{4} 
	\iint_{\bm{r}_{1}, \bm{r}_{2}} 
	\frac{1}{|\bm{r}|}
	\big( b_{m_{q}}(\bm{r}_{1}, \bm{r}_{2}) d_{m_{q}}(\bm{r}_{2}, \bm{r}_{1})
	\\
	+ b_{m_{q}}(\bm{r}_{2}, \bm{r}_{1}) d_{m_{q}}(\bm{r}_{1}, \bm{r}_{2})
	\big) .
\end{multline}
Taking $b_{m_{q}}(\bm{r}_{1}, \bm{r}_{2})$ from Eq.\eqref{b_m} and
$d_{m_{q}}(\bm{r}_{1}, \bm{r}_{2})$ from Eq.\eqref{d_m} and plugging in the 
above equation,
\begin{multline}
	\mathcal{I}_{2} = 
	\frac{1}{(2 \pi)^{2}} 
	\frac{1}{32 \pi} \iint_{s_{1}, s_{2}}
	\frac{\me^{-s_{1} m_{q}^{2}}}{\sqrt{s_{1}} \sinh^{2}(s_{1})}
	\frac{\me^{-s_{2} m_{q}^{2}}}{\sqrt{s_{2}} \sinh^{2}(s_{2})} \\
	\iint_{\bm{r}_{1}, \bm{r}_{2}} |\bm{r}| \, 
	\me^{-\frac{1}{4}|\bm{r}|^{2} 
	(\coth(s_{1}) + \coth(s_{2}))} .
\end{multline}
Once again the spatial integration is done by transforming to the center of
mass and relative coordinates and is done analytically using gamma functions.
The integrand has a leading contribution independent of variational parameter,
a similar situation was seen in the computation of the expectation value of the
kinetic term.  We drop the contributions that are independent of the
variational parameter, 
\begin{equation}
	\mathcal{I}_{2}= 
	\frac{\mathrm{V}}{2 \pi \ell_{c}^{2}} \, \eta_{bd}(m_{q}^{2}) .
\end{equation}
\begin{multline}
	\eta_{bd}(m_{q}^{2}) = 
	\iint_{s_{1}, s_{2}}
	\frac{(\me^{-(s_{1}+s_{2}) m_{q}^{2}} - 1)}
	{\sqrt{s_{1}} \sqrt{s_{2}} \sinh^{2}(s_{1}) \sinh^{2}(s_{2})} \\
	\frac{1} {(\coth(s_{1}) + \coth(s_{2}))^{\frac{3}{2}}} .
\end{multline}

The net variational parameter dependence of Coulomb term energy density is 
expressed as,
\begin{equation} 
	\mathcal{E}_{C} 
	= \frac{1}{2 \pi} \kappa_{C} \sum_{q=1}^{4} \eta_{C_{2}}(m_{q}^{2}) .
\end{equation}
where
\begin{equation}
	\eta_{C_{2}}(m_{q}^{2}) = \eta_{fg}(m_{q}^{2}) + \eta_{bd}(m_{q}^{2}) .
\end{equation}

\section{Order parameters and lattice dictionary}
\label{appx:dict}
In this section, we provide the simplest lattice representation for the 32
order parameters discussed in Sec.\ref{subsec-op}. The lattice vectors for
the nearest neighbour sites are: $\bm{b}_{1} = 0$, $\bm{b}_{2} = \bm{e}_{2}$,
$\bm{b}_{3} = \bm{e}_{1} + \bm{e}_{2}$,  and the next nearest neighbour vectors
are: $\bm{a}_{1} = \bm{e}_{1}$, $\bm{a}_{2} = \bm{e}_{2}$, $\bm{a}_{3} =
\bm{e}_{1} + \bm{e}_{2}$. The table lists the order parameters and the 
corresponding lattice representation. Here $\sigma^{\nu}$ are $2 \times 2$
identity and Pauli matrices for index $\nu = 0,x,y,z$ respectively.

\begin{widetext}
\begin{tabular}{|c|c|}
	\hline 
	Order parameter   & Lattice operator \\
	\hline \hline 
	$\mathcal{T}^{0 \nu}(\bm{x}) $ &
	$ \Big(c_{1, \bm{n}}^{\dagger} \, \sigma^{\nu} \, c_{1, \bm{n}} 
	+ c_{2, \bm{n}}^{\dagger} \, \sigma^{\nu} \, c_{2, \bm{n}} \Big) 
	+ \textrm{h.c} $ \\
	\hline
	$\mathcal{T}^{3 \nu}(\bm{x}) $ &
	$ -\mi \displaystyle{ \Big( c_{1, \bm{n}}^{\dagger} \, \sigma^{\nu} 
		\big( \sum_{j=1}^{3} c_{1, \bm{n} + \bm{a}_{j}} \big) 
		+ c_{2, \bm{n}}^{\dagger} \, \sigma^{\nu}  
	\big( \sum_{j=1}^{3} c_{2, \bm{n} + \bm{a}_{j}} \big) \Big) }
	+ \textrm{h.c} $ \\
	\hline
	$\mathcal{T}^{1 \nu}(\bm{x}) $ &
	$ \Big( 
	\displaystyle{ \me^{\mi \bm{K}_{+} \cdot \bm{n}} c^{\dagger}_{\bm{n},1}
	\sigma^{\nu}
	\sum_{j=1}^{3} \me^{-\mi \bm{K}_{-} \cdot (\bm{n}+\bm{b}_{j})} 
	c_{\bm{n}+\bm{b}_{j},2} 
	- \me^{-\mi \bm{K}_{+} \cdot \bm{n}} c^{\dagger}_{\bm{n},1}
	\sigma^{\nu}
	\sum_{j=1}^{3} \me^{\mi \bm{K}_{-} \cdot (\bm{n}+\bm{b}_{j})} 
	c_{\bm{n}+\bm{b}_{j},2} } \Big) + \textrm{h.c} $ \\
	\hline
	$\mathcal{T}^{2 \nu}(\bm{x}) $ &
	$ -\mi \Big( 
	\displaystyle{ \me^{\mi \bm{K}_{+} \cdot \bm{n}} c^{\dagger}_{\bm{n},1}
	\sigma^{\nu}
	\sum_{j=1}^{3} \me^{-\mi \bm{K}_{-} \cdot (\bm{n}+\bm{b}_{j})} 
	c_{\bm{n}+\bm{b}_{j},2} 
	- \me^{-\mi \bm{K}_{+} \cdot \bm{n}} c^{\dagger}_{\bm{n},1}
	\sigma^{\nu}
	\sum_{j=1}^{3} \me^{\mi \bm{K}_{-} \cdot (\bm{n}+\bm{b}_{j})} 
	c_{\bm{n}+\bm{b}_{j},2} } \Big) + \textrm{h.c} $ \\
	\hline  \hline
	$\widetilde{\mathcal{T}}^{0 \nu}(\bm{x}) $ &
	$ -\mi \displaystyle{ \Big( c_{1, \bm{n}}^{\dagger} \, \sigma^{\nu} 
		( \sum_{j=1}^{3} c_{1, \bm{n} + \bm{a}_{j}} ) 
		- c_{2, \bm{n}}^{\dagger} \, \sigma^{\nu} 
	( \sum_{j=1}^{3} c_{2, \bm{n} + \bm{a}_{j}} ) \Big) }
	- \textrm{h.c} $ \\
	\hline
	$\widetilde{\mathcal{T}}^{3 \nu}(\bm{x}) $ &
	$ \Big(c^{\dagger}_{\bm{n},1}  \, \sigma^{\nu} \, c_{\bm{n},1} 
	- c^{\dagger}_{\bm{n},2} \, \sigma^{\nu} \, c_{\bm{n},2} \Big) 
	+ \textrm{h.c} $ \\
	\hline
	$\widetilde{\mathcal{T}}^{1 \nu}(\bm{x}) $ &
	$ \Big( 
	\displaystyle{ \me^{\mi \bm{K}_{+} \cdot \bm{n}} c^{\dagger}_{\bm{n},1}
	\sigma^{\nu}
	\sum_{j=1}^{3} \me^{-\mi \bm{K}_{-} \cdot (\bm{n}+\bm{b}_{j})} 
	c_{\bm{n}+\bm{b}_{j},2} 
	+ \me^{-\mi \bm{K}_{+} \cdot \bm{n}} c^{\dagger}_{\bm{n},1}
	\sigma^{\nu}
	\sum_{j=1}^{3} \me^{\mi \bm{K}_{-} \cdot (\bm{n}+\bm{b}_{j})} 
	c_{\bm{n}+\bm{b}_{j},2} } \Big) + \textrm{h.c} $ \\
	\hline
	$\widetilde{\mathcal{T}}^{2 \nu}(\bm{x}) $ &
	$ -\mi \Big( 
	\displaystyle{ \me^{\mi \bm{K}_{+} \cdot \bm{n}} c^{\dagger}_{\bm{n},1}
	\sigma^{\nu}
	\sum_{j=1}^{3} \me^{-\mi \bm{K}_{-} \cdot (\bm{n}+\bm{b}_{j})} 
	c_{\bm{n}+\bm{b}_{j},2} 
	+ \me^{-\mi \bm{K}_{+} \cdot \bm{n}} c^{\dagger}_{\bm{n},1}
	\sigma^{\nu}
	\sum_{j=1}^{3} \me^{\mi \bm{K}_{-} \cdot (\bm{n}+\bm{b}_{j})} 
	c_{\bm{n}+\bm{b}_{j},2} }  \Big) + \textrm{h.c} $ \\
	\hline  \hline
\end{tabular}
\end{widetext}

\bibliography{su4-sb}

\end{document}